\documentclass[11pt,a4paper]{article}
\pdfoutput=1
\usepackage{jheppub}
\usepackage[T1]{fontenc}

\usepackage{amsmath}
\usepackage{mathtools}
\usepackage{bm}
\usepackage{graphicx}
\usepackage{esvect}
\usepackage{fancyhdr}
\usepackage{graphicx}
\usepackage{graphics}
\usepackage{amssymb}
\usepackage{enumitem}
\usepackage{braket}
\usepackage{xcolor}
\usepackage[colorlinks = true]{hyperref}
\usepackage{subfigure}
\usepackage{tikz}
\usetikzlibrary{decorations.pathreplacing} 

\definecolor{egg}{rgb}{.98,.97,.92}
\definecolor{astroorange}{rgb}{1,.93,.79}
\definecolor{darkorange}{rgb}{1,.89,.6}
\definecolor{dullblue}{rgb}{.29,.47,.77}
\definecolor{grayblue}{rgb}{.98,.98,.98}
\definecolor{fadedblue}{rgb}{.78,.86,.92}
\definecolor{tiffanyblue}{rgb}{.96,1,1}
\definecolor{charcoal}{rgb}{.247,.259,.27}
\definecolor{dullred}{rgb}{.831,.333,.333}
\definecolor{softred}{rgb}{.929,.698,.698}
\definecolor{fadedred}{rgb}{.96,.847,.847}

\newcommand{\Hr}{\mathrm{Haar}}
\newcommand{\tr}[1]{\mathrm{Tr}\left\{#1\right\}}
\newcommand{\sw}{\mathbb{S}}

\newcommand{\ot}{\mathrm{OTOC}}
\newcommand{\av}[1]{\underset{\tiny{#1}}{\mathbb{E}}}
\newcommand{\norm}[1]{\left|\left| #1 \right|\right|}
\newcommand{\abs}[1]{\left| #1 \right|}

\newcommand{\xpos}{0} 
\newcommand{\ypos}{0} 
\newcommand{\xrel}{1.1} 
\newcommand{\yrel}{.5} 
\newcommand{\height}{2em} 
\newcommand{\width}{2em} 
\newcommand{\name}{} 
\newcommand{\nodenum}{0} 
\newcommand{\heightsingle}{2em} 
\newcommand{\heightdouble}{4.6em} 
\newcommand{\heighttrace}{1.3*\heightsingle} 
\newcommand{\widthsingle}{2em} 
\newcommand{\rowspace}{2*\yrel} 

\newtheorem{theorem}{Theorem}
\newtheorem{lemma}{Lemma}
\newtheorem{prop}{Proposition}
\newtheorem{corollary}{Corollary}

\title{Quantifying scrambling in quantum neural networks}

\author{Roy J. Garcia,}
\author{Kaifeng Bu}
\author{and Arthur Jaffe}
\affiliation{Harvard University, \\Cambridge, Massachusetts 02138, USA}

\emailAdd{roygarcia@g.harvard.edu}
\emailAdd{kfbu@fas.harvard.edu}
\emailAdd{arthur\textunderscore jaffe@harvard.edu}

\date{\today}

\abstract{
We quantify the role of scrambling in quantum machine learning. We characterize a quantum neural network's (QNNs) error in terms of the network's scrambling properties via the out-of-time-ordered correlator (OTOC). A network can be trained by minimizing a loss function. We show that the loss function can be bounded by the OTOC. We prove that the gradient of the loss function can be bounded by the gradient of the OTOC. This demonstrates that the OTOC landscape regulates the trainability of  a QNN. We show numerically that this landscape is flat for maximally scrambling QNNs, which can pose a challenge to training. Our results pave the way for the exploration of quantum chaos in quantum neural networks.
}


 \arxivnumber{2112.01440}
 
\begin{document}
\maketitle
\flushbottom

\section{Introduction}
A quantum neural network (QNN) \cite{Schuld_2015, Wan_2017,Biamonte_2017, farhi2018classification, Cong_2019,Beer_2020} is a  quantum generalization of a classical neural network \cite{LeCunDeepLearning,tompson2014joint,10.1145/3065386} used to learn or optimize functions. QNNs are of growing interest because of their potential to provide a quantum speed-up \cite{Paparo_2014} and they present a promising application for near-term intermediate scale quantum devices \cite{Preskill_2018}. The role of classical chaos in classical neural networks has long been established  \cite{PhysRevLett.61.259,Wang9467, POTAPOV2000310, PhysRevLett.120.024102}, with more recent results linking chaos to expressivity \cite{poole2016exponential}. 

As the body of literature on QNNs grows, one question remains: what role does quantum chaos play in QNNs? Bridging these currently disjoint fields is necessary to understand the capabilities of QNNs in learning the chaotic properties of many-body systems, as recently demonstrated in \cite{PhysRevB.101.064406}. The connection between chaos and the generalization capability of QNNs has been explored in \cite{Choudhury_2021}, and the relation between chaos and approximation properties of QNNs has been studied via the Loschmidt echo in \cite{wu2021expressivity}. There is a growing interest in quantifying the role of quantum chaos in QNNs and further investigation is needed to rigorously establish this connection. 

In this work, we relate chaos to QNNs by establishing upper and lower bounds on training error in terms of quantum scrambling. Scrambling measures the delocalization of quantum information arising from chaotic evolution and is hence a measure of quantum chaos~\cite{Shenker_2014, Lewis_Swan_2019_Review,Lewis_Swan_2019,Swingle_2016,PhysRevX.9.031048,LiuJHEP,Leone2021}. It was recently shown that QNNs encounter barren plateaus, exponentially vanishing gradients in the cost function, when learning scrambling unitaries \cite{Holmes_2021}. Hence, scrambling plays an important role in training. However, QNNs themselves may also have chaotic properties which characterize their learning ability.
These properties have been investigated through scrambling measures such as the tripartite mutual information~\cite{Shen_2020} and operator size~\cite{wu2020scrambling}. Numerical evidence correlating the tripartite mutual information to the network's empirical training error has been demonstrated in \cite{Shen_2020}. One contribution of our work involves relating the tripartite mutual information to the network's true error via an inequality. The scrambling ability of QNNs has also been numerically linked to the design of efficient network architectures  \cite{wu2020scrambling}. However,  much needed analytic relations between scrambling measures and QNN training error largely remain unestablished. Our contribution is to establish a number of inequalities relating the two.

We demonstrate that training error can be bounded by the out-of-time-ordered correlator (OTOC), defined in the following section. This correlator is an essential tool in the study of chaos, as it can characterize fast scramblers \cite{PhysRevLett.70.3339,Kitaev2015,Susskind_FastScramblers,Maldacena_2016} and has even been used to decode the Hayden-Preskill protocol \cite{Hayden_2007,yoshida2017efficient}. In our context, we use the OTOC to quantify how well a QNN architecture scrambles information. We show that learning a unitary requires learning its scrambling properties.

Our main result, given in Theorem~\ref{TheoremOTOC}, shows that training is regulated by the gradient of the OTOC. In other words, trainability is regulated by the OTOC landscape. Hence, training depends on how a network's scrambling ability changes as its training parameters are perturbed. We provide numerical simulations to support the relevance of our analytic bounds. We show that when the QNN is maximally scrambling, the OTOC landscape is flat, which can pose a challenge to training. 

\subsection{Background on scrambling}
Here, we introduce the out-of-time-ordered correlator as a scrambling measure. A common definition of scrambling is the growth of the Hilbert-Schmidt norm of the commutator between two local Pauli strings that initially commute, as one operator evolves under the action of the  unitary Heisenberg group $U(t)$~\cite{Roberts_2015,PhysRevB.95.060201,PhysRevD.96.065005,PRXQuantum.2.020339}. A Pauli string is the tensor product of local Pauli operators.

Let $O_A$ and $O_D$ be two commuting, local Pauli strings on systems $A$ and $D$, respectively. Define $O_D(t)=U^\dagger(t)O_D U(t)$. The Hilbert-Schmidt norm of the commutator can be expressed as
\begin{equation}\label{Eq:Commutator}
\begin{split}
||[O_D(t)&,O_A]||_{\mathrm{HS}}=\sqrt{2d_\mathrm{tot}\left(1-\langle O_D(t)O_AO_D(t)O_A\rangle\right)}\;.
\end{split}
\end{equation}
The expectation value $\langle \bm{\cdot}\rangle$ is taken with respect to the $N$-qubit maximally mixed state $\frac{I}{d_\mathrm{tot}}$, and $d_\mathrm{tot}=2^N$. The quantity $\langle O_D(t)O_A O_D(t)O_A\rangle$ in Eq.~\eqref{Eq:Commutator} is an out-of-time-ordered correlator.
Although troublesome to measure, protocols to do this have been constructed~\cite{G_rttner_2017,PhysRevX.7.031011,Landsman_2019,PhysRevLett.124.240505, PhysRevResearch.3.033155}. 

To simplify notation, we suppress the variable $t$ and write $U$ for $U(t)$. We remove the dependence on the choice of $O_A$ and $O_D$ by redefining  the out-of-time-ordered correlator as the average 
\begin{equation}\label{Eq:OTOC}
	\overline{\ot}(U)=\av{O_A}\av{O_D}\langle UO_AU^\dagger O_D UO_AU^\dagger O_D \rangle \;,
\end{equation}
where (unless otherwise state) $\av{O_A}$ (or $\av{O_D}$) denotes an average over the Pauli group on system $A$ (or $D$). A signature of chaos is that the OTOC decays to a floor value and the Hilbert-Schmidt norm reaches a maximum at large time; see Corollary~\ref{Corollary:Us}. We define $U$ to be maximally scrambling if the OTOC decays to this floor value.

\section{Preliminaries}\label{Sec:Prelims}

\begin{figure}[t]
\centering
\scalebox{1}{
\begin{tikzpicture}

    \renewcommand{\nodenum}{v1}
    \renewcommand{\name}{}
	\renewcommand{\xpos}{2.5*\xrel}
    \renewcommand{\ypos}{-\yrel}
    \renewcommand{\height}{13.5em}
    \renewcommand{\width}{12.5em}
    \node[rectangle, dashed, line width=.35mm, fill=white,  rounded corners=1.1em, minimum width=\width, minimum height=\height, draw=charcoal] (\nodenum) at (\xpos,\ypos) {\name};
    
    \draw [thick,color=charcoal]
    (0,\yrel+\rowspace)--(5*\xrel,\yrel+\rowspace)
    (5*\xrel,1.1*\yrel+\rowspace)--(6*\xrel,1.1*\yrel+\rowspace)
    (5*\xrel,0.9*\yrel+\rowspace)--(6*\xrel,0.9*\yrel+\rowspace)
    (0,\yrel)--(5*\xrel,\yrel)
    (5*\xrel,1.1*\yrel)--(6*\xrel,1.1*\yrel)
    (5*\xrel,0.9*\yrel)--(6*\xrel,0.9*\yrel)
    (0,-\yrel)--(5*\xrel,-\yrel)
    (5*\xrel,-1.1*\yrel)--(6*\xrel,-1.1*\yrel)
    (5*\xrel,-0.9*\yrel)--(6*\xrel,-0.9*\yrel)
    (0,-\yrel-\rowspace)--(6*\xrel,-\yrel-\rowspace)
    (0,-\yrel-2*\rowspace)--(6*\xrel,-\yrel-2*\rowspace);
    
	\renewcommand{\xpos}{-.2*\xrel}
	\draw [decorate,line width=.75pt,color=charcoal,decoration={brace,amplitude=5pt},xshift=\xrel,yshift=-\rowspace]	(\xpos,-\yrel-2*\rowspace) -- (\xpos,\yrel) node [black,midway,xshift=9pt] {};

  	\renewcommand{\nodenum}{v2}
    \renewcommand{\name}{$A$}
	\renewcommand{\xpos}{-1.55*\xrel}
    \renewcommand{\ypos}{3*\yrel}
    \renewcommand{\height}{\heightsingle}
    \renewcommand{\width}{\widthsingle}
    \node[rectangle,, line width=.35mm, fill=egg, rounded corners, minimum width=\width, minimum height=\height, draw=dullblue] (\nodenum) at (\xpos,\ypos) {\name}; 
    
	\renewcommand{\nodenum}{v3}
    \renewcommand{\name}{$B$}
	\renewcommand{\xpos}{-1.55*\xrel}
    \renewcommand{\ypos}{-\rowspace}
    \renewcommand{\height}{9.8em}
    \renewcommand{\width}{\widthsingle}
    \node[rectangle,, line width=.35mm, fill=egg, rounded corners, minimum width=\width, minimum height=\height, draw=dullblue] (\nodenum) at (\xpos,\ypos) {\name}; 
    
	\renewcommand{\nodenum}{v4}
    \renewcommand{\name}{ $\rho_{d,A}$}
	\renewcommand{\xpos}{-.75*\xrel}
    \renewcommand{\ypos}{3*\yrel}
    \renewcommand{\height}{\heightsingle}
    \renewcommand{\width}{\widthsingle}
    \node[](\nodenum) at (\xpos,\ypos) {\name};
    
	\renewcommand{\nodenum}{v5}
    \renewcommand{\name}{$\rho_{B}$}
	\renewcommand{\xpos}{-.75*\xrel}
    \renewcommand{\ypos}{-\rowspace}
    \renewcommand{\height}{9.7em}
    \renewcommand{\width}{\widthsingle}
    \node[](\nodenum) at (\xpos,\ypos) {\name};  
	
	\renewcommand{\nodenum}{v6}
    \renewcommand{\name}{$U_{1,1}$}
	\renewcommand{\xpos}{\xrel}
    \renewcommand{\ypos}{\rowspace}
    \renewcommand{\height}{\heightdouble}
    \renewcommand{\width}{\widthsingle}
    \node[rectangle, fill=white,  line width =.3mm,rounded corners, minimum width=\width, minimum height=  	    \height, draw=charcoal] (\nodenum) at (\xpos,\ypos) {\name};
    
    \renewcommand{\nodenum}{v7}
    \renewcommand{\name}{$U_{1,2}$}
	\renewcommand{\xpos}{\xrel}
    \renewcommand{\ypos}{-\rowspace}
    \renewcommand{\height}{\heightdouble}
    \renewcommand{\width}{\widthsingle}
    \node[rectangle, fill=white,  line width =.3mm,rounded corners, minimum width=\width, minimum height=  	    \height, draw=charcoal] (\nodenum) at (\xpos,\ypos) {\name};
    
    \renewcommand{\nodenum}{v8}
    \renewcommand{\name}{$U_{2,1}$}
	\renewcommand{\xpos}{2*\xrel}
    \renewcommand{\ypos}{0}
    \renewcommand{\height}{\heightdouble}
    \renewcommand{\width}{\widthsingle}
    \node[rectangle, fill=white,  line width =.3mm,rounded corners, minimum width=\width, minimum height=  	    \height, draw=charcoal] (\nodenum) at (\xpos,\ypos) {\name};
    
    \renewcommand{\nodenum}{v9}
    \renewcommand{\name}{$U_{2,2}$}
	\renewcommand{\xpos}{2*\xrel}
    \renewcommand{\ypos}{-2*\rowspace}
    \renewcommand{\height}{\heightdouble}
    \renewcommand{\width}{\widthsingle}
    \node[rectangle, fill=white,  line width =.3mm,rounded corners, minimum width=\width, minimum height=  	    \height, draw=charcoal] (\nodenum) at (\xpos,\ypos) {\name};

	\renewcommand{\nodenum}{v10}
    \renewcommand{\name}{$U_{3,1}$}
	\renewcommand{\xpos}{3*\xrel}
    \renewcommand{\ypos}{\rowspace}
    \renewcommand{\height}{\heightdouble}
    \renewcommand{\width}{\widthsingle}
    \node[rectangle, fill=white,  line width =.3mm,rounded corners, minimum width=\width, minimum height=  	    \height, draw=charcoal] (\nodenum) at (\xpos,\ypos) {\name};
    
    \renewcommand{\nodenum}{v11}
    \renewcommand{\name}{$U_{3,2}$}
	\renewcommand{\xpos}{3*\xrel}
    \renewcommand{\ypos}{-\rowspace}
    \renewcommand{\height}{\heightdouble}
    \renewcommand{\width}{\widthsingle}
    \node[rectangle, fill=white,  line width =.3mm,rounded corners, minimum width=\width, minimum height=  	    \height, draw=charcoal] (\nodenum) at (\xpos,\ypos) {\name};
    
    \renewcommand{\nodenum}{v12}
    \renewcommand{\name}{$U_{4,1}$}
	\renewcommand{\xpos}{4*\xrel}
    \renewcommand{\ypos}{0}
    \renewcommand{\height}{\heightdouble}
    \renewcommand{\width}{\widthsingle}
    \node[rectangle, fill=white,  line width =.3mm,rounded corners, minimum width=\width, minimum height=  	    \height, draw=charcoal] (\nodenum) at (\xpos,\ypos) {\name};
    
    \renewcommand{\nodenum}{v13}
    \renewcommand{\name}{$U_{4,2}$}
	\renewcommand{\xpos}{4*\xrel}
    \renewcommand{\ypos}{-2*\rowspace}
    \renewcommand{\height}{\heightdouble}
    \renewcommand{\width}{\widthsingle}
    \node[rectangle, fill=white,  line width =.3mm,rounded corners, minimum width=\width, minimum height=  	    \height, draw=charcoal] (\nodenum) at (\xpos,\ypos) {\name};    
    
    \renewcommand{\nodenum}{v14}
    \renewcommand{\name}{\includegraphics[width=.025\textwidth]{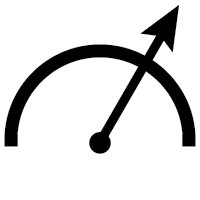}}
	\renewcommand{\xpos}{5.25*\xrel}
    \renewcommand{\ypos}{\yrel+\rowspace}
    \renewcommand{\height}{\heightsingle}
    \renewcommand{\width}{\widthsingle}
    \node[rectangle, fill=white,  line width =.3mm,rounded corners, minimum width=\width, minimum height=  	    \height, draw=charcoal] (\nodenum) at (\xpos,\ypos) {\name};
    
    \renewcommand{\nodenum}{v15}
    \renewcommand{\name}{\includegraphics[width=.025\textwidth]{Measure.png}}
	\renewcommand{\xpos}{5.25*\xrel}
    \renewcommand{\ypos}{\yrel}
    \renewcommand{\height}{\heightsingle}
    \renewcommand{\width}{\widthsingle}
    \node[rectangle, fill=white,  line width =.3mm,rounded corners, minimum width=\width, minimum height=  	    \height, draw=charcoal] (\nodenum) at (\xpos,\ypos) {\name};
    
    \renewcommand{\nodenum}{v16}
    \renewcommand{\name}{\includegraphics[width=.025\textwidth]{Measure.png}}
	\renewcommand{\xpos}{5.25*\xrel}
    \renewcommand{\ypos}{-\yrel}
    \renewcommand{\height}{\heightsingle}
    \renewcommand{\width}{\widthsingle}
    \node[rectangle, fill=white,  line width =.3mm,rounded corners, minimum width=\width, minimum height=  	    \height, draw=charcoal] (\nodenum) at (\xpos,\ypos) {\name};      
    
	\renewcommand{\nodenum}{v17}
    \renewcommand{\name}{$C$}
	\renewcommand{\xpos}{6.55*\xrel}
    \renewcommand{\ypos}{\yrel}
    \renewcommand{\height}{7.3em}
    \renewcommand{\width}{\widthsingle}
    \node[rectangle,, line width=.35mm, fill=egg, rounded corners, minimum width=\width, minimum height=\height, draw=dullblue] (\nodenum) at (\xpos,\ypos) {\name};
    
	\renewcommand{\nodenum}{v18}
    \renewcommand{\name}{$D$}
	\renewcommand{\xpos}{6.55*\xrel}
    \renewcommand{\ypos}{-2*\rowspace}
    \renewcommand{\height}{\heightdouble}
    \renewcommand{\width}{\widthsingle}
    \node[rectangle,, line width=.35mm, fill=egg, rounded corners, minimum width=\width, minimum height=\height, draw=dullblue] (\nodenum) at (\xpos,\ypos) {\name};

    \renewcommand{\nodenum}{v20}
    \renewcommand{\name}{$U$}
	\renewcommand{\xpos}{2.5*\xrel}
    \renewcommand{\ypos}{2.5*\rowspace}
    \renewcommand{\height}{\heightsingle}
    \renewcommand{\width}{\widthsingle}
    \node[] (\nodenum) at (\xpos,\ypos) {\name};
    
    \draw [-stealth,thick,color=charcoal](\xrel,-3.5*\rowspace) -- (4*\xrel,-3.5*\rowspace);
    
    \renewcommand{\nodenum}{v21}
    \renewcommand{\name}{Depth}
	\renewcommand{\xpos}{2.5*\xrel}
    \renewcommand{\ypos}{-4*\rowspace}
    \renewcommand{\height}{\heightsingle}
    \renewcommand{\width}{\widthsingle}
    \node[] (\nodenum) at (\xpos,\ypos) {\name};
\end{tikzpicture}
}
\caption{Quantum neural network with brick-wall architecture.  Data is encoded in state $\rho_{d,A}$, while system $B$ is initialized in the maximally mixed state, $\rho_B$.  The QNN unitary $U$ is composed of 2-qubit, parameterized unitaries, $U_{i,j}$.  Output system $C$ is measured after evolution by $U$.}
\label{Fig:QNN}

\end{figure}
In this section, we briefly review some basics of QNNs. A QNN is a parameterized quantum circuit with unitary $U(\bm{\theta})$ and parameters $\bm{\theta}$. There are two disjoint input subsystems $A$, $B$ and two disjoint output subsystems $C$, $D$ each with $N_{S'}$  qubits and Hilbert space dimension $d_{S'}=2^{N_{S'}}$, where $S'\in\{A,B,C,D\}$ (see Fig.~\ref{Fig:QNN}). A Pauli string acting on system $S'$ is denoted as $O_{S'}$. $N$ denotes the total number of qubits in the system, and $d_{\mathrm{tot}}=2^N$ is the corresponding Hilbert space dimension.

The parameters of the QNN are tuned to train the network to either learn a target unitary, $U_S$, or optimize a cost function. We focus on unitary learning in this section and refer to Appendix~\ref{Sec:CostMain} for a cost function treatment. We train the QNN with data $S=\Big\{\rho_d,y_d\Big\}_{d=1}^{n_S}$, where $d$ denotes each data point and $n_s$ is the total number of points. State $\rho_d$ encodes the input data and $y_d$ is the corresponding target function. $U(\bm{\theta})$ approximates $y_d$ by computing the output function $\tilde{y}_d$. We use the standard notation for the target function and output function: 
\begin{equation}\label{Eq:TargetFunctions}
	y_d=\tr{U_S\rho_dU_S^\dagger O_C}, \quad 	\tilde{y}_d=\tr{U\rho_d U^\dagger O_C}.
\end{equation}
The functions are expectation values of Pauli string $O_C$ on system $C$ with respect to the input state evolved with either $U_S$ or $U$, respectively. 

To determine the accuracy with which $U$ approximates $U_S$, we define the loss function:
\begin{equation}\label{Eq:LossDef}
L_d=\av{O_C}\abs{\tilde{y}_d-y_d}^2.
\end{equation}
Although it is common to measure the loss function with respect to one observable $O_C$, we take the average over all Pauli strings, as this will help establish a connection between $L_d$ and the OTOC. The case where the loss function is defined with respect to one observable is considered in Appendix~\ref{Sec:GeneralizeError}. The true error $L$ is defined as the average of the loss function over the data set $S$: 
\begin{equation}
	L=\av{d}L_d.
\end{equation} 
The empirical error is defined as the average of the loss function over a sample set, which is a finite subset of $S$. Practically, the QNN is trained by optimizing its parameters to minimize the empirical error. We assume $L$ is approximated sufficiently well by the empirical error.

Assume $\rho_d$ is local in system $A$, such that
\begin{equation}\label{Eq:rho}
	\rho_d=\rho_{d,A}\otimes \rho_B.
\end{equation}
The input data is encoded in pure state ${\rho_{d,A}=\ket{\psi_{d,A}}\bra{\psi_{d,A}}=U_A\ket{0}\bra{0}_AU^\dagger_A}$ using unitary $U_A$, while system $B$ is prepared in the maximally mixed state, $\rho_B=\frac{I_B}{d_B}$. Take $U_A$ to be a Haar random unitary sampled from the unitary group on $A$. The true error becomes an average over all uniformly distributed input states, $\rho_{d,A}$: 
 $L=\int_{\Hr}dU_AL_d$. Since only a 2-design is required, the average over $U_A$ can be taken over the Clifford group, which forms a 3-design \cite{10.5555/3179439.3179447,PhysRevA.96.062336}. 
 
In this work, we quantify the role of quantum chaos in QNNs. 
Chaos in QNNs has been explored  through the fidelity OTOC \cite{wu2021expressivity}, which has the general form $\abs{\bra{\psi}U^\dagger M U\ket{\psi}}^2$ \cite{PhysRevLett.124.160603,Chenu_2018,Lewis_Swan_2019}.  However, it was recently proposed that higher-point correlators can reveal the finer-grained dynamics of chaos \cite{Shenker_2014_shocks,PhysRevResearch.3.033155, PRXQuantum.2.010329, Roberts_2017}. Since the fidelity OTOC carries the same information as the 2-point correlator $\abs{\bra{\psi}U^\dagger M U\ket{\psi}}$, it may not reveal the finer scrambling dynamics available to the 4-point OTOC in Eq.~\ref{Eq:OTOC}. Therefore, we rely on $\overline{\ot}(U)$ to study scrambling in QNNs.

\section{Main results}
Let us first introduce the connection between the training error and the out-of-time-ordered correlator. In the following proposition, we write both the loss function and true error in terms of out-of-time-ordered correlators.
\begin{prop}\label{Prop:LossandError}

The loss function can be written as
\begin{equation}\label{Eq:LossCorr}
    L_d=d_D^2(C_d(U,U)+C_d(U_S,U_S)-2C_d(U,U_S))\;,
\end{equation}
where $C_d(U_1,U_2)$ is 
\begin{align}
    C_d(U_1,U_2)&=\av{O_D}\langle U_1\rho_d U_1^\dagger O_D^\dagger U_2 \rho_d U_2^\dagger O_D\rangle\;\label{Eq:LossOTOCdef}.
\end{align}
It follows that the true error can be written in terms of OTOCs:
\begin{align}
	L&=G\Big[\overline{\ot}(U)+\overline{\ot}(U_S)-2\mathrm{OP}(U,U_S) \Big]\;,\label{Eq:LossOTOC}
\end{align}
where  $\mathrm{OP}(U,U_S)$ and $G$ are defined by
\begin{align}
 \mathrm{OP}(U,U_S)&=\av{O_A}\av{O_D}\langle UO_AU^\dagger O_DU_S O_AU_S^\dagger O_D\rangle,\\
	G&=\frac{d_A^2}{(d_A+1)d_C^2}.
\end{align} 
\end{prop}

We give the proof of this proposition in Appendix~\ref{ProofofLossandError}. The expressions $C_d(U,U)$ and $C_d(U_S,U_S)$ have the form of out-of-time-ordered correlators, but they are sub-optimal scrambling measures since $\rho_d$ is non-unitary. Hence, we will later rely on the trusted scrambling measure $\overline{\mathrm{OTOC}}(U)$ to bound $L_d$ in Proposition~\ref{Prop:LossBound}. The first two terms in Eq.~\eqref{Eq:LossOTOC} depend only on the scrambling ability of $U$ and $U_S$, respectively. The \textit{optimization correlator}, $\mathrm{OP}(U,U_S)$, reflects the optimization of $U$ with respect to $U_S$ when learning. In Proposition~\ref{Prop:true-error}, we will establish bounds on $L$ which are independent of $\mathrm{OP}(U,U_S)$.

In the following corollary, we focus on the special case where the target unitary is maximally scrambling, as this is physically relevant when the QNN learns the large-time dynamics of a chaotic many-body system. As $U_S$ becomes more scrambling, the true error approaches the value obtained by integrating $U_S$ over the Haar measure on the unitary group \cite{Roberts_2017}, denoted as $L_{\mathrm{scram}} =\int_{\Hr} dU_S L$. When both $U$ and $U_S$ are maximally scrambling and independent of each other, the true error reaches a floor value of $L_{\mathrm{floor}}=\int_{\Hr}dUdU_S L$. We prove the following corollary in Appendix~\ref{Sec:AvgLoss}.

\begin{corollary}\label{Corollary:Us}

When $U$ is untrained and hence independent of $U_S$, $L_\mathrm{scram}$ and $L_{\mathrm{floor}}$ satisfy the following:
\begin{align}
	L_{\mathrm{scram}}&=G\left[\overline{\ot}(U) +\overline{\mathrm{\ot}}_{\mathrm{scram}}-\frac{2}{d_A^2}\right]\;,\label{Eq:AvgLoss}\\
	L_{\mathrm{floor}}&=2G\left[\overline{\ot}_{\mathrm{scram}}-\frac{1}{d_A^2}\right]\;.\label{Eq:LossFloor}
\end{align}
Here, the OTOC for a maximally scrambling unitary is 
\begin{equation}
    \overline{\mathrm{\ot}}_{\mathrm{scram}}= \frac{1}{(d_\mathrm{tot}^2-1)}\left[\frac{d_\mathrm{tot}^2}{d_A^2}-1+d_C^2\left(1-\frac{1}{d_A^2}\right)\right].\label{Eq:OTOCScram}
\end{equation}
\end{corollary}

When $d_\mathrm{tot}$ is large and $d_A>>1$, $L_{\mathrm{floor}}\rightarrow\frac{2}{d_Bd_\mathrm{tot}}$. For fixed $N_B$, the true error in this limit vanishes exponentially with the total number of qubits $N$. This is a relevant limit when learning many-body unitaries.

\subsection{Error bounds}\label{Sec:TrueErrorBounds}

In this subsection, we bound the true error and the loss function using OTOCs.

\begin{prop}\label{Prop:true-error}
The true error $L$ can be  bounded by OTOCs:
\begin{equation}
	L_{-}\leq L \leq L_+\;,
\end{equation}
where
\begin{equation}
    L_{\pm}(U,U_S)=G\left[\sqrt{\overline{\ot}(U) }\pm\sqrt{\overline{\ot}(U_S)}\right]^2\;.\label{Eq:LossBound}
\end{equation}
Also 
\begin{equation}
    \abs{L-L_{\pm}}\leq 4G\sqrt{\overline{\ot}(U)\overline{\ot}(U_S)}\;.\label{Eq:LpmIneq}
\end{equation}
\end{prop}

We prove Proposition \ref{Prop:true-error} in Appendix~\ref{Sec:OTOCProb}. $\overline{\ot}(U)$ decays as $U$ becomes more scrambling. Hence, the upper bound $L_{+}$ decays as $U$ or $U_S$ become more scrambling. The lower bound $L_-$ depends on the distance between the OTOCs of $U$ and $U_S $. This implies that $L_-$ vanishes as the QNN learns the scrambling properties of the target unitary. A mismatch in the scrambling abilities of $U$ and $U_S$ may therefore inhibit the optimization of $L$. Hence, learning a target unitary requires learning its scrambling properties. The bound in Ineq.~ \eqref{Eq:LpmIneq} decays with scrambling, causing $L$ to approach $L_{\pm}$.

At the start of training, the QNN is initialized as a random parameterized quantum circuit (RPQC) with unitary $U_0$.  Sufficiently deep random circuits are scrambling \cite{PhysRevE.99.052212,PhysRevX.8.031057} and lead to OTOC decay with circuit depth \cite{PhysRevX.8.021014}. Therefore, the initialized QNN becomes increasingly scrambling with circuit depth, causing $\overline{\ot}(U_0) $ and $L_{+}(U_0,U_S)$ to decay. 
Random local quantum circuits form approximate polynomial-designs \cite{Brand_o_2016, Harrow_2009}. Hence, when the initialized QNN becomes sufficiently deep, $\overline{\ot}(U_0)$ tends to $\overline{\mathrm{\ot}}_{\mathrm{scram}}$ in Eq.~\eqref{Eq:OTOCScram}, the value found by integrating $U_0$ over the Haar measure on the unitary group \cite{Roberts_2017};  see Appendix~\ref{Sec:AvgLoss}. The corresponding true error bounds are found by setting  ${\overline{\ot}(U_0)\rightarrow\overline{\mathrm{\ot}}_{\mathrm{scram}}}$  in Eq.~\eqref{Eq:LossBound}. 

Aside from the OTOC, the second R\'enyi entropy and the tripartite mutual information are also common measures for scrambling. We define the Choi isomorphism of $U$ as ${\rho(U)=\ket{U}\bra{U}}$, where
$
    \ket{U}=\frac{1}{\sqrt{d_{\mathrm{tot}}}}\sum_{i=1}^{d_{\mathrm{tot}}}\ket{i}\otimes U\ket{i}.
$
For a system $S'$ and its complement $S'_c$ in system $ABCD$, define the reduced density operator ${\rho_{S'}(U)=\mathrm{Tr}_{S'_c}\{\rho(U)\}}$. For example, $\rho_{AC}(U)=\mathrm{Tr}_{BD}\{\rho(U)\}$. The second R\'enyi entropy is defined as 
${S^{(2)}_{AC}(U)=-\log_2\{\rho_{AC}^2(U)\}}$. The tripartite mutual information is defined as \linebreak ${
	I_{3,U}(A{:}C{:}D)=I_{U}(A{:}C)+I_{U}(A{:}D)-I_{U}(A{:}CD)},$
where the mutual information is, for example, ${I_{U}(A:C)=S_A(U)+S_C(U)-S_{AC}(U)}$ and the entanglement entropy is \linebreak${S_{AC}(U)=-\tr{\rho_{AC}(U)\log_2\rho_{AC}(U)}}$.

The negativity of $I_{3,U}(A{:}C{:}D)$ measures how much information in $A$ is shared by $C$ and $D$ after evolution by $U$. As the system scrambles, the magnitude of the tripartite mutual information increases. 
A numerical correlation between the empirical error and the tripartite mutual information was found in ~\cite{Shen_2020}. In the following corollary, we relate the true error to the second R\'enyi entropy and the tripartite mutual information.

\begin{corollary}
The true error bounds $L_{\pm}$ can be expressed in terms of the second R\'enyi entropy $S^{(2)}_{AC}(U_i)$
with  $U_i\in\{U,U_S\}$ as follows,
\begin{equation}
    L_{\pm}= 2^{N-N_A-N_D}G\left[2^{-S^{(2)}_{AC}(U)/2} \pm 2^{-S^{(2)}_{AC}(U_S)/2}\right]^2.\label{Eq:RenyiBound}
\end{equation}
Moreover, 
the true error can be bounded by the tripartite mutual information $I_{3,U_i}(A{:}C{:}D)$:
\begin{equation}
\begin{split}\label{Eq:MutualEq}
	L\geq G\Big[2^{(I_{3,U}(A{:}C{:}D)-2N_A)/2}+2^{(I_{3,U_S}(A{:}C{:}D)-2N_A)/2}-2\mathrm{OP}(U,U_S) \Big]\;.
\end{split}
\end{equation}
\end{corollary}

The OTOC for $U$ is related to the second R\'enyi  entropy, $S^{(2)}_{AC}(U)$, through \linebreak ${\overline{\ot}(U)=2^{N-N_A-N_D-S^{(2)}_{AC}(U)}}$ \cite{Hosur_2016,yoshida2017efficient}. We use this relation along with Eq.~\eqref{Eq:LossBound} to prove Eq.~ \eqref{Eq:RenyiBound}. It has been  shown  that
$\overline{\ot}(U)\geq 2^{(I_{3,U}(A{:}C{:}D)-2N_A)/2}$~\cite{Hosur_2016}, which
implies that the tripartite mutual information lower bounds the true error via Eq.~\eqref{Eq:LossOTOC}. This yields Ineq.~\eqref{Eq:MutualEq}.


In Proposition~\ref{Prop:LossBound}, we establish a bound on the loss function in terms of $\overline{\ot}(U)$. In the proof (see Appendix~\ref{Sec:LevyBound}), we use  Levy's lemma to construct a concentration inequality for the loss function. When working with Levy's lemma, it is useful to define the following function
\begin{equation}\label{Eq:Ffunction}
    f(\epsilon)=\sqrt{\frac{9\pi^3}{2d_A}\ln\frac{2}{\epsilon}}.
\end{equation}

\begin{prop}\label{Prop:LossBound}
Let $\ket{\psi_{d,A}}$ be sampled from the Haar measure on the Hilbert space of system $A$. With probability at least $1-\epsilon$, the loss function $L_d(\ket{\psi_{d,A}})$ satisfies
\begin{equation}
	\abs{L_d-L_{\pm}}
	\leq \eta f(\epsilon)+4G\sqrt{\overline{\ot}(U)\overline{\ot}(U_S)},\label{Eq:ConcentrationBoundLpm}
\end{equation}
and 
\begin{equation}
	\abs{L_d-L}\leq \eta f(\epsilon),\label{Eq:ConcentrationBoundL}
\end{equation}
where the Lipschitz constant $\eta$ is
\begin{equation}
    \eta=8\sqrt{d_A(d_A+1)L_{+}}.
\end{equation}
\end{prop}

The Lipschitz constant decays as $U$ or $U_S$ become more scrambling, since $L_+$ is a function of OTOCs, which decay. This decay concentrates $L_d$ about $L_{\pm}$ and $L$ via Ineqs.~\eqref{Eq:ConcentrationBoundLpm} and~\eqref{Eq:ConcentrationBoundL}, respectively. Concentration occurs as a RPQC QNN becomes deeper. In the case where $U$ and $U_S$ are maximally scrambling, it is shown in Appendix~\ref{Sec:LevyBound} that for large $d_\mathrm{tot}$ and $d_D>>d_A$, the right-hand side of  Ineq.~\eqref{Eq:ConcentrationBoundL} becomes 
$\eta f(\epsilon)\rightarrow \frac{16}{d_C}\sqrt{\frac{9\pi^3}{2}\ln\frac{2}{\epsilon}}$. 
This implies that performing measurements on a larger system $C$ will concentrate $L_d$ about $L$. Also assuming $d_A>>1$, $L\rightarrow\frac{2}{d_Bd_\mathrm{tot}}$ by Eq.~\eqref{Eq:LossFloor}. This decays exponentially with $N$, so $L_d$ can be made to concentrate near 0. The loss function when learning many-body unitaries can therefore vanish.

\subsection{Gradient of loss function}\label{Sec:GradofLoss}
We now introduce our principal result regarding training. We bound the gradient of the loss function with the gradient of the OTOC. Doing so helps in performing gradient descent \cite{JMLR:v12:duchi11a, reddi2019convergence}, one of the widely used methods for training neural networks. Let us consider a QNN as in ~\cite{McClean_2018} given by $U(\bm{\theta})=\prod_{l=1}^{L}U_l(\theta_l)W_l$ where $U_l(\theta_l)=e^{-i\theta_l V_l}$, $V_l$ is  Hermitian, $W_l$ is a constant unitary, and $L$ is the circuit depth. An application of Levy's lemma yields the following theorem; see Appendix~\ref{Sec:ConcentrationGrad}.

\begin{theorem}\label{TheoremOTOC}
Let $\ket{\psi_{d,A}}$ be sampled from the Haar measure on the Hilbert space of system $A$. With probability at least $1-\epsilon$, $\partial_{\theta_l}L_d(\ket{\psi_{d,A}})$ satisfies
\begin{equation}\label{Eq:GradConcentrationIneq}
	\abs{\partial_{\theta_l}L_d-\partial_{\theta_l}L}\leq \eta_g f(\epsilon),
\end{equation}
where
\begin{equation}\label{Eq:TrueErrorGrad}
		\partial_{\theta_l}L=G\Big[\partial_{\theta_l}\overline{\ot}(U(\bm{\theta}))-2\partial_{\theta_l}\mathrm{OP}(U(\bm{\theta}),U_S) \Big],
\end{equation}
and the Lipschitz constant $\eta_g$ is
\begin{equation} \eta_g=8\norm{V_l}_\infty\left(\sqrt{d_A(d_A+1)L_{+}}+2\right).
\end{equation}
For a maximally scrambling target unitary and an untrained $U$, 
\begin{equation}\label{Eq:GradLScram}
    \partial_{\theta_l}L_{\mathrm{scram}}=G\partial_{\theta_l}\overline{\ot}(U(\bm{\theta})).
\end{equation}
\end{theorem}

The quantity 
$\partial_{\theta_l}L_d$ is probabilistically bounded by the OTOC and its gradient via Ineq.~\eqref{Eq:GradConcentrationIneq}. It is shown in Appendix~\ref{Sec:ConcentrationGrad} that \begin{equation}
\abs{\partial_{\theta_l}L}\leq\frac{8d_A^2}{(d_A+1)d_C^2} \norm{V_l}_{\infty}\;.
\end{equation} 
This bound decays exponentially with $N_C$, indicating that non-local measurements can produce vanishing gradients. This is consistent with the presence of exponentially vanishing gradients of globally defined functions found in~\cite{Cerezo_2021, liu2021presence}.

The Lipschitz constant $\eta_g$ decays as $U$ or $U_S$ become more scrambling. When $U$ and $U_S$ are maximally scrambling unitaries, Ineq.~\eqref{Eq:GradConcentrationIneq} simplifies to  \linebreak ${\abs{\partial_{\theta_l}L_d-\partial_{\theta_l}L}\leq 16\norm{V_l}_\infty\left(\frac{1}{d_C}+\frac{1}{\sqrt{d_A}}\right)\sqrt{\frac{9\pi^3}{2}\ln\frac{2}{\epsilon}} }$. This bound tightens as $d_A$ and $d_C$ increase, causing $\partial_{\theta_l}L_d$ to concentrate near $\partial_{\theta_l}L$. This corresponds to encoding the data using more qubits and performing measurements on a larger subsystem $C$. 

Eq.~\ref{Eq:GradLScram} shows that the true error gradient is given by the gradient of the OTOC. The OTOC landscape therefore regulates trainability when implementing gradient descent. The OTOC reaches a floor value for a maximally scrambling QNN and no further chaotic evolution by $U$ changes the OTOC appreciably. One therefore expects that an infinitesimal change in $\bm{\theta}$ should not perturb the OTOC, and hence $L_{\mathrm{scram}}$, from its fixed value. We present numerical simulations in Sec.~\ref{Sec:Simulations} to demonstrate this. This hints that barren plateaus, exponentially vanishing gradients \cite{McClean_2018, wang2021noiseinduced, Holmes_2021}, may potentially arise due to the flat OTOC landscape of maximally scrambling QNNs. This also suggests that weakly scrambling QNNs with $\bm{\theta}$-sensitive OTOCs may hold potential in avoiding barren plateaus. Indeed, shallow QNN architectures have been shown to circumvent barren plateaus \cite{Cerezo_2021,PhysRevX.11.041011}; these architectures are typically weak scramblers.

\section{Numerical Simulations}\label{Sec:Simulations}

\begin{figure}[h!]
    \centering
   \subfigure{\includegraphics[scale=.5]{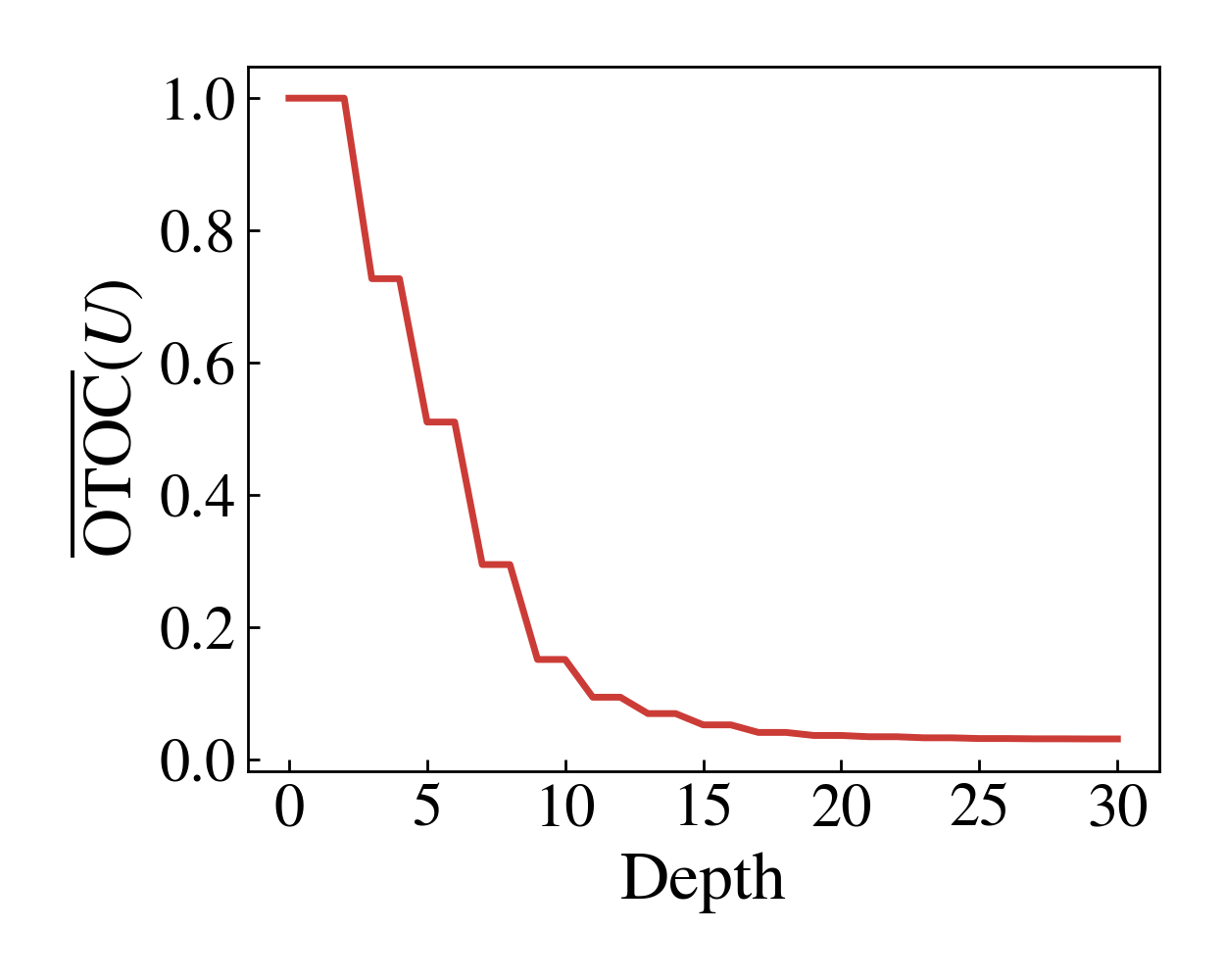}}
    \subfigure{\includegraphics[scale=.51]{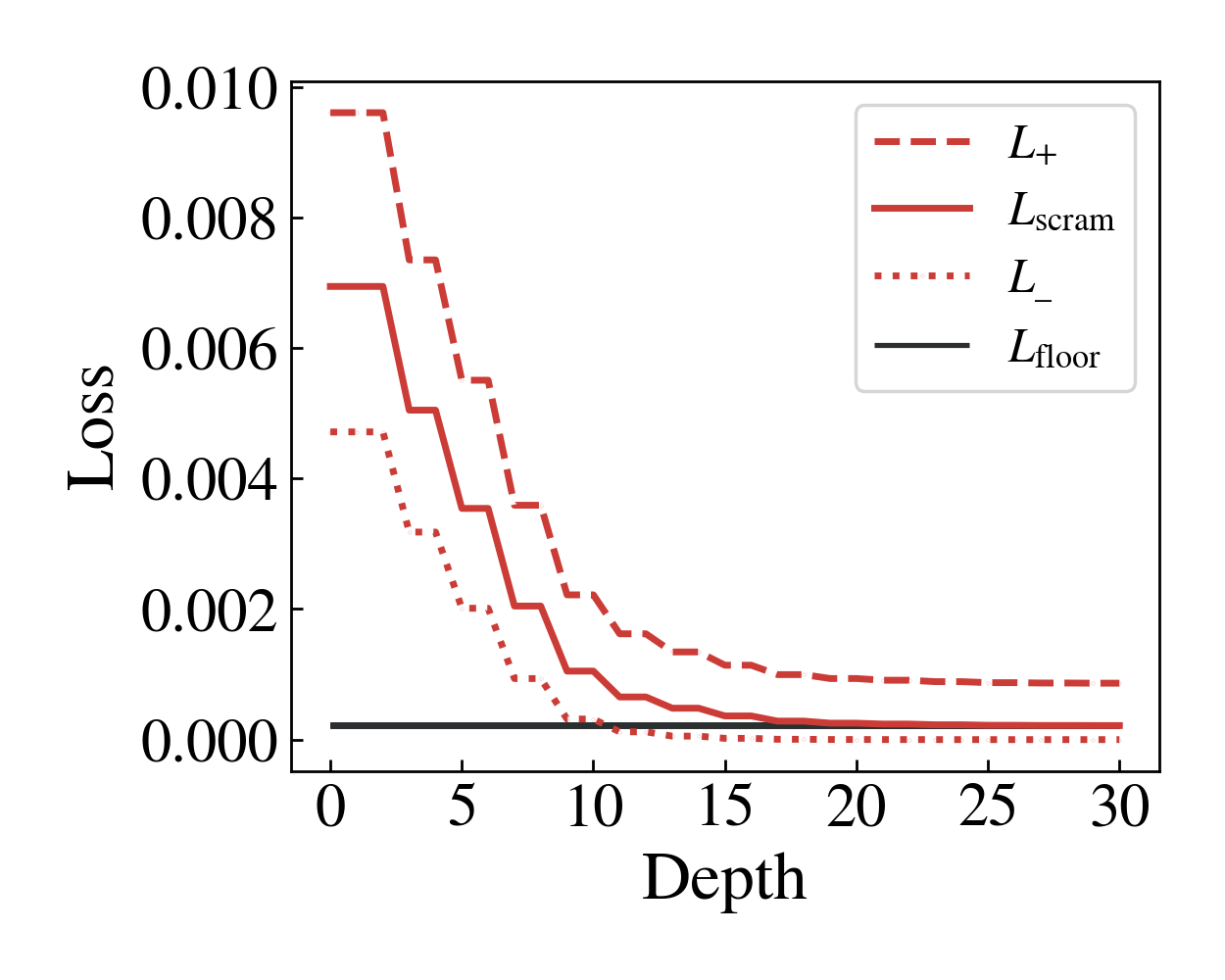}}
    \caption{(Left) Plot of $\overline{\ot}(U)$ against circuit depth at initialization for $N=8$, $N_A=N_D=3$. Systems $A$ and $D$ are disjoint (see Fig.~\ref{Fig:QNN} for example), so the commutator between any $O_A$ and $O_D$ vanishes, yielding $\overline{\ot}(U)=1$. The support of $UO_AU^\dagger$ reaches $D$ after three QNN layers, causing the OTOC to decay. The system becomes maximally scrambled once the OTOC reaches its floor value. (Right) The corresponding $L_{\mathrm{scram}}$ for a maximally scrambling target unitary is plotted along with $L_{+}$, $L_{-}$, and $L_{\mathrm{floor}}$.}
    \label{Fig:OTOC}
\end{figure}

For concreteness, we adopt the brick-wall network architecture (see Fig.~\ref{Fig:QNN}) to numerically simulate the decay of the true error with circuit depth. At layer $i$ in the QNN, a 2-qubit unitary $U_{i,j}$ is applied to each pair of neighboring qubits. Index $j$ labels the unitaries in a given layer. Odd layers are staggered by one qubit with respect to even layers, forming a brick-wall geometry. Each unitary takes on the form $U_{i,j}=e^{-i \theta_{i,j}V_{i,j}}$, where $V_{i,j}$ is Hermitian and $\theta_{i,j}$ is a training parameter.

We numerically simulate the true error for a QNN at initialization when the target unitary is maximally scrambling. We choose each parameter $\theta_{i,j}$ randomly such that we can assume $U_{i,j}$ is effectively a Haar random unitary. $L_{\mathrm{scram}}$ is plotted in Fig.~\ref{Fig:OTOC} (right) along with $L_{+}$, $L_{-}$, and $L_{\mathrm{floor}}$ as a function of circuit depth. $\overline{\ot}(U)$, and hence $L_{\mathrm{scram}}$, decays with circuit depth. The bounds $L_{+}$ and $L_-$ tighten with circuit depth; this is consistent with Ineq.~\eqref{Eq:LpmIneq}. $L_{\mathrm{scram}}$ decays to $L_{\mathrm{floor}}$ at sufficiently large depth, since the QNN becomes maximally scrambling.

\begin{figure}[h!]
    \centering
    \subfigure{\includegraphics[scale=.51]{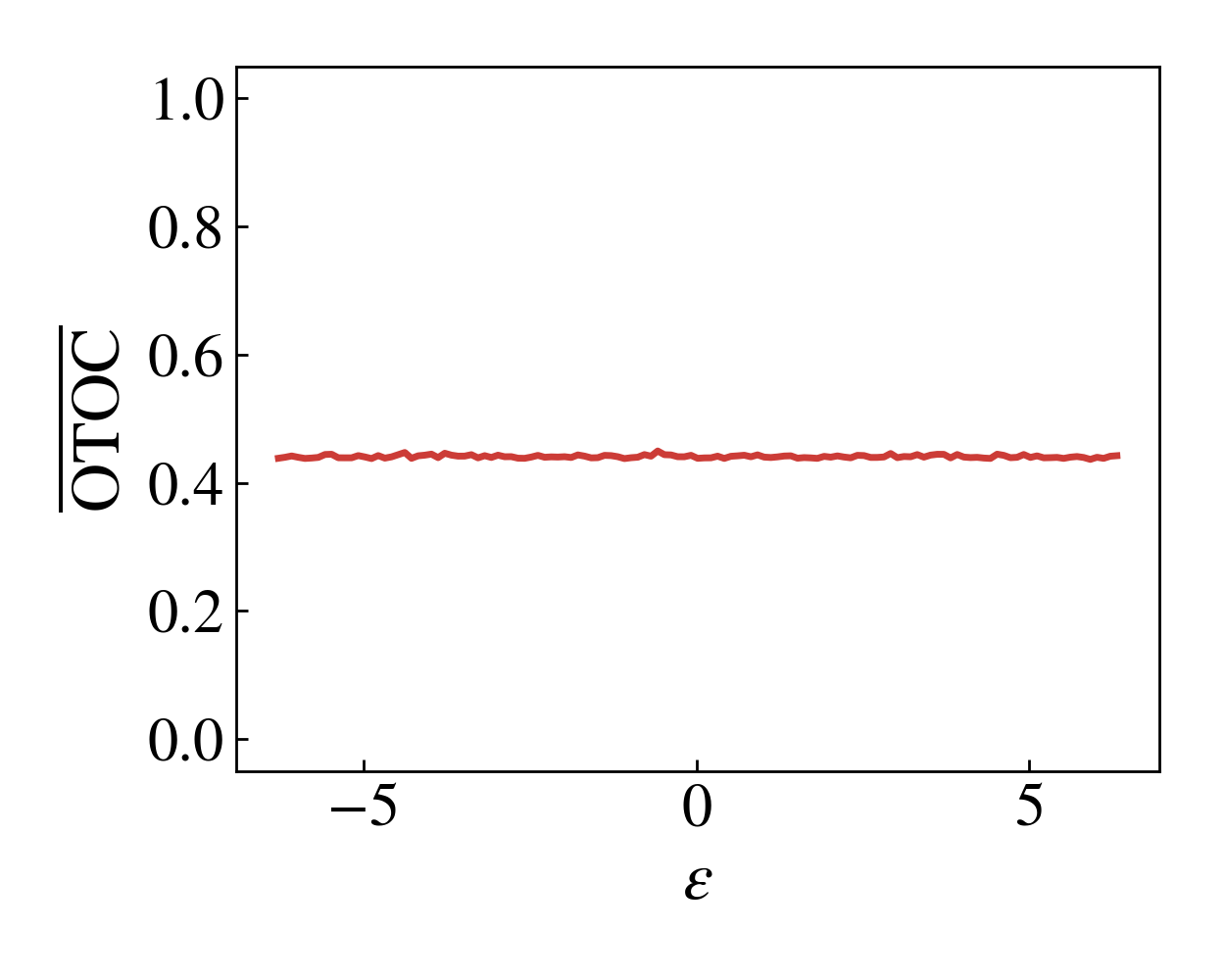}}
    \caption{
    OTOC landscape with respect to perturbation parameter $\varepsilon\in [-2\pi,2\pi]$. Each training parameter $\theta_{i,j}$ is randomly initialized. The OTOC is computed using updated parameters $\theta_{i,j}\rightarrow \theta_{i,j}+\varepsilon$. The QNN has a circuit depth of 30 layers and $N=8, N_A=1,N_C=N-1$, ensuring the QNN is in the maximally scrambling regime. }
    \label{Fig:Landscape}
\end{figure}

In Figure~\ref{Fig:Landscape}, we numerically simulate the OTOC landscape generated by varying each training parameter $\theta_{i,j}$ using perturbation parameter $\varepsilon \in [-2\pi,2\pi]$. Each $\theta_{i,j}$ is randomly initialized. A circuit depth of 30 layers is used with  $N=8, N_A=1, N_C=N-1$. This choice of circuit parameters ensures that, at initialization, the QNN is maximally scrambling and the OTOC has attained its floor value. The flat landscape indicates that even after perturbing the parameters, the QNN remains maximally scrambling and the OTOC retains its floor value. From Theorem~\ref{TheoremOTOC}, we see that $\partial_{\theta_l}L_{\mathrm{scram}}=G\partial_{\theta_l}\overline{\ot}(U(\bm{\theta}))$. That is, the OTOC landscape determines the $L_{\mathrm{scram}}$ landscape. Hence, the flat OTOC landscapes arising from maximally scrambling QNNs may pose challenges to training via gradient descent algorithms.

\section{Discussion}
We have shown that training error is bounded by the OTOC, a scrambling measure. Our results demonstrate that learning a unitary necessitates learning its scrambling properties. Training of the QNN via gradient descent requires computing the gradient of a loss function. We establish an inequality relating this gradient to the gradient of the OTOC. As a result, training is regulated by the OTOC landscape. We show that maximally scrambling QNNs can produce flat OTOC landscapes, which may present a roadblock to training. An open question, which we pose, is to prove whether weakly scrambling QNN architectures can remove barren plateaus in the training landscape.

\acknowledgments
We thank Jordan Cotler for discussions about Levy's lemma. We thank Liyuan Chen for discussions about R\'enyi entropy. This work was supported in part by the ARO Grant W911NF-19-1-0302 and the ARO MURI Grant W911NF-20-1-0082.

\appendix

\section{Training with cost functions}\label{Sec:CostMain}
In the main text, we considered the training of QNNs by examining the loss function and its gradient. This method is useful when training with a data set. However, QNNs can also be trained to optimize a problem-specific cost function, which does not involve a data set. Let $\rho_{d'}$ be a fixed, random state of the form Eq.~\eqref{Eq:rho}. We define the cost function as
\begin{equation}\label{Eq:CostFun}
	\mathcal{C}=\av{O_C}\tilde{y}_{d'}^2,
\end{equation}
where $\tilde{y}_{d'}$ has the form of Eq.~\eqref{Eq:TargetFunctions}. Eq.~\eqref{Eq:CostFun} does not include a target function, since the QNN is not learning a target unitary. It is shown in Appendix~\ref{Sec:CostFunction} that the cost function and its gradient can be written in terms of the out-of-time-ordered correlator $C_{d'}(U,U)$ defined in Eq.~\eqref{Eq:LossOTOCdef}:
\begin{lemma}\label{Lemma:Cost}
The cost function and its gradient can be expressed in terms of an out-of-time-ordered correlator, $C_{d'}(U,U)$, as follows
\begin{align}
	\mathcal{C}&=d_D^2C_{d'}(U,U),\\
	\partial_{\theta_l}\mathcal{C}&=d_D^2\partial_{\theta_l}C_{d'}(U,U).\label{Eq:CostGradExact}
\end{align}
\end{lemma}

To understand how the network's scrambling properties affect training via gradient descent algorithms, we bound the cost function and its gradient using the OTOC from Eq.~\eqref{Eq:OTOC}. We consider a QNN architecture as in Sec.~\ref{Sec:GradofLoss}. The following proposition bounds $\mathcal{C}$ and its gradient when $\rho_{d',A}$ is a random state (see Appendix~\ref{Proof:CostConcentrateProof} for a proof).

\begin{prop}\label{Prop:CostConcentrate}
Let $\ket{\psi_{d',A}}$ be sampled from the Haar measure on the Hilbert space of system $A$. With probability at least $1-\epsilon$, $\mathcal{C}(\ket{\psi_{d',A}})$ satisfies
\begin{equation}
	\abs{\mathcal{C}-\mathcal{C}_{\mathrm{av}}}\leq \eta_{\mathcal{C}}f(\epsilon),
\end{equation}
where the Lipschitz constant $\eta_\mathcal{C}$ is
\begin{equation}
    \eta_{\mathcal{C}}=\frac{4\sqrt{d_A}d_A}{d_C}\sqrt{\overline{\ot}(U)},
\end{equation}
and the average cost function is
\begin{equation}
    \mathcal{C}_{\mathrm{av}}=\int_{\Hr}dU_A \mathcal{C}=G\Big[\frac{1}{d_A}+\overline{\ot}(U) \Big].\label{Eq:CostAv}
\end{equation}
With probability at least $1-\epsilon$, $\partial_{\theta_l}\mathcal{C}(\ket{\psi_{d',A}})$ satisfies

\begin{equation}\label{Eq:CostGradConc}
	\abs{\partial_{\theta_l}\mathcal{C}-\partial_{\theta_l}\mathcal{C}_{\mathrm{av}}}\leq \eta_{\mathcal{C},g}f(\epsilon),
\end{equation}
where the Lipschitz constant $\eta_{\mathcal{C},g}$ is 
\begin{equation}
    \eta_{\mathcal{C},g}=8\norm{V_l}_\infty\left(\frac{\sqrt{d_A}d_A}{d_C}\sqrt{\overline{\ot}(U)}+1\right),
\end{equation}
and the gradient of $\mathcal{C}_{\mathrm{av}}$ is
\begin{equation}
    \partial_{\theta_l}\mathcal{C}_{\mathrm{av}}=G\partial_{\theta_l}\overline{\ot}(U).\label{Eq:CostGradAv}
\end{equation}
\end{prop}

$\mathcal{C}$ and $\partial_{\theta_l}\mathcal{C}$ are probabilistically bounded by $\overline{\ot}(U)$. Importantly, Ineq.~\eqref{Eq:CostGradConc} implies that $\partial_{\theta_l}\mathcal{C}$ can be made to concentrate about $\partial_{\theta_l}\mathcal{C}_{\mathrm{av}}$, which depends only on the gradient of the OTOC, as shown in Eq.~\eqref{Eq:CostGradAv}. The OTOC landscape therefore regulates the network's trainability when optimizing the cost function.
Eqs.~\eqref{Eq:CostGradExact} and \eqref{Eq:CostGradAv} suggest that maximally scrambling QNNs with flat out-of-time-ordered correlator landscapes may potentially account for barren plateaus.

\section{Supplementary numerical simulations}
\begin{figure}[h!]
    \centering
    \subfigure{\includegraphics[scale=.5]{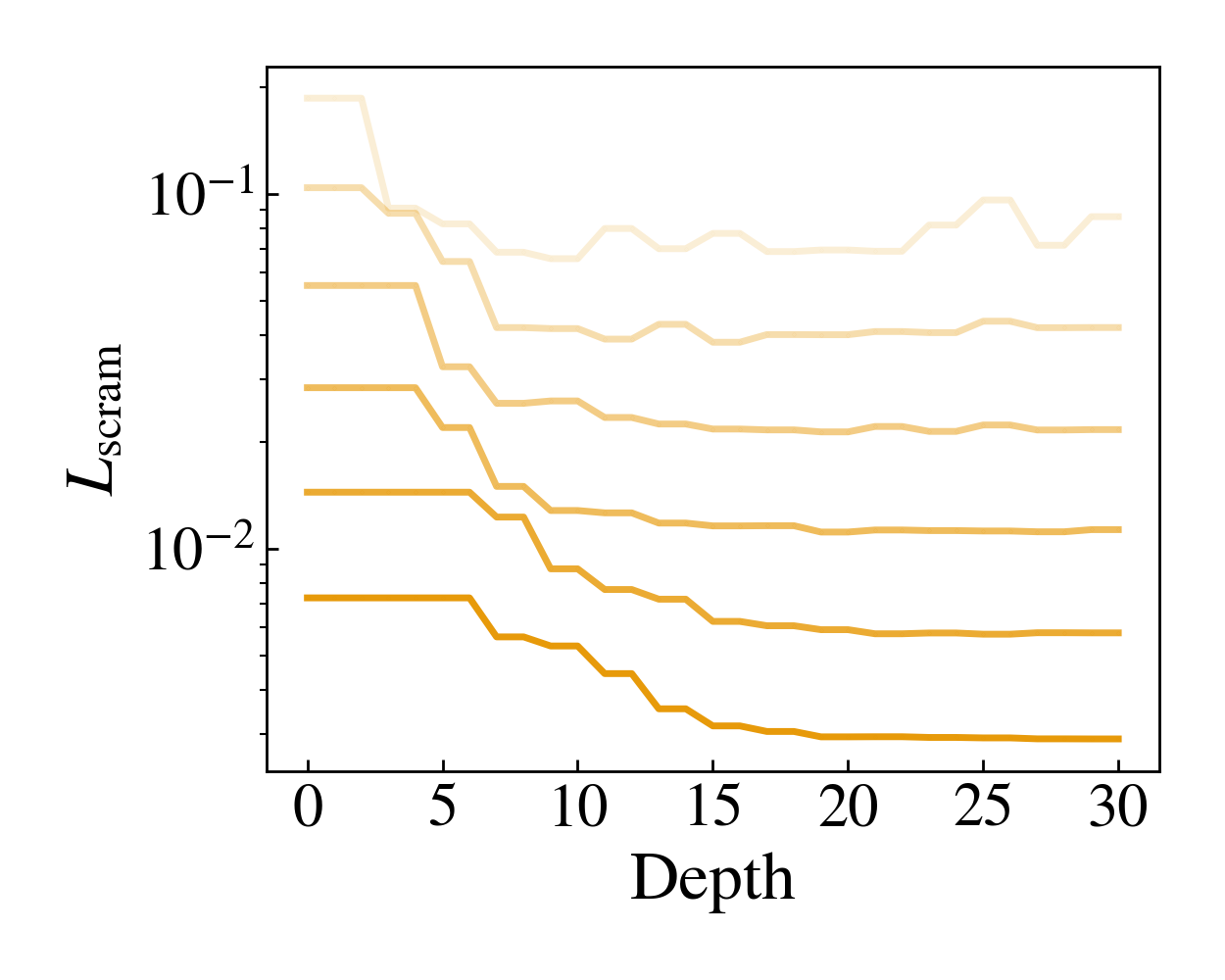}} 
    \subfigure{\includegraphics[scale=.5]{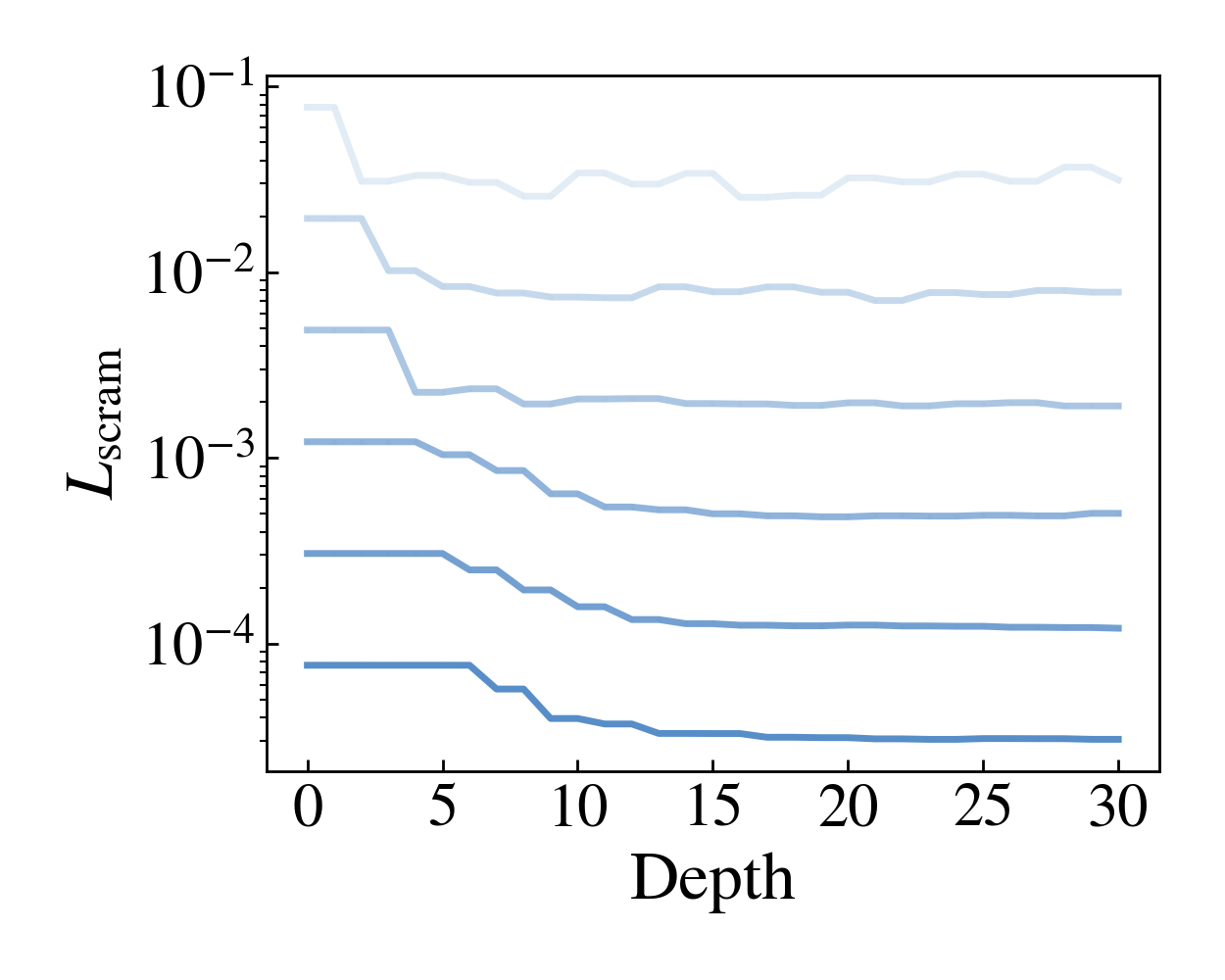}} 
    \caption{ Plots of $L_\mathrm{scram}$ for a randomly initialized QNN of varying depth for a maximally scrambling target unitary. $L_\mathrm{scram}$ is plotted for $N=3,4,\ldots, 8$ (darker colors indicate increasing $N$) for subsystem sizes: (left) $N_A=N-1, N_C=1$ and (right) $N_A=1, N_C=N-1$.}
    \label{Fig:Loss}
\end{figure}

Fig.~\ref{Fig:Loss} shows the decay of $L_{\mathrm{scram}}$  with respect to circuit depth for various $N $. In the left plot, the input data is encoded in a large state supported on $N_A=N-1$ qubits and a single-qubit measurement on $C$ is performed. In the right plot, the data is encoded in a single-qubit state and a measurement on $N_C=N-1$ qubits is performed. For a given $N$, the right plot produces a smaller true error. This is consistent with the behavior of $L_{\mathrm{floor}}$, which decreases as $N_A$ decreases (i.e. as $N_B$ increases) for a given $N$. Both plots indicate $L_{\mathrm{scram}}$ decays with circuit depth and width. 

\section{Generalizing true error}\label{Sec:GeneralizeError}
Although $L_d$ in Eq.~\eqref{Eq:LossDef} is defined with respect to an average of $O_C$ over $\mathcal{P}_C$, the Pauli group on $C$, variants of the true error may be defined with respect to a uniform average over a subset $S_C\subseteq \mathcal{P}_C $ with cardinality $\abs{S_C}$. Some useful variants include:
\begin{equation}
\begin{split}
L_{V1}&=\av{d}\av{O_C\in S_C}|\tilde{y}_d-y_d|^2, \\
 L_{V2}&=\av{d}\abs{\av{O_C\in S_C}(\tilde{y}_d-y_d)}^2, \\ L_{V3}&=\av{d}\abs{\av{O_C\in S_C}(\tilde{y}_d-y_d)},
\end{split}
\end{equation}
where $\av{d}$ implies the Haar average over $U_A$. Taking $S_C$ to consist of a single Pauli observable gives a common definition of the loss function. It can be shown that
\begin{equation}
L_{V3}^2\leq L_{V2}\leq L_{V1}\leq \frac{d_C^2}{|S_C|}L\leq \frac{d_C^2}{|S_C|}L_{+}.\label{Ineq:Generalloss}
\end{equation}
$L_{+}$ bounds a function of each variant. Hence, scrambling bounds a larger class of true error definitions. Variants of the cost function in Eq.~\eqref{Eq:CostFun} can be defined in a similar fashion.

To prove Ineq.~\eqref{Ineq:Generalloss}, first recall $L\leq L_{+}$, as established in the main text. Also note that $\sum_{O_C\in S_C}|\tilde{y}_d-y_d|^2\leq \sum_{O_C\in \mathcal{P}_C}|\tilde{y}_d-y_d|^2$. We can write
\begin{equation}
	\av{O_C\in S_C}|\tilde{y}_d-y_d|^2 =\frac{1}{|S_C|}\sum_{O_C\in S_C}|\tilde{y}_d-y_d|^2  \leq \frac{1}{|S_C|} \frac{d^2_C}{d^2_C}\sum_{O_C\in \mathcal{P}_C}|\tilde{y}_d-y_d|^2=\frac{d^2_C}{|S_C|} \av{O_C\in \mathcal{P}_C}|\tilde{y}_d-y_d|^2.
\end{equation}
This implies $L_{V1}\leq \frac{d_C^2}{|S_C|}L$. Since the variance is non-negative, $(\av{O_C\in S_C}(\cdot))^2\leq \av{O_C\in S_C}((\cdot)^2)$. This implies $L_{V2}\leq L_{V1}$. We also have $(\av{d}(\cdot))^2\leq\av{d}((\cdot)^2)$, which implies $L_{V3}^2\leq L_{V_2}$.

\section{Diagrammatic formalism}\label{Sec:Diagrams}
We can diagrammatically express unitary $U_i\in\{U,U_S\}$ as
\begin{equation}
\begin{tikzpicture}

    \draw [thick,color=dullblue]
    (-\xrel,\yrel)--(\xrel,\yrel)
    (-\xrel,-\yrel)--(\xrel,-\yrel);
    
	\renewcommand{\nodenum}{v0}
    \renewcommand{\name}{$U_i=$}
	\renewcommand{\xpos}{-2.5*\xrel}
    \renewcommand{\ypos}{0}
    \renewcommand{\height}{\heightsingle}
    \renewcommand{\width}{\widthsingle}
    \node[] (\nodenum) at (\xpos,\ypos) {\name};    
 
	\renewcommand{\nodenum}{v1}
    \renewcommand{\name}{$U_i$}
	\renewcommand{\xpos}{0}
    \renewcommand{\ypos}{0}
    \renewcommand{\height}{\heightdouble}
    \renewcommand{\width}{\widthsingle}
    \node[rectangle, fill=egg, rounded corners, minimum width=\width, minimum height=\height, draw] (\nodenum) at (\xpos,\ypos) {\name};
   
   	\renewcommand{\nodenum}{v1}
    \renewcommand{\name}{$A$}
	\renewcommand{\xpos}{1.5*\xrel}
    \renewcommand{\ypos}{\yrel}
    \renewcommand{\height}{\heightsingle}
    \renewcommand{\width}{\widthsingle}
    \node[] (\nodenum) at (\xpos,\ypos) {\name};

   	\renewcommand{\nodenum}{v1}
    \renewcommand{\name}{$B$}
	\renewcommand{\xpos}{1.5*\xrel}
    \renewcommand{\ypos}{-\yrel}
    \renewcommand{\height}{\heightsingle}
    \renewcommand{\width}{\widthsingle}
    \node[] (\nodenum) at (\xpos,\ypos) {\name};    
    
    \renewcommand{\nodenum}{v1}
    \renewcommand{\name}{$C$}
	\renewcommand{\xpos}{-1.5*\xrel}
    \renewcommand{\ypos}{\yrel}
    \renewcommand{\height}{\heightsingle}
    \renewcommand{\width}{\widthsingle}
    \node[] (\nodenum) at (\xpos,\ypos) {\name};    
    
    \renewcommand{\nodenum}{v1}
    \renewcommand{\name}{$D$}
	\renewcommand{\xpos}{-1.5*\xrel}
    \renewcommand{\ypos}{-\yrel}
    \renewcommand{\height}{\heightsingle}
    \renewcommand{\width}{\widthsingle}
    \node[] (\nodenum) at (\xpos,\ypos) {\name};    
\end{tikzpicture},
\end{equation}
where we label the input $(A, B)$ and output $(C, D)$ subsystems. Define the Bell state between systems $S_1$ and $S_2$, each of dimension $d_{S'}$, as
\begin{equation}
\begin{tikzpicture}

    \draw [thick,color=dullblue]
    (0,\yrel)--(.75*\xrel,\yrel)
    (0,-\yrel)--(.75*\xrel,-\yrel) ;
    
    \draw [thick,color=dullblue]
    (.75*\xrel,\yrel)-- (.75*\xrel,-\yrel);

    	\renewcommand{\nodenum}{v0}
    \renewcommand{\name}{$\ket{\rm{Bell}}_{S_1,S_2}=\frac{1}{\sqrt{d_{S'}}}\sum_{i=1}^{d_{S'}}\ket{i}_{S_1}\otimes \ket{i}_{S_2}=\frac{1}{\sqrt{d_{S'}}}$}
	\renewcommand{\xpos}{-3.55*\xrel}
    \renewcommand{\ypos}{0}
    \renewcommand{\height}{\heightsingle}
    \renewcommand{\width}{\widthsingle}
    \node[] (\nodenum) at (\xpos,\ypos) {\name};
    
    \renewcommand{\nodenum}{v1}
    \renewcommand{\name}{$S_1$}
	\renewcommand{\xpos}{1.15*\xrel}
    \renewcommand{\ypos}{\yrel}
    \renewcommand{\height}{\heightsingle}
    \renewcommand{\width}{\widthsingle}
    \node[] (\nodenum) at (\xpos,\ypos) {\name};

   	\renewcommand{\nodenum}{v1}
    \renewcommand{\name}{$S_2$}
	\renewcommand{\xpos}{1.15*\xrel}
    \renewcommand{\ypos}{-\yrel}
    \renewcommand{\height}{\heightsingle}
    \renewcommand{\width}{\widthsingle}
    \node[] (\nodenum) at (\xpos,\ypos) {\name};    
\end{tikzpicture}.
\end{equation}
The corresponding bra vector diagram is
\begin{equation}
\begin{tikzpicture}

    \draw [thick,color=dullblue]
    (0,\yrel)--(.75*\xrel,\yrel)
    (0,-\yrel)--(.75*\xrel,-\yrel) ;
    
    \draw [thick,color=dullblue]
    (0,\yrel)-- (0,-\yrel);

    	\renewcommand{\nodenum}{v0}
    \renewcommand{\name}{$\bra{\rm{Bell}}_{S_1,S_2}=\frac{1}{\sqrt{d_{S'}}}$}
	\renewcommand{\xpos}{-1.55*\xrel}
    \renewcommand{\ypos}{0}
    \renewcommand{\height}{\heightsingle}
    \renewcommand{\width}{\widthsingle}
    \node[] (\nodenum) at (\xpos,\ypos) {\name};
    
    \renewcommand{\nodenum}{v1}
    \renewcommand{\name}{$S_1$}
	\renewcommand{\xpos}{1.15*\xrel}
    \renewcommand{\ypos}{\yrel}
    \renewcommand{\height}{\heightsingle}
    \renewcommand{\width}{\widthsingle}
    \node[] (\nodenum) at (\xpos,\ypos) {\name};

   	\renewcommand{\nodenum}{v1}
    \renewcommand{\name}{$S_2$}
	\renewcommand{\xpos}{1.15*\xrel}
    \renewcommand{\ypos}{-\yrel}
    \renewcommand{\height}{\heightsingle}
    \renewcommand{\width}{\widthsingle}
    \node[] (\nodenum) at (\xpos,\ypos) {\name};    
\end{tikzpicture}.
\end{equation}
An identity relating an operator $Q$ to its transpose is
\begin{equation}\label{Eq:Tranpose}
\begin{tikzpicture}

    \draw [thick,color=dullblue]
    (0,\yrel)--(1.5*\xrel,\yrel)
    (0,-\yrel)--(1.5*\xrel,-\yrel);
    
    \draw [thick,color=dullblue]
    (0,\yrel)-- (0,-\yrel);

    \renewcommand{\nodenum}{v1}
    \renewcommand{\name}{$Q$}
	\renewcommand{\xpos}{.75*\xrel}
    \renewcommand{\ypos}{\yrel}
    \renewcommand{\height}{\heightsingle}
    \renewcommand{\width}{\widthsingle}
    \node[rectangle, fill=egg, rounded corners, minimum width=\width, minimum height=\height, draw] (\nodenum) at (\xpos,\ypos) {\name};
    
    \renewcommand{\nodenum}{v1}
    \renewcommand{\name}{$=$}
	\renewcommand{\xpos}{2*\xrel}
    \renewcommand{\ypos}{0}
    \renewcommand{\height}{\heightsingle}
    \renewcommand{\width}{\widthsingle}
    \node[] (\nodenum) at (\xpos,\ypos) {\name};
    
    \draw [thick,color=dullblue]
    (2.5*\xrel,\yrel)--(4*\xrel,\yrel)
    (2.5*\xrel,-\yrel)--(4*\xrel,-\yrel);
    
    \draw [thick,color=dullblue]
    (2.5*\xrel,\yrel)-- (2.5*\xrel,-\yrel);
    
    \renewcommand{\nodenum}{v1}
    \renewcommand{\name}{$Q^T$}
	\renewcommand{\xpos}{3.25*\xrel}
    \renewcommand{\ypos}{-\yrel}
    \renewcommand{\height}{\heightsingle}
    \renewcommand{\width}{\widthsingle}
    \node[rectangle, fill=egg, rounded corners, minimum width=\width, minimum height=\height, draw] (\nodenum) at (\xpos,\ypos) {\name};
\end{tikzpicture}.
\end{equation}

\section{Properties of the twirling channel}\label{Sec:Twirl}
Define the $k$-fold twirling channel for an operator $Q$ on the $k$-copy Hilbert space as
\begin{equation}
	\Phi^{(k)}_{\Hr(d')}(Q)=\int_{\Hr}dU U^{\dagger \otimes k}QU^{ \otimes k},
\end{equation}
where unitary $U$ is sampled from Haar measure on the unitary group of dimension $d'$, $\mathcal{U}(d')$. This can be expanded in terms of permutation operators $T_{\pi}$, where $\pi\in \mathcal{S}_k$ and $\mathcal{S}_k$ is the set of all permutations of the set $\{1,2,\ldots, k\}$:
\begin{equation}
	\Phi^{(k)}_{\Hr(d')}(Q)=\sum_{\pi,\sigma\in S_k}C_{\pi,\sigma}T_{\pi}\tr{T_{\sigma}Q}.
\end{equation}
The coefficients $C_{\pi,\sigma}$ form the Weingarten matrix. In the case of $k=1$,
\begin{equation}\label{Eq:OneTwirl}
	\Phi^{(1)}_{\Hr(d')}(Q)=\frac{1}{d'}I\tr{Q}.
\end{equation}
The case of $k=2$ gives
\begin{equation}\label{Eq:TwoTwirl}
\begin{split}
	\Phi^{(2)}_{\Hr(d')}(Q)
	=&\frac{1}{d'^2-1}\Big[I_{1,2}\tr{I_{1,2}Q}
	+\sw_{1,2}\tr{\sw_{1,2}Q}\\
	&\hspace{14mm}-\frac{1}{d'}I_{1,2}\tr{\sw_{1,2}Q}
	-\frac{1}{d'}\sw_{1,2}\tr{I_{1,2}Q}\Big],
\end{split}
\end{equation}
where $\mathbb{S}_{1,2}$ is the swap operator between the two Hilbert space copies. For a pure state $\ket{\psi}$, the following identity is useful
\begin{equation}\label{Eq:TwirlState}
		\Phi^{(k)}_{\Hr(d')}((\ket{\psi}\bra{\psi})^{\otimes k})=
		\frac{\sum_{\pi\in S_k}T_{\pi}}{d'(d'+1)\cdots(d'+k-1)}.
\end{equation}
Permutation operators are invariant under the twirling channel:
\begin{equation}
	\Phi^{(k)}_{\Hr(d')}(T_{\pi})=T_{\pi}.
\end{equation}
A $k$-design is a finite ensemble $
\mathcal{E}$ of unitaries which can replicate the first $k$ moments of the Haar measure:
\begin{equation}
	\av{U\sim\mathcal{E}} \left[U^{\dagger \otimes k}QU^{ \otimes k}\right]=\int_{\Hr}dU U^{\dagger \otimes k}QU^{ \otimes k}.
\end{equation}
The Pauli group forms a 1-design and the Clifford group forms a 3-design.

\section{Calculus identity}\label{Sec:Derivative}
We derive a useful calculus identity. Define state $\ket{\psi}\in \mathbb{C}^{d_A}$ and real vectors $\bm{v}_1,\bm{v}_2\in \mathbb{R}^{d_A}$. Any state can be written as
\begin{equation}
	\ket{\psi}=\bm{v}_1+i\bm{v}_2.
\end{equation}
Define $I_A$ as the $d_A\times d_A$-dimensional identity matrix. We can write
\begin{equation}
\begin{split}
	\ket{\psi}&= 
	\begin{bmatrix}
		I_A& iI_A
	\end{bmatrix}\bm{w}\\
	\bm{w}&=\begin{bmatrix}
		\bm{v}_1\\ \bm{v}_2
	\end{bmatrix}.
\end{split}
\end{equation}
For a Hermitian matrix $Q$, we compute the following derivative: 
\begin{equation}\label{Eq:DerivativeNorm}
\begin{split}
	\norm{\frac{d}{d \bm{w}}	\bra{\psi}Q\ket{\psi}}_2&=\norm{\frac{d}{d\bm{w}}
	\left(\bm{w}^T
	\begin{bmatrix}
		I_A \\ -iI_A
	\end{bmatrix}
	Q
	\begin{bmatrix}
		I_A & iI_A
	\end{bmatrix}\bm{w}\right)}_2\\
	&=\norm{\bm{w}^T
	\left(
	\begin{bmatrix}
		I_A\\ -iI_A
	\end{bmatrix}
	Q
	\begin{bmatrix}
		I_A & iI_A
	\end{bmatrix}
	+
\begin{bmatrix}
		I_A \\ iI_A
	\end{bmatrix}
	Q^T
	\begin{bmatrix}
		I_A & -iI_A
	\end{bmatrix}	
	\right)}_2\\
	&=\norm{\bra{\psi}
	Q
	\begin{bmatrix}
		I_A& iI_A
	\end{bmatrix}
	+
	\bra{\psi}^*
	Q^T
	\begin{bmatrix}
		I_A & -iI_A
	\end{bmatrix}	}_2\\ 
	&=\Bigg[\left(\bra{\psi}
	Q
	\begin{bmatrix}
		I_A & iI_A
	\end{bmatrix}
	+
	\bra{\psi}^*
	Q^T
	\begin{bmatrix}
		I_A& -iI_A
	\end{bmatrix}	\right)\\
	&\hspace{8mm}\cdot
	\left(
	\begin{bmatrix}
		I_A \\ -iI_A
	\end{bmatrix}Q^\dagger\ket{\psi}
	+
	\begin{bmatrix}
		I_A \\ iI_A
	\end{bmatrix}	Q^* \ket{\psi}^*\right) \Bigg]^{1/2}\\ 
	&= \sqrt{\bra{\psi}
	Q
	\begin{bmatrix}
		I_A & iI_A
	\end{bmatrix}
	\begin{bmatrix}
		I_A \\- iI_A
	\end{bmatrix}
	Q^\dagger \ket{\psi}
	+\bra{\psi}^*
	Q^T
	\begin{bmatrix}
		I_A & -iI_A
	\end{bmatrix}
	\begin{bmatrix}
		I_A\\ iI_A
	\end{bmatrix}
	Q^*\ket{\psi}^*}\\
&= \sqrt{\bra{\psi}
	Q
	(2I_A)
	Q^\dagger \ket{\psi}
	+\bra{\psi}^*
	Q^T (2I_A)
	Q^*\ket{\psi}^*}\\
&= \sqrt{2\bra{\psi}
	Q
	Q^\dagger \ket{\psi}
	+2(\bra{\psi}^*
	Q^T
	Q^*\ket{\psi}^*)^*}\\
&= \sqrt{2\bra{\psi}
	Q^\dagger 
	Q\ket{\psi}
	+2\bra{\psi}
	Q^\dagger
	Q\ket{\psi}}\\
&=2\norm{Q\ket{\psi}}_2.
\end{split}
\end{equation}
The second line follows from a standard calculus identity. 

\section{Proof of Proposition~\ref{Prop:LossandError} }\label{ProofofLossandError}
\subsection{Computing the loss function}
\label{Sec:OTOCtarget}
We will show that the loss function can be written as
\begin{equation}
L_d=d_D^2(C_d(U,U)+C_d(U_S,U_S)-2C_d(U,U_S)),
\end{equation}
where
\begin{equation}
	C_d(U_1,U_2)=\av{O_D}\langle U_1\rho_d U_1^\dagger O_D^\dagger U_2 \rho_d U_2^\dagger O_D\rangle.
\end{equation}
Let $Q$ be an arbitrary operator on the $d_{\mathrm{tot}}$-dimensional Hilbert space. We can define the more general correlator $C(Q,U_1,U_2)$ as
\begin{equation}\label{Cgeneral}
	C(Q,U_1,U_2)=\av{O_D}\langle U_1Q U_1^\dagger O_D^\dagger U_2 Q U_2^\dagger O_D\rangle.
\end{equation}
Note that $C_d(U_1,U_2)=C(\rho_d,U_1,U_2)$. We express $C(Q,U_1,U_2)$ diagrammatically:

\begin{equation}
\begin{tikzpicture}

    
	\renewcommand{\nodenum}{v0}
    \renewcommand{\name}{$C(Q,U_1,U_2)=\frac{1}{d_\mathrm{tot}}\av{O_D}$}
	\renewcommand{\xpos}{-2.2*\xrel}
    \renewcommand{\ypos}{-\rowspace}
    \renewcommand{\height}{\heightsingle}
    \renewcommand{\width}{\widthsingle}
    \node[] (\nodenum) at (\xpos,\ypos) {\name};

    \renewcommand{\nodenum}{v1}
    \renewcommand{\name}{$Trace1$}
	\renewcommand{\xpos}{4.5*\xrel}
    \renewcommand{\ypos}{-\rowspace}
    \renewcommand{\height}{\heighttrace}
    \renewcommand{\width}{25em}
    \node[rectangle,rounded corners, thick, minimum width=\width, minimum height=\height, draw=dullblue] (\nodenum) at (\xpos,\ypos) {};

    \renewcommand{\nodenum}{v1}
    \renewcommand{\name}{$Trace2$}
	\renewcommand{\xpos}{4.5*\xrel}
    \renewcommand{\ypos}{-\rowspace}
    \renewcommand{\height}{3*\heighttrace}
    \renewcommand{\width}{27em}
    \node[rectangle,rounded corners, thick, minimum width=\width, minimum height=\height, draw=dullblue] (\nodenum) at (\xpos,\ypos) {};     
 
	\renewcommand{\nodenum}{v1}
    \renewcommand{\name}{$U_1$}
	\renewcommand{\xpos}{\xrel}
    \renewcommand{\ypos}{0}
    \renewcommand{\height}{\heightdouble}
    \renewcommand{\width}{\widthsingle}
    \node[rectangle, fill=egg, rounded corners, minimum width=\width, minimum height=\height, draw] (\nodenum) at (\xpos,\ypos) {\name};

	\renewcommand{\nodenum}{v2}
    \renewcommand{\name}{$Q$}
	\renewcommand{\xpos}{2*\xrel}
    \renewcommand{\ypos}{0}
    \renewcommand{\height}{\heightdouble}
    \renewcommand{\width}{\widthsingle}
    \node[rectangle, fill=egg, rounded corners, minimum width=\width, minimum height=\height, draw] (\nodenum) at (\xpos,\ypos) {\name};
 
 \renewcommand{\nodenum}{v3}
    \renewcommand{\name}{$U_1^\dagger$}
	\renewcommand{\xpos}{3*\xrel}
    \renewcommand{\ypos}{0}
    \renewcommand{\height}{\heightdouble}
    \renewcommand{\width}{\widthsingle}
    \node[rectangle, fill=egg, rounded corners, minimum width=\width, minimum height=\height, draw] (\nodenum) at (\xpos,\ypos) {\name};

	\renewcommand{\nodenum}{v4}
    \renewcommand{\name}{$O_D^\dagger$}
	\renewcommand{\xpos}{4*\xrel}
    \renewcommand{\ypos}{-\yrel}
    \renewcommand{\height}{\heightsingle}
    \renewcommand{\width}{\widthsingle}
    \node[rectangle, fill=egg, rounded corners, minimum width=\width, minimum height=\height, draw] (\nodenum) at (\xpos,\ypos) {\name};
    
	 \renewcommand{\nodenum}{v5}
    \renewcommand{\name}{$U_2$}
	\renewcommand{\xpos}{5*\xrel}
    \renewcommand{\ypos}{0}
    \renewcommand{\height}{\heightdouble}
    \renewcommand{\width}{\widthsingle}
    \node[rectangle, fill=egg, rounded corners, minimum width=\width, minimum height=\height, draw] (\nodenum) at (\xpos,\ypos) {\name};

	\renewcommand{\nodenum}{v6}
    \renewcommand{\name}{$Q$}
	\renewcommand{\xpos}{6*\xrel}
    \renewcommand{\ypos}{0}
    \renewcommand{\height}{\heightdouble}
    \renewcommand{\width}{\widthsingle}
    \node[rectangle, fill=egg, rounded corners, minimum width=\width, minimum height=\height, draw] (\nodenum) at (\xpos,\ypos) {\name};
    
     \renewcommand{\nodenum}{v7}
    \renewcommand{\name}{$U_2^\dagger$}
	\renewcommand{\xpos}{7*\xrel}
    \renewcommand{\ypos}{0}
    \renewcommand{\height}{\heightdouble}
    \renewcommand{\width}{\widthsingle}
    \node[rectangle, fill=egg, rounded corners, minimum width=\width, minimum height=\height, draw] (\nodenum) at (\xpos,\ypos) {\name};

	\renewcommand{\nodenum}{v8}
    \renewcommand{\name}{$O_D$}
	\renewcommand{\xpos}{8*\xrel}
    \renewcommand{\ypos}{-\yrel}
    \renewcommand{\height}{\heightsingle}
    \renewcommand{\width}{\widthsingle}
    \node[rectangle, fill=egg, rounded corners, minimum width=\width, minimum height=\height, draw] (\nodenum) at (\xpos,\ypos) {\name};
    
\end{tikzpicture}.
\end{equation}
Using the transpose identity from Eq.~\ref{Eq:Tranpose},

\begin{equation}
\begin{tikzpicture}

    
    
    \renewcommand{\nodenum}{v1}
    \renewcommand{\name}{$Trace1$}
	\renewcommand{\xpos}{2*\xrel}
    \renewcommand{\ypos}{-\rowspace}
    \renewcommand{\height}{\heighttrace}
    \renewcommand{\width}{16em}
    \node[rectangle,rounded corners, thick, minimum width=\width, minimum height=\height, draw=dullblue] (\nodenum) at (\xpos,\ypos) {}; 
    
    \renewcommand{\nodenum}{v1}
    \renewcommand{\name}{$Trace2$}
	\renewcommand{\xpos}{2*\xrel}
    \renewcommand{\ypos}{-\rowspace}
    \renewcommand{\height}{3*\heighttrace}
    \renewcommand{\width}{18em}
    \node[rectangle,rounded corners, thick, minimum width=\width, minimum height=\height, draw=dullblue] (\nodenum) at (\xpos,\ypos) {}; 
    
	\renewcommand{\nodenum}{v0}
    \renewcommand{\name}{$C(Q,U_1,U_2)=\frac{1}{d_\mathrm{tot}}\av{O_D}$}
	\renewcommand{\xpos}{-3*\xrel}
    \renewcommand{\ypos}{-\rowspace}
    \renewcommand{\height}{\heightsingle}
    \renewcommand{\width}{\widthsingle}
    \node[] (\nodenum) at (\xpos,\ypos) {\name};

     \renewcommand{\nodenum}{v5}
    \renewcommand{\name}{$O_D^\dagger$}
	\renewcommand{\xpos}{0}
    \renewcommand{\ypos}{-\yrel}
    \renewcommand{\height}{\heightsingle}
    \renewcommand{\width}{\widthsingle}
    \node[rectangle, fill=egg, rounded corners, minimum width=\width, minimum height=\height, draw] (\nodenum) at (\xpos,\ypos) {\name};

	\renewcommand{\nodenum}{v1}
    \renewcommand{\name}{$U_1^*$}
	\renewcommand{\xpos}{\xrel}
    \renewcommand{\ypos}{-2*\rowspace}
    \renewcommand{\height}{\heightdouble}
    \renewcommand{\width}{\widthsingle}
    \node[rectangle, fill=egg, rounded corners, minimum width=\width, minimum height=\height, draw] (\nodenum) at (\xpos,\ypos) {\name};

	\renewcommand{\nodenum}{v2}
    \renewcommand{\name}{$Q$}
	\renewcommand{\xpos}{2*\xrel}
    \renewcommand{\ypos}{0}
    \renewcommand{\height}{\heightdouble}
    \renewcommand{\width}{\widthsingle}
    \node[rectangle, fill=egg, rounded corners, minimum width=\width, minimum height=\height, draw] (\nodenum) at (\xpos,\ypos) {\name};
 
 \renewcommand{\nodenum}{v3}
    \renewcommand{\name}{$U_1^T$}
	\renewcommand{\xpos}{3*\xrel}
    \renewcommand{\ypos}{-2*\rowspace}
    \renewcommand{\height}{\heightdouble}
    \renewcommand{\width}{\widthsingle}
    \node[rectangle, fill=egg, rounded corners, minimum width=\width, minimum height=\height, draw] (\nodenum) at (\xpos,\ypos) {\name};

	\renewcommand{\nodenum}{v4}
    \renewcommand{\name}{$O_D$}
	\renewcommand{\xpos}{4*\xrel}
    \renewcommand{\ypos}{-\yrel}
    \renewcommand{\height}{\heightsingle}
    \renewcommand{\width}{\widthsingle}
    \node[rectangle, fill=egg, rounded corners, minimum width=\width, minimum height=\height, draw] (\nodenum) at (\xpos,\ypos) {\name};

    	\renewcommand{\nodenum}{v6}
    \renewcommand{\name}{$U_2$}
	\renewcommand{\xpos}{\xrel}
    \renewcommand{\ypos}{0}
    \renewcommand{\height}{\heightdouble}
    \renewcommand{\width}{\widthsingle}
    \node[rectangle, fill=egg, rounded corners, minimum width=\width, minimum height=\height, draw] (\nodenum) at (\xpos,\ypos) {\name};

	\renewcommand{\nodenum}{v7}
    \renewcommand{\name}{$Q^T$}
	\renewcommand{\xpos}{2*\xrel}
    \renewcommand{\ypos}{-2*\rowspace}
    \renewcommand{\height}{\heightdouble}
    \renewcommand{\width}{\widthsingle}
    \node[rectangle, fill=egg, rounded corners, minimum width=\width, minimum height=\height, draw] (\nodenum) at (\xpos,\ypos) {\name};
 
 \renewcommand{\nodenum}{v8}
    \renewcommand{\name}{$U_2^\dagger$}
	\renewcommand{\xpos}{3*\xrel}
    \renewcommand{\ypos}{0}
    \renewcommand{\height}{\heightdouble}
    \renewcommand{\width}{\widthsingle}
    \node[rectangle, fill=egg, rounded corners, minimum width=\width, minimum height=\height, draw] (\nodenum) at (\xpos,\ypos) {\name};

\end{tikzpicture}.
\end{equation}
Perform the average of $O_D$ over the Pauli group (which forms a 1-design) using the identity from Eq.~\eqref{Eq:OneTwirl}: $\av{O_D}O_D^\dagger Q'O_D=\frac{1}{d_D}I_D\mathrm{Tr}_D\{Q'\}$ where $Q'=U_2QU_2^\dagger$. This identity produces

\begin{equation}
\begin{tikzpicture}

    

    \renewcommand{\nodenum}{v1}
    \renewcommand{\name}{$Trace1$}
	\renewcommand{\xpos}{2*\xrel}
    \renewcommand{\ypos}{0}
    \renewcommand{\height}{\heighttrace}
    \renewcommand{\width}{12em}
    \node[rectangle,rounded corners, thick, minimum width=\width, minimum height=\height, draw=dullblue] (\nodenum) at (\xpos,\ypos) {}; 
     
    \renewcommand{\nodenum}{v1}
    \renewcommand{\name}{$Trace2$}
	\renewcommand{\xpos}{2*\xrel}
    \renewcommand{\ypos}{-2*\rowspace}
    \renewcommand{\height}{\heighttrace}
    \renewcommand{\width}{12em}
    \node[rectangle,rounded corners, thick, minimum width=\width, minimum height=\height, draw=dullblue] (\nodenum) at (\xpos,\ypos) {};

	\renewcommand{\nodenum}{v1}
    \renewcommand{\name}{$Trace3$}
	\renewcommand{\xpos}{2*\xrel}
    \renewcommand{\ypos}{-\rowspace}
    \renewcommand{\height}{6.5*\heightsingle}
    \renewcommand{\width}{14em}
    \node[rectangle,rounded corners, thick, minimum width=\width, minimum height=\height, draw=dullblue] (\nodenum) at (\xpos,\ypos) {}; 
    
	\renewcommand{\nodenum}{v0}
    \renewcommand{\name}{$C(Q,U_1,U_2)=\frac{1}{d_\mathrm{tot}d_D}$}
	\renewcommand{\xpos}{-2.5*\xrel}
    \renewcommand{\ypos}{-\rowspace}
    \renewcommand{\height}{\heightsingle}
    \renewcommand{\width}{\widthsingle}
    \node[] (\nodenum) at (\xpos,\ypos) {\name};

	\renewcommand{\nodenum}{v1}
    \renewcommand{\name}{$U_1^*$}
	\renewcommand{\xpos}{\xrel}
    \renewcommand{\ypos}{-3*\rowspace}
    \renewcommand{\height}{\heightdouble}
    \renewcommand{\width}{\widthsingle}
    \node[rectangle, fill=egg, rounded corners, minimum width=\width, minimum height=\height, draw] (\nodenum) at (\xpos,\ypos) {\name};

	\renewcommand{\nodenum}{v2}
    \renewcommand{\name}{$Q^T$}
	\renewcommand{\xpos}{2*\xrel}
    \renewcommand{\ypos}{-3*\rowspace}
    \renewcommand{\height}{\heightdouble}
    \renewcommand{\width}{\widthsingle}
    \node[rectangle, fill=egg, rounded corners, minimum width=\width, minimum height=\height, draw] (\nodenum) at (\xpos,\ypos) {\name};
 
 \renewcommand{\nodenum}{v3}
    \renewcommand{\name}{$U_1^T$}
	\renewcommand{\xpos}{3*\xrel}
    \renewcommand{\ypos}{-3*\rowspace}
    \renewcommand{\height}{\heightdouble}
    \renewcommand{\width}{\widthsingle}
    \node[rectangle, fill=egg, rounded corners, minimum width=\width, minimum height=\height, draw] (\nodenum) at (\xpos,\ypos) {\name};

    	\renewcommand{\nodenum}{v4}
    \renewcommand{\name}{$U_2$}
	\renewcommand{\xpos}{\xrel}
    \renewcommand{\ypos}{\rowspace}
    \renewcommand{\height}{\heightdouble}
    \renewcommand{\width}{\widthsingle}
    \node[rectangle, fill=egg, rounded corners, minimum width=\width, minimum height=\height, draw] (\nodenum) at (\xpos,\ypos) {\name};

	\renewcommand{\nodenum}{v5}
    \renewcommand{\name}{$Q$}
	\renewcommand{\xpos}{2*\xrel}
    \renewcommand{\ypos}{\rowspace}
    \renewcommand{\height}{\heightdouble}
    \renewcommand{\width}{\widthsingle}
    \node[rectangle, fill=egg, rounded corners, minimum width=\width, minimum height=\height, draw] (\nodenum) at (\xpos,\ypos) {\name};
 
 \renewcommand{\nodenum}{v6}
    \renewcommand{\name}{$U_2^\dagger$}
	\renewcommand{\xpos}{3*\xrel}
    \renewcommand{\ypos}{\rowspace}
    \renewcommand{\height}{\heightdouble}
    \renewcommand{\width}{\widthsingle}
    \node[rectangle, fill=egg, rounded corners, minimum width=\width, minimum height=\height, draw] (\nodenum) at (\xpos,\ypos) {\name};

\end{tikzpicture}.
\end{equation}
Introduce a 1-design on system $C$ by using the identity $\av{O_C}O_C\otimes O_C^*=\ket{\rm{Bell}}\bra{\rm{Bell}}_{C,C'}$,

\begin{equation}
\begin{tikzpicture}

    

    \renewcommand{\nodenum}{v1}
    \renewcommand{\name}{$Trace1$}
	\renewcommand{\xpos}{2.5*\xrel}
    \renewcommand{\ypos}{\yrel}
    \renewcommand{\height}{\heighttrace}
    \renewcommand{\width}{14em}
    \node[rectangle,rounded corners, thick, minimum width=\width, minimum height=\height, draw=dullblue] (\nodenum) at (\xpos,\ypos) {}; 
     
    \renewcommand{\nodenum}{v1}
    \renewcommand{\name}{$Trace2$}
	\renewcommand{\xpos}{2.5*\xrel}
    \renewcommand{\ypos}{-\yrel-2*\rowspace}
    \renewcommand{\height}{\heighttrace}
    \renewcommand{\width}{14em}
    \node[rectangle,rounded corners, thick, minimum width=\width, minimum height=\height, draw=dullblue] (\nodenum) at (\xpos,\ypos) {};

	\renewcommand{\nodenum}{v1}
    \renewcommand{\name}{$Trace3$}
	\renewcommand{\xpos}{2.5*\xrel}
    \renewcommand{\ypos}{-1.5*\yrel-2*\rowspace}
    \renewcommand{\height}{2.5*\heighttrace}
    \renewcommand{\width}{16em}
    \node[rectangle,rounded corners, thick, minimum width=\width, minimum height=\height, draw=dullblue] (\nodenum) at (\xpos,\ypos) {}; 
    
    \renewcommand{\nodenum}{v1}
    \renewcommand{\name}{$Trace4$}
	\renewcommand{\xpos}{2.5*\xrel}
    \renewcommand{\ypos}{1.5*\yrel}
    \renewcommand{\height}{2.5*\heighttrace}
    \renewcommand{\width}{16em}
    \node[rectangle,rounded corners, thick, minimum width=\width, minimum height=\height, draw=dullblue] (\nodenum) at (\xpos,\ypos) {}; 
    
	\renewcommand{\nodenum}{v0}
    \renewcommand{\name}{$C(Q,U_1,U_2)=\frac{d_C}{d_\mathrm{tot}d_D}\av{O_C}$}
	\renewcommand{\xpos}{-2.5*\xrel}
    \renewcommand{\ypos}{-\rowspace}
    \renewcommand{\height}{\heightsingle}
    \renewcommand{\width}{\widthsingle}
    \node[] (\nodenum) at (\xpos,\ypos) {\name};

	\renewcommand{\nodenum}{v1}
    \renewcommand{\name}{$U_1^*$}
	\renewcommand{\xpos}{\xrel}
    \renewcommand{\ypos}{-\yrel-3*\rowspace}
    \renewcommand{\height}{\heightdouble}
    \renewcommand{\width}{\widthsingle}
    \node[rectangle, fill=egg, rounded corners, minimum width=\width, minimum height=\height, draw] (\nodenum) at (\xpos,\ypos) {\name};

	\renewcommand{\nodenum}{v2}
    \renewcommand{\name}{$Q^T$}
	\renewcommand{\xpos}{2*\xrel}
    \renewcommand{\ypos}{\yrel-4*\rowspace}
    \renewcommand{\height}{\heightdouble}
    \renewcommand{\width}{\widthsingle}
    \node[rectangle, fill=egg, rounded corners, minimum width=\width, minimum height=\height, draw] (\nodenum) at (\xpos,\ypos) {\name};
 
 \renewcommand{\nodenum}{v3}
    \renewcommand{\name}{$U_1^T$}
	\renewcommand{\xpos}{3*\xrel}
    \renewcommand{\ypos}{-\yrel-3*\rowspace}
    \renewcommand{\height}{\heightdouble}
    \renewcommand{\width}{\widthsingle}
    \node[rectangle, fill=egg, rounded corners, minimum width=\width, minimum height=\height, draw] (\nodenum) at (\xpos,\ypos) {\name};

	\renewcommand{\nodenum}{v2}
    \renewcommand{\name}{$O_C^*$}
	\renewcommand{\xpos}{4*\xrel}
    \renewcommand{\ypos}{-4*\rowspace}
    \renewcommand{\height}{\heightsingle}
    \renewcommand{\width}{\widthsingle}
    \node[rectangle, fill=egg, rounded corners, minimum width=\width, minimum height=\height, draw] (\nodenum) at (\xpos,\ypos) {\name}; 
    
    	\renewcommand{\nodenum}{v4}
    \renewcommand{\name}{$U_2$}
	\renewcommand{\xpos}{\xrel}
    \renewcommand{\ypos}{\yrel+\rowspace}
    \renewcommand{\height}{\heightdouble}
    \renewcommand{\width}{\widthsingle}
    \node[rectangle, fill=egg, rounded corners, minimum width=\width, minimum height=\height, draw] (\nodenum) at (\xpos,\ypos) {\name};

	\renewcommand{\nodenum}{v5}
    \renewcommand{\name}{$Q$}
	\renewcommand{\xpos}{2*\xrel}
    \renewcommand{\ypos}{\yrel+\rowspace}
    \renewcommand{\height}{\heightdouble}
    \renewcommand{\width}{\widthsingle}
    \node[rectangle, fill=egg, rounded corners, minimum width=\width, minimum height=\height, draw] (\nodenum) at (\xpos,\ypos) {\name};
 
 \renewcommand{\nodenum}{v6}
    \renewcommand{\name}{$U_2^\dagger$}
	\renewcommand{\xpos}{3*\xrel}
    \renewcommand{\ypos}{\yrel+\rowspace}
    \renewcommand{\height}{\heightdouble}
    \renewcommand{\width}{\widthsingle}
    \node[rectangle, fill=egg, rounded corners, minimum width=\width, minimum height=\height, draw] (\nodenum) at (\xpos,\ypos) {\name};

	\renewcommand{\nodenum}{v5}
    \renewcommand{\name}{$O_C$}
	\renewcommand{\xpos}{4*\xrel}
    \renewcommand{\ypos}{2*\rowspace}
    \renewcommand{\height}{\heightsingle}
    \renewcommand{\width}{\widthsingle}
    \node[rectangle, fill=egg, rounded corners, minimum width=\width, minimum height=\height, draw] (\nodenum) at (\xpos,\ypos) {\name};
	
\end{tikzpicture}.
\end{equation}
$O_C$ is a Pauli string. Using $d_\mathrm{tot}=d_Cd_D$, we write this as
\begin{equation}
	C(Q,U_1,U_2)=\frac{1}{d_D^2}\av{O_C}\tr{U_1^*Q^TU_1^TO^*_C}\tr{U_2 Q U_2^\dagger O_C}.
\end{equation}
The transpose leaves the trace invariant, so
\begin{equation}
	\tr{U_1^*Q^TU_1^TO^*_C}=\tr{(U_1^*Q^TU_1^TO^*_C)^T}=\tr{O_C^\dagger U_1 Q U_1^\dagger}=\tr{U_1 Q U_1^\dagger O_C},
\end{equation}
where we used the Hermiticity of $O_C$. This produces
\begin{equation}\label{Cidentity}
	d_D^2C(Q,U_1,U_2)=\av{O_C}\tr{U _1Q U_1^\dagger O_C}\tr{U _2Q U_2^\dagger O_C}.
\end{equation}
Setting $Q=\rho_d$, we obtain an identity for $C_d(U_1,U_2)$,
\begin{equation}\label{Eq:CdTarget}
	d_D^2C_d(U_1,U_2)=\av{O_C}\tr{U _1\rho_d U_1^\dagger O_C}\tr{U _2\rho_d U_2^\dagger O_C}.
\end{equation}
 $L_d$ can be written as
\begin{equation}
L_d=\av{O_C}\tilde{y}^2_d+\av{O_C}y_d^2-2\av{O_C}y_d\tilde{y}_d.
\end{equation}
From Eq.~\eqref{Eq:CdTarget}, we readily see that 
\begin{equation}
\begin{split}
	\av{O_C} \tilde{y}_d^2&=d_D^2 C_d(U,U),\\
	\av{O_C} y_d^2&=d_D^2 C_d(U_S,U_S),\\
	\av{O_C} y_d\tilde{y}_d&=d_D^2 C_d(U,U_S).
\end{split}
\end{equation}
This produces
\begin{equation}
L_d=d_D^2(C_d(U,U)+C_d(U_S,U_S)-2C_d(U,U_S)).
\end{equation}

\subsection{Computing true error}\label{Sec:LossRandom}
We will prove that 
\begin{equation}
	 L=\frac{d_A^2}{(d_A+1)d_C^2}\Big[\overline{\ot}(U)+\overline{\ot}(U_S)-2\mathrm{OP}(U,U_S) \Big].
\end{equation}
Begin by writing $L$ using Eq.~\eqref{Eq:LossCorr}:
\begin{equation}\label{Eq:LossDefAppendix}
\begin{split}
	L
	&=\int_{\Hr}dU_A L_d\\
	&=d_D^2\int_{\Hr}dU_A(C_d(U,U)+C_d(U_S,U_S)-2C_d(U,U_S)).
\end{split}
\end{equation}
For unitaries $U_1,U_2\in \{U,U_S\}$, we compute the following average
\begin{equation}\label{Eq:Cd}
\begin{split}
\int_{\Hr} &dU_A C_d(U_1,U_2)\\
	&=\frac{1}{d_\mathrm{tot}}\av{O_D}\int_{\Hr}dU_A\tr{U_1\rho_d U_1^\dagger O_DU_2\rho_dU_2^\dagger O_D}\\
	&=
\frac{1}{d_\mathrm{tot}}\av{O_D}\int_{\Hr}dU_A\mathrm{Tr}\Big\{U_1(U_A \ket{0}\bra{0}_AU_A^\dagger\otimes \rho_B)U_1^\dagger O_D\\
&\hspace{10.5em}\cdot U_2(U_A \ket{0}\bra{0}_AU_A^\dagger\otimes \rho_B)U_2^\dagger O_D\Big\}\\
	&=\frac{1}{d_\mathrm{tot}}\av{O_D}\int_{\Hr}dU_A\tr{(U_1\otimes U_2)(U_A \ket{0}\bra{0}_AU_A^\dagger\otimes \rho_B)^{\otimes 2}(U_1^{\dagger }\otimes U_2^\dagger) O_D^{\otimes 2}\sw_{1,2}}\\
	&=\frac{1}{d_\mathrm{tot}}\av{O_D}\tr{(U_1\otimes U_2)\Big[\int_{\Hr}dU_A(U_A \ket{0}\bra{0}_AU_A^\dagger\otimes \rho_B)^{\otimes 2}\Big](U_1^{\dagger }\otimes U_2^\dagger) O_D^{\otimes 2}\sw_{1,2}}.
\end{split}
\end{equation}
In the second line, we introduce $\rho_d=U_A \ket{0}\bra{0}_AU_A^\dagger\otimes \rho_B$. In the third line, we introduce the swap operator $\sw_{1,2}$ on the doubled Hilbert space.
Compute the average over $U_A$ and keep track of the doubled systems (e.g. $A_1$ and $A_2$):
\begin{equation}\label{Eq:UAaverage}
\begin{split}
	\int_{\Hr} dU_A &(U_A \ket{0}\bra{0}_{A}U_A^\dagger\otimes \rho_B)^{\otimes 2}\\
	&=\int_{\Hr}dU_A(U_A \ket{0}\bra{0}_{A_1}U_A^\dagger\otimes \rho_{B_1})\otimes (U_A \ket{0}\bra{0}_{A_2}U_A^\dagger\otimes \rho_{B_2})\\
&=\int_{\Hr}dU_A(U_A^{\otimes 2} (\ket{0}\bra{0}_{A_1}\otimes \ket{0}\bra{0}_{A_2})U_A^{\dagger\otimes 2})\otimes(\rho_{B_1}\otimes \rho_{B_2})\\
&=\Phi_{\Hr(d_A)}^{(2)} (\ket{0}\bra{0}_{A_1}\otimes \ket{0}\bra{0}_{A_2})\otimes(\rho_{B_1}\otimes \rho_{B_2})\\
&=\frac{1}{d_A(d_A+1)}\Big[I_{A_1,A_2}+\sw_{A_1,A_2}\Big]\otimes(\rho_{B_1}\otimes \rho_{B_2})\\
&=\frac{1}{d_A(d_A+1)}\Big[I_{A_1,A_2}+d_A\av{O_{A}}O_{A}\otimes O_{A}^\dagger\Big]\otimes(\rho_{B_1}\otimes \rho_{B_2})\\
&=\frac{1}{d_A(d_A+1)d_B^2}\Big[I_{A,B}\otimes I_{A,B}+d_A\av{O_{A}}((O_{A}\otimes I_B)\otimes (O_{A}^\dagger\otimes I_B))\Big]\\
&=\frac{1}{d_A(d_A+1)d_B^2}\Big[I_{A,B}^{\otimes  2}+d_A\av{O_{A}}(O_{A}\otimes O_{A}^\dagger)\Big].
\end{split}
\end{equation}
In the second line, we switch the order of the tensor product on systems $B_1$ and $A_2$ for ease of computation. We switch back to the correct order in line six. In the third line, we introduce the 2-fold twirling channel, $\Phi_{\Hr(d_A)}^{(2)}(\cdot)$ (see Appendix~\ref{Sec:Twirl}). In the fourth line, we use Eq.~\eqref{Eq:TwirlState}. In the fifth line we introduce the 1-design identity: \linebreak
$\sw_{A_1,A_2}=d_A\av{O_A}O_{A}\otimes O_{A}^\dagger$. In the sixth line, we use $\rho_B=\frac{1}{d_B}I_B$, reorganize the tensor product and drop the $A_i$ and $B_i$ labels. In line seven, we redefine notation $O_A\otimes I_B\rightarrow O_A$. Plugging into Eq.~\ref{Eq:Cd},
\begin{equation}\label{Eq:CdandF}
\begin{split}
\int_{\Hr}dU_A&C_d(U_1,U_2)\\
&=\frac{1}{d_\mathrm{tot}}\frac{1}{d_A(d_A+1)d_B^2}\\
&\hspace{2em}\cdot\av{O_D}\tr{(U_1\otimes U_2)\Big[I_{A,B}^{\otimes  2}+d_A\av{O_{A}}(O_{A}\otimes O_{A}^\dagger)\Big](U_1^{\dagger }\otimes U_2^\dagger) O_D^{\otimes 2}\sw_{1,2 }}\\
&=\frac{1}{d_\mathrm{tot}}\frac{1}{d_A(d_A+1)d_B^2}\\
&\hspace{2em}\cdot\av{O_D}\tr{O_D^{\otimes 2}\sw_{1,2} +d_A\av{O_{A}}(U_1\otimes U_2)(O_{A}\otimes O_{A}^\dagger)(U_1^{\dagger }\otimes U_2^\dagger)O_D^{\otimes 2}\sw_{1,2} }\\
&=\frac{1}{d_\mathrm{tot}}\frac{1}{d_A(d_A+1)d_B^2}\\
&\hspace{2em}\cdot\av{O_D}\Big[\tr{O_D^2} +d_A\av{O_{A}}\tr{(U_1O_{A}U_1^\dagger O_D\otimes U_2 O_{A}^\dagger U_2^{\dagger }O_D)\sw_{1,2} }\Big]\\
	&=\frac{1}{d_\mathrm{tot}}\frac{1}{d_A(d_A+1)d_B^2}\av{O_D}\Big[d_\mathrm{tot}+d_A\av{O_{A}}\tr{U_1O_AU_1^\dagger O_DU_2 O_AU_2^\dagger O_D}\Big]\\
	&=\frac{1}{d_A(d_A+1)d_B^2}\Big[1+d_AF(U_1,U_2) \Big].
\end{split}
\end{equation}
In the fourth line, we use the Hermiticity of $O_A$ and $\tr{O_D^2}=\tr{I_{C,D}}=d_\mathrm{tot}$. In line five, we define the correlation function
\begin{equation}\label{Eq:F}
	F(U_1,U_2)=\av{O_A}\av{O_D}\langle U_1 O_A U_1^\dagger O_D U_2 O_A U_2^\dagger O_D\rangle.
\end{equation}
In the case where $U_1=U_2$, $F(U_1,U_1)=\overline{\ot}(U_1)$.
In the case where $U_1=U$ and $U_2=U_S$, we retrieve the optimization correlator  $F(U,U_S)=\mathrm{OP}(U,U_S)$. From Eq.~\eqref{Eq:CdandF}, we readily find
\begin{align}
	\int_{\Hr}dU_AC_d(U_1,U_1)&=\frac{1}{d_A(d_A+1)d_B^2}\Big[1+d_A\overline{\ot}(U_1) \Big],\label{Eq:Cdand OTOC}\\
	\int_{\Hr}dU_AC_d(U,U_S)&=\frac{1}{d_A(d_A+1)d_B^2}\Big[1+d_A\mathrm{OP}(U,U_S) \Big].\label{Eq:Cdand OP}
\end{align}
Eq.~\eqref{Eq:LossDefAppendix} is then

\begin{equation}
\begin{split}
	L&=\frac{d_D^2}{d_A(d_A+1)d_B^2}\Big[(1+d_A\overline{\ot}(U) )+(1+d_A\overline{\ot}(U_S))-2(1+d_A\mathrm{OP}(U,U_S) ) \Big]\\
	&=\frac{d_D^2d_A}{d_A(d_A+1)d_B^2}\Big[\overline{\ot}(U)+\overline{\ot}(U_S)-2\mathrm{OP}(U,U_S)\Big]\\
&=G\Big[\overline{\ot}(U)+\overline{\ot}(U_S)-2\mathrm{OP}(U,U_S) \Big].
\end{split}
\end{equation}
In the third line, we use $d_Ad_B=d_Cd_D$ and define $G=\frac{d_A^2}{(d_A+1)d_C^2}$.

\section{Proof of Corollary~\ref{Corollary:Us}}\label{Sec:AvgLoss}
We show that the true error for a maximally scrambling target unitary is
\begin{equation}
		L_{\mathrm{scram}}=G\left(\overline{\ot}(U) +\overline{\mathrm{\ot}}_{\mathrm{scram}}-\frac{2}{d_A^2}\right).
\end{equation}
This is computed from $L$ by integrating $U_S$ over the Haar measure on the unitary group, which is valid under the assumption $U_S$ is scrambling:
\begin{equation}
\begin{split}
	L_{\mathrm{scram}}
	&=\int_{\Hr}dU_S L\\
	&=G\int_{\Hr}dU_S\Big[\overline{\ot}(U) +\overline{\ot}(U_S)-2\mathrm{OP}(U,U_S)  \Big].
\end{split}
\end{equation}
First compute the integral over the OTOC:
\begin{equation}\label{Eq:AvgOTOC}
\begin{split}
	\int_{\Hr}dU_S\overline{\ot}(U_S)&=\int_{\Hr}dU_S\av{O_A}\av{O_D}\langle U_SO_AU_S^\dagger O_DU_S O_AU_S^\dagger O_D\rangle\\
	&=\frac{1}{d_\mathrm{tot}}\av{O_A}\av{O_D}\int_{\Hr}dU_S\tr{ U_SO_AU_S^\dagger O_DU_S O_AU_S^\dagger O_D}\\
	&=\frac{1}{d_\mathrm{tot}}\av{O_A}\av{O_D}\tr{ \left[\int_{\Hr}dU_SU_S^{\otimes 2}O_A^{\otimes 2}U_S^{\dagger\otimes 2} \right]O_D^{\otimes 2}\sw_{1,2}}\\
	&=\frac{1}{d_\mathrm{tot}}\av{O_D}\tr{ \av{O_A}\Phi_{\Hr(d_{\mathrm{tot}})}^{(2)}(O_A^{\otimes 2})O_D^{\otimes 2}\sw_{1,2}}.
\end{split}
\end{equation}
The average over $U_S$ is just the 2-fold twirling channel, $\Phi_{\Hr(d_\mathrm{tot})}^{(2)}(O_A^{\otimes 2})$ (see Appendix~\ref{Sec:Twirl}). The swap operator  $\sw_{1,2}$ acts over the doubled Hilbert space. Using Eq.~\eqref{Eq:TwoTwirl} we compute the average of the twirling channel,
\begin{equation}
\begin{split}
	\Phi_{\Hr(d_\mathrm{tot})}^{(2)}(O_A^{\otimes 2})
	&=\frac{1}{d_\mathrm{tot}^2-1}\Big[I_{1,2}\tr{O_A^{\otimes 2}}+\sw_{1,2}\tr{O_A^{\otimes 2}\sw_{1,2}}\\
	&\hspace{20mm}-\frac{1}{d_\mathrm{tot}}I_{1,2}\tr{O_A^{\otimes 2}\sw_{1,2}}-\frac{1}{d_\mathrm{tot}}\sw_{1,2}\tr{O_A^{\otimes 2}}\Big]\\
	& =\frac{1}{d_\mathrm{tot}^2-1}\Big[I_{1,2}\left(\tr{O_A}^2-\frac{1}{d_\mathrm{tot}}\tr{O_A^2}\right)\\
	&\hspace{20mm}+\sw_{1,2}\left(\tr{O_A^2}-\frac{1}{d_\mathrm{tot}}\tr{O_A}^2\right)\Big]\\
&=\frac{1}{d_\mathrm{tot}^2-1}\left[I_{1,2}\left(d_\mathrm{tot}^2\delta_{O_A,I_A}-1\right)+\sw_{1,2} \left(d_\mathrm{tot}-d_\mathrm{tot}\delta_{O_A,I_A}\right)\right].
\end{split}
\end{equation}
In the above, $O_A$ actually denotes $O_A\otimes I_B$, so ${\tr{O_A}=d_\mathrm{tot}\delta_{O_A,I_A}}$. We also use \linebreak ${\tr{O_A^2}=\tr{I_{A,B}}=d_{\mathrm{tot}}}$. Now average the twirling channel over $O_A$, using \linebreak ${\av{O_A}\delta_{O_A,I_A}=\frac{1}{d_A^2}\sum_{O_A}\delta_{O_A,I_A}=\frac{1}{d^2_A}}$:
\begin{equation}\label{Eq:TwirlPauli}
\begin{split}
	\av{O_A}\Phi_{\Hr(d_\mathrm{tot})}^{(2)}(O_A^{\otimes 2}) &=\frac{1}{d_\mathrm{tot}^2-1}\left[I_{1,2}\left(d_\mathrm{tot}^2\av{O_A}\delta_{O_A,I_A}-1\right)+\sw_{1,2} \left(d_\mathrm{tot}-d_\mathrm{tot}\av{O_A}\delta_{O_A,I_A}\right)\right]\\
	 &=\frac{1}{d_\mathrm{tot}^2-1}\left[I_{1,2}\left(\frac{d_\mathrm{tot}^2}{d_A^2}-1\right)+\sw_{1,2}\left(d_\mathrm{tot}-\frac{d_\mathrm{tot}}{d_A^2}\right)\right].
\end{split}
\end{equation}
Plug this into Eq.~\eqref{Eq:AvgOTOC}
\begin{equation}
\begin{split}
	\int_{\Hr} &dU_S\overline{\ot}(U_S)\\
	&=\frac{1}{d_\mathrm{tot}(d_\mathrm{tot}^2-1)}\av{O_D}\tr{ \left[I_{1,2}\left(\frac{d_\mathrm{tot}^2}{d_A^2}-1\right)+\sw_{1,2}\left(d_\mathrm{tot}-\frac{d_\mathrm{tot}}{d_A^2}\right)\right]O_D^{\otimes 2}\sw_{1,2}}\\
	&=\frac{1}{d_\mathrm{tot}(d_\mathrm{tot}^2-1)}\av{O_D} \left[\tr{O_D^{\otimes 2}\sw_{1,2}}\left(\frac{d_\mathrm{tot}^2}{d_A^2}-1\right)+\tr{\sw_{1,2} O_D^{\otimes 2}\sw_{1,2}} \left(d_\mathrm{tot}-\frac{d_\mathrm{tot}}{d_A^2}\right)\right]\\
	&=\frac{1}{d_\mathrm{tot}(d_\mathrm{tot}^2-1)}\av{O_D} \left[\tr{O_D^2}\left(\frac{d_\mathrm{tot}^2}{d_A^2}-1\right)+\tr{O_D}^2 \left(d_\mathrm{tot}-\frac{d_\mathrm{tot}}{d_A^2}\right)\right]\\
	&=\frac{1}{d_\mathrm{tot}(d_\mathrm{tot}^2-1)}\av{O_D} \left[d_\mathrm{tot}\left(\frac{d_\mathrm{tot}^2}{d_A^2}-1\right)+d_\mathrm{tot}^2\delta_{O_D,I_D} \left(d_\mathrm{tot}-\frac{d_\mathrm{tot}}{d_A^2}\right)\right]\\
	&=\frac{1}{d_\mathrm{tot}(d_\mathrm{tot}^2-1)}\left[d_\mathrm{tot}\left(\frac{d_\mathrm{tot}^2}{d_A^2}-1\right)+d_C^2\left(d_\mathrm{tot}-\frac{d_\mathrm{tot}}{d_A^2}\right)\right]\\
	&=\overline{\ot}_{\mathrm{scram}},
\end{split}
\end{equation}
where we define
\begin{equation}
	\overline{\ot}_{\mathrm{scram}}=\frac{1}{(d_\mathrm{tot}^2-1)}\left[\left(\frac{d_\mathrm{tot}^2}{d_A^2}-1\right)+d_C^2\left(1-\frac{1}{d_A^2}\right)\right].
\end{equation}
In the above, we use $\tr{O_D}=d_\mathrm{tot}\delta_{O_D,I_D}$, $\tr{O_D^2}=d_\mathrm{tot}$, and $\av{O_D}\delta_{O_D,I_D}=\frac{1}{d_D^2}$. For large $d_\mathrm{tot}$,
$\overline{\ot}_{\mathrm{scram}}\rightarrow \frac{1}{d_A^2}+\frac{1}{d_D^2}-\frac{1}{d_A^2d_D^2}$.
 This result was originally shown in \cite{PhysRevX.9.011006}. 
 
 Now we compute the integral over the optimization correlator, assuming $U$ is independent of $U_S$ (i.e. $U$ is not yet trained):
\begin{equation}
\begin{split}
	\int_{\Hr}dU_S\mathrm{OP}(U,U_S)  &=\frac{1}{d_\mathrm{tot}}\int_{\Hr}dU_S\av{O_A}\av{O_D}\tr{UO_AU^\dagger O_DU_S O_AU_S^\dagger O_D}\\
	&=\frac{1}{d_\mathrm{tot}}\av{O_A}\av{O_D}\tr{UO_AU^\dagger O_D \Phi_{\Hr(d_\mathrm{tot})}^{(1)}(O_A) O_D}\\
	&=\frac{1}{d_\mathrm{tot}^2}\av{O_A}\av{O_D}\tr{UO_AU^\dagger  O_D^2}\tr{O_A}\\
	&=\frac{1}{d_\mathrm{tot}^2}\av{O_A}\av{O_D}\tr{O_A}^2\\
	&=\frac{1}{d_\mathrm{tot}^2}\av{O_A}\av{O_D}(d_\mathrm{tot}^2\delta_{O_A,I_A})\\
	&=\frac{1}{d_A^2}.
	\end{split}
\end{equation}
$L_{\mathrm{scram}}$ becomes
\begin{equation}
	L_{\mathrm{scram}}=G\left(\overline{\ot}(U) +\overline{\ot}_{\mathrm{scram}}-\frac{2}{d_A^2}\right).
\end{equation}
In the case where $U$ is maximally scrambling, we obtain the floor value
\begin{equation}
	L_{\mathrm{floor}}=2G\left(\overline{\ot}_{\mathrm{scram}}-\frac{1}{d_A^2}\right).
\end{equation}
In the large $d_\mathrm{tot}$ limit, this floor value becomes
\begin{equation}
\begin{split}
	\lim_{d_\mathrm{tot}\rightarrow \infty}L_{\mathrm{floor}}	&=2G\left(\lim_{d_\mathrm{tot}\rightarrow \infty}\overline{\ot}_{\mathrm{scram}}-\frac{1}{d_A^2}\right)\\
	&=2G\left(\frac{1}{d_A^2}+\frac{1}{d_D^2}-\frac{1}{d_A^2d_D^2}-\frac{1}{d_A^2}\right)\\
	&=\frac{2d_A^2}{(d_A+1)d_C^2}\left(\frac{1}{d_D^2}-\frac{1}{d_A^2d_D^2}\right)\\
		&=\frac{2}{(d_A+1)d_B^2}\left(1-\frac{1}{d_A^2}\right).
\end{split}
\end{equation}
For $d_A>>1$,
\begin{equation}
	\lim_{d_\mathrm{tot}\rightarrow \infty}L_{\mathrm{floor}}=\frac{2}{d_Bd_\mathrm{tot}}.
\end{equation}

\section{Proof of Proposition~\ref{Prop:true-error}}\label{Sec:OTOCProb}
We will prove that $L$ is bounded by
\begin{equation}
	L_{\pm}=G\Big[\sqrt{\overline{\ot}(U) }\pm\sqrt{\overline{\ot}(U_S)}\Big]^2.
\end{equation}
By noting $L$ is non-negative, we apply the triangle inequality to the true error in Eq.~\eqref{Eq:LossOTOC}:
\begin{equation}\label{Eq:LIneq}
	L\leq G\Big[\abs{\overline{\ot}(U)} +\abs{\overline{\ot}(U_S)}+2|\mathrm{OP}(U,U_S) | \Big].
\end{equation}
We now bound the third term. Begin by rewriting $F(U_1,U_2)$ from Eq.~\eqref{Eq:F} diagrammatically, taking $O_A=O_A^\dagger$ and $O_D=O_D^\dagger$ wherever convenient:
\begin{equation}
\begin{tikzpicture}

    
	\renewcommand{\nodenum}{v0}
    \renewcommand{\name}{$F(U_1,U_2)=\frac{1}{d_\mathrm{tot}}\av{O_A}\av{O_D}$}
	\renewcommand{\xpos}{-2.5*\xrel}
    \renewcommand{\ypos}{-\rowspace}
    \renewcommand{\height}{\heightsingle}
    \renewcommand{\width}{\widthsingle}
    \node[] (\nodenum) at (\xpos,\ypos) {\name};

    \renewcommand{\nodenum}{v1}
    \renewcommand{\name}{$Trace1$}
	\renewcommand{\xpos}{4.5*\xrel}
    \renewcommand{\ypos}{-\rowspace}
    \renewcommand{\height}{\heighttrace}
    \renewcommand{\width}{25em}
    \node[rectangle,rounded corners, thick, minimum width=\width, minimum height=\height, draw=dullblue] (\nodenum) at (\xpos,\ypos) {};

    \renewcommand{\nodenum}{v1}
    \renewcommand{\name}{$Trace2$}
	\renewcommand{\xpos}{4.5*\xrel}
    \renewcommand{\ypos}{-\rowspace}
    \renewcommand{\height}{3*\heighttrace}
    \renewcommand{\width}{27em}
    \node[rectangle,rounded corners, thick, minimum width=\width, minimum height=\height, draw=dullblue] (\nodenum) at (\xpos,\ypos) {};

	\renewcommand{\nodenum}{v1}
    \renewcommand{\name}{$U_1$}
	\renewcommand{\xpos}{\xrel}
    \renewcommand{\ypos}{0}
    \renewcommand{\height}{\heightdouble}
    \renewcommand{\width}{\widthsingle}
    \node[rectangle, fill=egg, rounded corners, minimum width=\width, minimum height=\height, draw] (\nodenum) at (\xpos,\ypos) {\name};

	\renewcommand{\nodenum}{v2}
    \renewcommand{\name}{$O_A^\dagger$}
	\renewcommand{\xpos}{2*\xrel}
    \renewcommand{\ypos}{\yrel}
    \renewcommand{\height}{\heightsingle}
    \renewcommand{\width}{\widthsingle}
    \node[rectangle, fill=egg, rounded corners, minimum width=\width, minimum height=\height, draw] (\nodenum) at (\xpos,\ypos) {\name};
 
 \renewcommand{\nodenum}{v3}
    \renewcommand{\name}{$U_1^\dagger$}
	\renewcommand{\xpos}{3*\xrel}
    \renewcommand{\ypos}{0}
    \renewcommand{\height}{\heightdouble}
    \renewcommand{\width}{\widthsingle}
    \node[rectangle, fill=egg, rounded corners, minimum width=\width, minimum height=\height, draw] (\nodenum) at (\xpos,\ypos) {\name};

	\renewcommand{\nodenum}{v4}
    \renewcommand{\name}{$O_D$}
	\renewcommand{\xpos}{4*\xrel}
    \renewcommand{\ypos}{-\yrel}
    \renewcommand{\height}{\heightsingle}
    \renewcommand{\width}{\widthsingle}
    \node[rectangle, fill=egg, rounded corners, minimum width=\width, minimum height=\height, draw] (\nodenum) at (\xpos,\ypos) {\name};
    
	 \renewcommand{\nodenum}{v5}
    \renewcommand{\name}{$U_2$}
	\renewcommand{\xpos}{5*\xrel}
    \renewcommand{\ypos}{0}
    \renewcommand{\height}{\heightdouble}
    \renewcommand{\width}{\widthsingle}
    \node[rectangle, fill=egg, rounded corners, minimum width=\width, minimum height=\height, draw] (\nodenum) at (\xpos,\ypos) {\name};

	\renewcommand{\nodenum}{v6}
    \renewcommand{\name}{$O_A$}
	\renewcommand{\xpos}{6*\xrel}
    \renewcommand{\ypos}{\yrel}
    \renewcommand{\height}{\heightsingle}
    \renewcommand{\width}{\widthsingle}
    \node[rectangle, fill=egg, rounded corners, minimum width=\width, minimum height=\height, draw] (\nodenum) at (\xpos,\ypos) {\name};
    
     \renewcommand{\nodenum}{v7}
    \renewcommand{\name}{$U_2^\dagger$}
	\renewcommand{\xpos}{7*\xrel}
    \renewcommand{\ypos}{0}
    \renewcommand{\height}{\heightdouble}
    \renewcommand{\width}{\widthsingle}
    \node[rectangle, fill=egg, rounded corners, minimum width=\width, minimum height=\height, draw] (\nodenum) at (\xpos,\ypos) {\name};

	\renewcommand{\nodenum}{v8}
    \renewcommand{\name}{$O_D^\dagger$}
	\renewcommand{\xpos}{8*\xrel}
    \renewcommand{\ypos}{-\yrel}
    \renewcommand{\height}{\heightsingle}
    \renewcommand{\width}{\widthsingle}
    \node[rectangle, fill=egg, rounded corners, minimum width=\width, minimum height=\height, draw] (\nodenum) at (\xpos,\ypos) {\name};
    
\end{tikzpicture}.
\end{equation}
Using the transpose identity from Eq.~\eqref{Eq:Tranpose},
\begin{equation}
\begin{tikzpicture}

    

    \renewcommand{\nodenum}{v1}
    \renewcommand{\name}{$Trace1$}
	\renewcommand{\xpos}{2*\xrel}
    \renewcommand{\ypos}{-\rowspace}
    \renewcommand{\height}{\heighttrace}
    \renewcommand{\width}{16em}
    \node[rectangle,rounded corners, thick, minimum width=\width, minimum height=\height, draw=dullblue] (\nodenum) at (\xpos,\ypos) {}; 
    
    \renewcommand{\nodenum}{v1}
    \renewcommand{\name}{$Trace2$}
	\renewcommand{\xpos}{2*\xrel}
    \renewcommand{\ypos}{-\rowspace}
    \renewcommand{\height}{3*\heighttrace}
    \renewcommand{\width}{18em}
    \node[rectangle,rounded corners, thick, minimum width=\width, minimum height=\height, draw=dullblue] (\nodenum) at (\xpos,\ypos) {};

	\renewcommand{\nodenum}{v0}
    \renewcommand{\name}{$F(U_1,U_2)=\frac{1}{d_\mathrm{tot}}\av{O_A}\av{O_D}$}
	\renewcommand{\xpos}{-3.5*\xrel}
    \renewcommand{\ypos}{-\rowspace}
    \renewcommand{\height}{\heightsingle}
    \renewcommand{\width}{\widthsingle}
    \node[] (\nodenum) at (\xpos,\ypos) {\name};

     \renewcommand{\nodenum}{v5}
    \renewcommand{\name}{$O_A^\dagger$}
	\renewcommand{\xpos}{0}
    \renewcommand{\ypos}{\yrel}
    \renewcommand{\height}{\heightsingle}
    \renewcommand{\width}{\widthsingle}
    \node[rectangle, fill=egg, rounded corners, minimum width=\width, minimum height=\height, draw] (\nodenum) at (\xpos,\ypos) {\name};

	\renewcommand{\nodenum}{v1}
    \renewcommand{\name}{$U_1^T$}
	\renewcommand{\xpos}{\xrel}
    \renewcommand{\ypos}{-2*\rowspace}
    \renewcommand{\height}{\heightdouble}
    \renewcommand{\width}{\widthsingle}
    \node[rectangle, fill=egg, rounded corners, minimum width=\width, minimum height=\height, draw] (\nodenum) at (\xpos,\ypos) {\name};

	\renewcommand{\nodenum}{v2}
    \renewcommand{\name}{$O_D^*$}
	\renewcommand{\xpos}{2*\xrel}
    \renewcommand{\ypos}{\yrel-2*\rowspace}
    \renewcommand{\height}{\heightsingle}
    \renewcommand{\width}{\widthsingle}
    \node[rectangle, fill=egg, rounded corners, minimum width=\width, minimum height=\height, draw] (\nodenum) at (\xpos,\ypos) {\name};
 
 \renewcommand{\nodenum}{v3}
    \renewcommand{\name}{$U_2^*$}
	\renewcommand{\xpos}{3*\xrel}
    \renewcommand{\ypos}{-2*\rowspace}
    \renewcommand{\height}{\heightdouble}
    \renewcommand{\width}{\widthsingle}
    \node[rectangle, fill=egg, rounded corners, minimum width=\width, minimum height=\height, draw] (\nodenum) at (\xpos,\ypos) {\name};

	\renewcommand{\nodenum}{v4}
    \renewcommand{\name}{$O_A$}
	\renewcommand{\xpos}{4*\xrel}
    \renewcommand{\ypos}{\yrel}
    \renewcommand{\height}{\heightsingle}
    \renewcommand{\width}{\widthsingle}
    \node[rectangle, fill=egg, rounded corners, minimum width=\width, minimum height=\height, draw] (\nodenum) at (\xpos,\ypos) {\name};

    	\renewcommand{\nodenum}{v6}
    \renewcommand{\name}{$U_1^\dagger$}
	\renewcommand{\xpos}{\xrel}
    \renewcommand{\ypos}{0}
    \renewcommand{\height}{\heightdouble}
    \renewcommand{\width}{\widthsingle}
    \node[rectangle, fill=egg, rounded corners, minimum width=\width, minimum height=\height, draw] (\nodenum) at (\xpos,\ypos) {\name};

	\renewcommand{\nodenum}{v7}
    \renewcommand{\name}{$O_D$}
	\renewcommand{\xpos}{2*\xrel}
    \renewcommand{\ypos}{-\yrel}
    \renewcommand{\height}{\heightsingle}
    \renewcommand{\width}{\widthsingle}
    \node[rectangle, fill=egg, rounded corners, minimum width=\width, minimum height=\height, draw] (\nodenum) at (\xpos,\ypos) {\name};
 
 \renewcommand{\nodenum}{v8}
    \renewcommand{\name}{$U_2$}
	\renewcommand{\xpos}{3*\xrel}
    \renewcommand{\ypos}{0}
    \renewcommand{\height}{\heightdouble}
    \renewcommand{\width}{\widthsingle}
    \node[rectangle, fill=egg, rounded corners, minimum width=\width, minimum height=\height, draw] (\nodenum) at (\xpos,\ypos) {\name};

\end{tikzpicture}.
\end{equation}
Let $\Pi_{D_1,D_2}$ be the projection onto the Bell state between systems $D_1$ and $D_2$
\begin{equation}
\Pi_{D_1,D_2}=\ket{\rm{Bell}}\bra{\rm{Bell}}_{D_1,D_2}.
\end{equation}
We average over $O_A$ and $O_D$, and use the 1-design identity $\av{O_D}O_D\otimes O_D^*=\Pi_{D_1,D_2}$:
\begin{equation}\label{Eq:Fdiagram}
\begin{tikzpicture}

    
    
	\renewcommand{\nodenum}{v0}
    \renewcommand{\name}{$F(U_1,U_2)=\frac{1}{d_\mathrm{tot}d_A}$}
	\renewcommand{\xpos}{-2*\xrel}
    \renewcommand{\ypos}{-\rowspace}
    \renewcommand{\height}{\heightsingle}
    \renewcommand{\width}{\widthsingle}
    \node[] (\nodenum) at (\xpos,\ypos) {\name};

    \renewcommand{\nodenum}{v1}
    \renewcommand{\name}{$Trace1$}
	\renewcommand{\xpos}{2*\xrel}
    \renewcommand{\ypos}{\rowspace}
    \renewcommand{\height}{\heighttrace}
    \renewcommand{\width}{12em}
    \node[rectangle,rounded corners, thick, minimum width=\width, minimum height=\height, draw=dullblue] (\nodenum) at (\xpos,\ypos) {}; 
     
    \renewcommand{\nodenum}{v1}
    \renewcommand{\name}{$Trace2$}
	\renewcommand{\xpos}{2*\xrel}
    \renewcommand{\ypos}{-\rowspace}
    \renewcommand{\height}{\heighttrace}
    \renewcommand{\width}{12em}
    \node[rectangle,rounded corners, thick, minimum width=\width, minimum height=\height, draw=dullblue] (\nodenum) at (\xpos,\ypos) {};

	\renewcommand{\nodenum}{v1}
    \renewcommand{\name}{$Trace3$}
	\renewcommand{\xpos}{2*\xrel}
    \renewcommand{\ypos}{-3*\rowspace}
    \renewcommand{\height}{1.3*\heightsingle}
    \renewcommand{\width}{12em}
    \node[rectangle,rounded corners, thick, minimum width=\width, minimum height=\height, draw=dullblue] (\nodenum) at (\xpos,\ypos) {};

	\renewcommand{\nodenum}{v1}
    \renewcommand{\name}{$U_1^T$}
	\renewcommand{\xpos}{.75*\xrel}
    \renewcommand{\ypos}{-2*\rowspace}
    \renewcommand{\height}{\heightdouble}
    \renewcommand{\width}{\widthsingle}
    \node[rectangle, fill=egg, rounded corners, minimum width=\width, minimum height=\height, draw] (\nodenum) at (\xpos,\ypos) {\name};

 \renewcommand{\nodenum}{v2}
    \renewcommand{\name}{$U_2^*$}
	\renewcommand{\xpos}{3.25*\xrel}
    \renewcommand{\ypos}{-2*\rowspace}
    \renewcommand{\height}{\heightdouble}
    \renewcommand{\width}{\widthsingle}
    \node[rectangle, fill=egg, rounded corners, minimum width=\width, minimum height=\height, draw] (\nodenum) at (\xpos,\ypos) {\name};

    	\renewcommand{\nodenum}{v3}
    \renewcommand{\name}{$U_1^\dagger$}
	\renewcommand{\xpos}{.75*\xrel}
    \renewcommand{\ypos}{0}
    \renewcommand{\height}{\heightdouble}
    \renewcommand{\width}{\widthsingle}
    \node[rectangle, fill=egg, rounded corners, minimum width=\width, minimum height=\height, draw] (\nodenum) at (\xpos,\ypos) {\name};

 \renewcommand{\nodenum}{v4}
    \renewcommand{\name}{$U_2$}
	\renewcommand{\xpos}{3.25*\xrel}
    \renewcommand{\ypos}{0}
    \renewcommand{\height}{\heightdouble}
    \renewcommand{\width}{\widthsingle}
    \node[rectangle, fill=egg, rounded corners, minimum width=\width, minimum height=\height, draw] (\nodenum) at (\xpos,\ypos) {\name};

	\renewcommand{\nodenum}{v5}
    \renewcommand{\name}{$\Pi_{D_1,D_2}$}
	\renewcommand{\xpos}{2*\xrel}
    \renewcommand{\ypos}{-\rowspace}
    \renewcommand{\height}{\heightdouble}
    \renewcommand{\width}{\widthsingle}
    \node[rectangle, fill=egg, rounded corners, minimum width=\width, minimum height=\height, draw] (\nodenum) at (\xpos,\ypos) {\name};

     \renewcommand{\nodenum}{v18}
    \renewcommand{\name}{$R_1$}
	\renewcommand{\xpos}{5*\xrel}
    \renewcommand{\ypos}{\yrel+\rowspace}
    \renewcommand{\height}{\heightsingle}
    \renewcommand{\width}{\widthsingle}
    \node[] (\nodenum) at (\xpos,\ypos) {\name};
    
     \renewcommand{\nodenum}{v17}
    \renewcommand{\name}{$A_1$}
	\renewcommand{\xpos}{5*\xrel}
    \renewcommand{\ypos}{\yrel}
    \renewcommand{\height}{\heightsingle}
    \renewcommand{\width}{\widthsingle}
    \node[] (\nodenum) at (\xpos,\ypos) {\name};
    
	\renewcommand{\nodenum}{v17}
    \renewcommand{\name}{$B_1$}
	\renewcommand{\xpos}{5*\xrel}
    \renewcommand{\ypos}{-\yrel}
    \renewcommand{\height}{\heightsingle}
    \renewcommand{\width}{\widthsingle}
    \node[] (\nodenum) at (\xpos,\ypos) {\name};
    
	\renewcommand{\nodenum}{v18}
    \renewcommand{\name}{$B_2$}
	\renewcommand{\xpos}{5*\xrel}
    \renewcommand{\ypos}{\yrel-2*\rowspace}
    \renewcommand{\height}{\heightsingle}
    \renewcommand{\width}{\widthsingle}
    \node[] (\nodenum) at (\xpos,\ypos) {\name};
    
	\renewcommand{\nodenum}{v18}
    \renewcommand{\name}{$A_2$}
	\renewcommand{\xpos}{5*\xrel}
    \renewcommand{\ypos}{\yrel-3*\rowspace}
    \renewcommand{\height}{\heightsingle}
    \renewcommand{\width}{\widthsingle}
    \node[] (\nodenum) at (\xpos,\ypos) {\name};    
            
    \renewcommand{\nodenum}{v18}
    \renewcommand{\name}{$R_2$}
	\renewcommand{\xpos}{5*\xrel}
    \renewcommand{\ypos}{-\yrel-3*\rowspace}
    \renewcommand{\height}{\heightsingle}
    \renewcommand{\width}{\widthsingle}
    \node[] (\nodenum) at (\xpos,\ypos) {\name};
    
\end{tikzpicture}.
\end{equation}
The labels on the right-hand side refer to the systems in the tensor network. Reference system $R_i$ has dimension $d_A$, where $i\in\{1,2\}$. 

Define the states
\begin{equation}
\begin{split}
	\ket{\psi(U_i)}&=I_{R_1}\otimes U_i \otimes U_i^*\otimes I_{R_2}\ket{\psi_0},\\
	\ket{\psi_0}&=\ket{\rm{Bell}}_{R_1,A_1}\otimes \ket{\rm{Bell}}_{B_1B_2}\otimes \ket{\rm{Bell}}_{A_2R_2}.
\end{split}
\end{equation}
Eq.~\eqref{Eq:Fdiagram} can be written as
\begin{equation}
	F(U_1,U_2)=\bra{\psi(U_1)}\Pi_{D_1,D_2}\ket{\psi(U_2)}.
\end{equation}
$\Pi_{D_1,D_2}$ actually denotes $I_{R_1,A_1,A_2,R_2}\otimes\Pi_{D_1,D_2}$, but the identity operator is omitted for convenience. Setting $U_1=U_2$ yields $F(U_1,U_1)=\overline{\ot}(U_1)$, as shown in Appendix~\ref{Sec:LossRandom}. $F(U_1,U_1)$ also gives the probability $P(U_1)$ of projecting a Bell state onto $D_1D_2$ while in state $\ket{\psi(U_1)}$ \cite{PhysRevX.9.011006}. We have the relation
\begin{equation}
	\overline{\ot}(U_1)=\bra{\psi(U_1)}\Pi_{D_1,D_2}\ket{\psi(U_1)}=P(U_1).
\end{equation}
Since the OTOC represents a probability,  $0\leq \overline{\ot}(U_1)\leq 1$. In the case where $U_1=U$ and $U_2=U_S$, $F(U,U_S)=\mathrm{OP}(U,U_S)$, which produces
\begin{equation}
	\mathrm{OP}(U,U_S)=\bra{\psi(U)}\Pi_{D_1,D_2}\ket{\psi(U_S)}.
\end{equation}

We now bound $\abs{\mathrm{OP}(U,U_S)}$. Begin by writing
\begin{equation}
\begin{split}
	\abs{\mathrm{OP}(U,U_S)}&=\abs{\bra{\psi(U)}\Pi_{D_1,D_2}^2\ket{\psi(U_S)}}\\
	&=\abs{\bra{\psi(U,\Pi_{D_1,D_2})}\psi(U_S,\Pi_{D_1,D_2})\rangle},
\end{split}
\end{equation}
where we define the unnormalized state $\ket{\psi(U_i,\Pi_{D_1,D_2})}=\Pi_{D_1,D_2}\ket{\psi(U_i)}$ and use the fact $\Pi_{D_1,D_2}=\Pi_{D_1,D_2}^2$. Apply the Cauchy-Schwarz inequality,
\begin{equation}
\begin{split}
	\abs{\mathrm{OP}(U,U_S)}&\leq\sqrt{ \bra{\psi(U,\Pi_{D_1,D_2})}\psi(U,\Pi_{D_1,D_2})\rangle\bra{\psi(U_S,\Pi_{D_1,D_2})}\psi(U_S,\Pi_{D_1,D_2})\rangle}\\
	&=\sqrt{\bra{\psi(U)}\Pi_{D_1,D_2}^2\ket{\psi(U)}\bra{\psi(U_S)}\Pi_{D_1,D_2}^2\ket{\psi(U_S)}}\\
	&=\sqrt{\bra{\psi(U)}\Pi_{D_1,D_2}\ket{\psi(U)}\bra{\psi(U_S)}\Pi_{D_1,D_2}\ket{\psi(U_S)}}\\
	&=\sqrt{\overline{\ot}(U)\overline{\ot}(U_S)}.
\end{split}
\end{equation}
Ineq.~\ref{Eq:LIneq} becomes
\begin{equation}
	L\leq G\Big[\overline{\ot}(U) +\overline{\ot}(U_S)+2\sqrt{\overline{\ot}(U)\overline{\ot}(U_S)}\Big].
\end{equation}
The right-hand side is defined as the upper bound on $L$:
\begin{equation}
	L_{+}=G\Big[\sqrt{\overline{\ot}(U) }+\sqrt{\overline{\ot}(U_S)}\Big]^2.
\end{equation}
We also construct a lower bound on $L$. We first bound $\mathrm{OP}(U,U_S)$:
\begin{equation}
\begin{split}
\mathrm{OP}(U,U_S)&\leq \abs{\mathrm{OP}(U,U_S)}\\
	&\leq \sqrt{\overline{\ot}(U)\overline{\ot}(U_S)}.
\end{split}
\end{equation}
We now bound the true error:
\begin{equation}
\begin{split}
	L&=G\Big[\overline{\ot}(U)+\overline{\ot}(U_S)-2\mathrm{OP}(U,U_S) \Big]\\
	&\geq G\Big[\overline{\ot}(U)+\overline{\ot}(U_S)-2\sqrt{\overline{\ot}(U)\overline{\ot}(U_S)}\Big].
\end{split}
\end{equation}
We define the right-hand side as the lower bound on $L$:
\begin{equation}
	L_{-}=G\Big[\sqrt{\overline{\ot}(U)}-\sqrt{\overline{\ot}(U_S)}\Big]^2.
\end{equation}
Using $\abs{L-L_{\pm}}\leq L_{+}-L_{-}$, we arrive at the bound
\begin{equation}
	\abs{L-L_{\pm}}\leq 4G\sqrt{\overline{\ot}(U)\overline{\ot}(U_S)}.
\end{equation}

\section{Proof of Proposition~\ref{Prop:LossBound}}\label{Sec:LevyBound}
We derive a concentration inequality for the loss function by using Levy's lemma. Assume input state $\rho_d$ is given by Eq.~\eqref{Eq:rho}. It will be convenient to introduce the notation $L_d= L_d(\ket{\psi_{d,A}})$. Levy's lemma in this context states:
\begin{lemma}[Levy's lemma]
Let $L_d : S^{2d_{\mathrm{tot}}-1}\rightarrow \mathbb{R}$ satisfy $\abs{L_d(\ket{\psi_{1,A}})-L_d(\ket{\psi_{2,A}})}\leq \eta\norm{\ket{\psi_{1,A}}-\ket{\psi_{2,A}}}_2$. Then $\forall$ $\delta\geq 0$,
\begin{equation}\label{Eq:Probability}
	\underset{\ket{\psi_{d,A}}\sim \Hr(d_A)}{\mathrm{Prob}}\left[\abs{L_d(\ket{\psi_{d,A}})-L}\geq \delta \right]\leq \epsilon,
\end{equation}
where $\epsilon=2\mathrm{exp}\left(-\frac{2d_A\delta^2}{9\pi^3\eta^2}\right)$ and $\eta$ is the Lipschitz constant. By definition, \linebreak $L=\av{\ket{\psi_{d,A}}\sim \Hr(d_A)}\left[L_d(\ket{\psi_{d,A}})\right].$ The average is over the uniform distribution on the Hilbert space of system $A$.
\end{lemma}
Parameter $\delta$ can be written in terms of $\epsilon$: $\delta=\eta f(\epsilon)$, where we define $
f(\epsilon)=\sqrt{\frac{9\pi^3}{2d_A}\ln\frac{2}{\epsilon}}$, as in Eq.~\eqref{Eq:Ffunction}.
Levy's lemma implies that with probability at least $1-\epsilon$,
\begin{equation}
	\abs{L_d(\ket{\psi_{d,A}})-L}\leq \eta f(\epsilon).
\end{equation}
With probability at least $1-\epsilon$, we can also construct an inequality for $\abs{L_d(\ket{\psi_{d,A}})-L_{\pm}}$:
\begin{equation}
\begin{split}
	\abs{L_d(\ket{\psi_{d,A}})-L_{\pm}}&=\abs{L_d(\ket{\psi_{d,A}})-L+L-L_{\pm}}\\
	&\leq\abs{L_d(\ket{\psi_{d,A}})-L}+\abs{L-L_{\pm}}\\
	&\leq\eta f(\epsilon) +\abs{L-L_{\pm}}\\
	&\leq\eta f(\epsilon) +4G\sqrt{\overline{\ot}(U)\overline{\ot}(U_S)}.
\end{split}
\end{equation}

\subsection{Lipschitz constant}
We now compute the Lipschitz constant. We can write state $\ket{\psi_{d,A}}$ in terms of real vectors $\bm{v}^d_1,\bm{v}^d_{2}\in \mathbb{R}^{d_A}$:
\begin{equation}
	\ket{\psi_{d,A}}=\bm{v}^d_1+i\bm{v}^d_2.
\end{equation}
The state can also be written in the form:
\begin{equation}\label{Eq:StateW}
\begin{split}
	\ket{\psi_{d,A}}&= 
	\begin{bmatrix}
		I_A & iI_A
	\end{bmatrix}\bm{w}_d\\
	\bm{w}_d&=\begin{bmatrix}
		\bm{v}^d_1\\ \bm{v}^d_2
	\end{bmatrix}.
\end{split}
\end{equation}
It is straightforward to show that
\begin{equation}
	\norm{\ket{\psi_{1,A}}-\ket{\psi_{2,A}}}_2=\norm{\bm{w}_1-\bm{w}_2}_2.
\end{equation}
Therefore, the Lipschitz continuity condition for $L_d(\ket{\psi_{d,A}})$ can be written as
\begin{equation}
	\abs{L_d(\ket{\psi_{1,A}})-L_d(\ket{\psi_{2,A}})}\leq \eta\norm{\bm{w}_1-\bm{w}_2}_2.
\end{equation}
We can therefore compute a Lipschitz constant by finding any $\eta$ such that
\begin{equation}
	\norm{\frac{d}{d\bm{w}_d}L_d(\ket{\psi_{d,A}})}_2 \leq \eta.
\end{equation}

 We compute
\begin{equation}\label{Eq:DerivativeLoss}
\begin{split}
	\norm{\frac{d}{d\bm{w}_d}L_d(\ket{\psi_{d,A}})}_2
	&=\norm{\frac{d}{d\bm{w}_d}\av{O_C}\abs{\tilde{y}_d-y_d}^2}_2\\
	&=\norm{\av{O_C}2(\tilde{y}_d-y_d)\frac{d}{d\bm{w}_d}(\tilde{y}_d-y_d)}_2\\
	&\leq 2\av{O_C}\norm{(\tilde{y}_d-y_d)\frac{d}{d\bm{w}_d}(\tilde{y}_d-y_d)}_2\\
	&=2\av{O_C}\abs{\tilde{y}_d-y_d}\norm{\frac{d}{d\bm{w}_d}(\tilde{y}_d-y_d)}_2\\
	&\leq2\av{O_C}(\abs{\tilde{y}_d}+\abs{y_d})\norm{\frac{d}{d\bm{w}_d}(\tilde{y}_d-y_d)}_2\\
	&\leq 4\av{O_C}\norm{\frac{d}{d\bm{w}_d}(\tilde{y}_d-y_d)}_2\\
	&\leq 4\av{O_C}\left[\norm{\frac{d}{d\bm{w}_d}\tilde{y}_d}_2+\norm{\frac{d}{d\bm{w}_d}y_d}_2\right].
\end{split}
\end{equation}
The sixth line follows from $\abs{\tilde{y}_d},\abs{y_d}\leq 1$. The remaining inequalities follow by the triangle inequality. Let $Q$ be a Hermitian operator on system $AB$. Compute the following norm: 
\begin{equation}\label{Eq:DerQ}
\begin{split}
	\norm{\frac{d}{d\bm{w}_d}\tr{\rho_d Q}}_2&=\norm{\frac{d}{d\bm{w}_d}\tr{(\ket{\psi_{d,A}}\bra{\psi_{d,A}}\otimes \rho_B) Q }}_2\\
	&=\norm{\frac{1}{d_B}\frac{d}{d\bm{w}_d}\tr{(\ket{\psi_{d,A}}\bra{\psi_{d,A}}\otimes I_B )Q }}_2\\
	&=\frac{1}{d_B}\norm{\frac{d}{d\bm{w}_d}\mathrm{Tr}_A\left\{\ket{\psi_{d,A}}\bra{\psi_{d,A}} \mathrm{Tr}_B\left\{Q\right\} \right\}}_2\\
	&=\frac{1}{d_B}\norm{\frac{d}{d\bm{w}_d}\bra{\psi_{d,A}}\mathrm{Tr}_B\left\{Q\right\}\ket{\psi_{d,A}}}_2 \\
	&=\frac{2}{d_B}\norm{\mathrm{Tr}_B\left\{Q\right\}\ket{\psi_{d,A}}}_2.
	\end{split}
\end{equation}
$\mathrm{Tr}_B\{\bm{\cdot}\}$ denotes the partial trace. 
Line five follows from Appendix \ref{Sec:Derivative}. Now we bound the average over the norm:
\begin{equation}
\begin{split}
	\av{O_C}\norm{\frac{d}{d\bm{w}_d}\tr{\rho_d Q}}_2
	&\leq\frac{2}{d_B}\av{O_C}\norm{\mathrm{Tr}_B\{Q\} }_\infty\\ 
	&\leq\frac{2}{d_B}\av{O_C}\norm{\mathrm{Tr}_B\{Q\} }_{\mathrm{HS}}\\ 
	&\leq\frac{2}{d_B}\sqrt{\av{O_C}\norm{\mathrm{Tr}_B\left\{Q\right\}}_\mathrm{HS}^2}.
	\end{split}
\end{equation}
The second line follows from the Hilbert-Schmidt norm upper bounding the operator norm. The third line follows from the variance being non-negative: $(\av{O_C}(\bm{\cdot}))^2\leq\av{O_C}(\bm{\cdot})^2$. Letting $Q=U^\dagger O_CU$ gives
\begin{equation}\label{Eq:AvgNormY}
	\av{O_C}\norm{\frac{d}{d\bm{w}_d}\tilde{y}_d}_2 \leq\frac{2}{d_B}\sqrt{\av{O_C}\norm{\mathrm{Tr}_B\left\{U^\dagger O_CU\right\}}_\mathrm{HS}^2}.
\end{equation} 

Now we compute
\begin{equation}
	\av{O_C}\norm{\mathrm{Tr}_B\left\{U^\dagger O_C U\right\}}_{\mathrm{HS}}^2=\av{O_C}\mathrm{Tr}_A\left\{\abs{\mathrm{Tr}_B\left\{U^\dagger O_C U\right\}}^2\right\}.
\end{equation}
Writing this diagrammatically,

\begin{equation}
\begin{tikzpicture}

    
	\renewcommand{\nodenum}{v0}
    \renewcommand{\name}{$\av{O_C}\norm{\mathrm{Tr}_B\left\{U^\dagger O_C U\right\}}_{\mathrm{HS}}^2=\av{O_C}$}
	\renewcommand{\xpos}{-2.5*\xrel}
    \renewcommand{\ypos}{0}
    \renewcommand{\height}{\heightsingle}
    \renewcommand{\width}{\widthsingle}
    \node[] (\nodenum) at (\xpos,\ypos) {\name};
    
    \renewcommand{\nodenum}{v1}
    \renewcommand{\name}{$Trace1$}
	\renewcommand{\xpos}{4*\xrel}
    \renewcommand{\ypos}{\rowspace}
    \renewcommand{\height}{\heighttrace}
    \renewcommand{\width}{22.5em}
    \node[rectangle,rounded corners, thick, minimum width=\width, minimum height=\height, draw=dullblue] (\nodenum) at (\xpos,\ypos) {};     
    
    \renewcommand{\nodenum}{v1}
    \renewcommand{\name}{$Trace2$}
	\renewcommand{\xpos}{2*\xrel}
    \renewcommand{\ypos}{-\rowspace}
    \renewcommand{\height}{\heighttrace}
    \renewcommand{\width}{10em}
    \node[rectangle,rounded corners, thick, minimum width=\width, minimum height=\height, draw=dullblue] (\nodenum) at (\xpos,\ypos) {};

    \renewcommand{\nodenum}{v1}
    \renewcommand{\name}{$Trace3$}
	\renewcommand{\xpos}{6*\xrel}
    \renewcommand{\ypos}{-\rowspace}
    \renewcommand{\height}{\heighttrace}
    \renewcommand{\width}{10em}
    \node[rectangle,rounded corners, thick, minimum width=\width, minimum height=\height, draw=dullblue] (\nodenum) at (\xpos,\ypos) {};     
 
	\renewcommand{\nodenum}{v1}
    \renewcommand{\name}{$U^\dagger$}
	\renewcommand{\xpos}{\xrel}
    \renewcommand{\ypos}{0}
    \renewcommand{\height}{\heightdouble}
    \renewcommand{\width}{\widthsingle}
    \node[rectangle, fill=egg, rounded corners, minimum width=\width, minimum height=\height, draw] (\nodenum) at (\xpos,\ypos) {\name};

	\renewcommand{\nodenum}{v2}
    \renewcommand{\name}{$O_C$}
	\renewcommand{\xpos}{2*\xrel}
    \renewcommand{\ypos}{\yrel}
    \renewcommand{\height}{\heightsingle}
    \renewcommand{\width}{\widthsingle}
    \node[rectangle, fill=egg, rounded corners, minimum width=\width, minimum height=\height, draw] (\nodenum) at (\xpos,\ypos) {\name};
 
 \renewcommand{\nodenum}{v3}
    \renewcommand{\name}{$U$}
	\renewcommand{\xpos}{3*\xrel}
    \renewcommand{\ypos}{0}
    \renewcommand{\height}{\heightdouble}
    \renewcommand{\width}{\widthsingle}
    \node[rectangle, fill=egg, rounded corners, minimum width=\width, minimum height=\height, draw] (\nodenum) at (\xpos,\ypos) {\name};
    
	 \renewcommand{\nodenum}{v5}
    \renewcommand{\name}{$U^\dagger$}
	\renewcommand{\xpos}{5*\xrel}
    \renewcommand{\ypos}{0}
    \renewcommand{\height}{\heightdouble}
    \renewcommand{\width}{\widthsingle}
    \node[rectangle, fill=egg, rounded corners, minimum width=\width, minimum height=\height, draw] (\nodenum) at (\xpos,\ypos) {\name};

	\renewcommand{\nodenum}{v6}
    \renewcommand{\name}{$O_C$}
	\renewcommand{\xpos}{6*\xrel}
    \renewcommand{\ypos}{\yrel}
    \renewcommand{\height}{\heightsingle}
    \renewcommand{\width}{\widthsingle}
    \node[rectangle, fill=egg, rounded corners, minimum width=\width, minimum height=\height, draw] (\nodenum) at (\xpos,\ypos) {\name};
    
     \renewcommand{\nodenum}{v7}
    \renewcommand{\name}{$U$}
	\renewcommand{\xpos}{7*\xrel}
    \renewcommand{\ypos}{0}
    \renewcommand{\height}{\heightdouble}
    \renewcommand{\width}{\widthsingle}
    \node[rectangle, fill=egg, rounded corners, minimum width=\width, minimum height=\height, draw] (\nodenum) at (\xpos,\ypos) {\name};
    
\end{tikzpicture}.
\end{equation}
By introducing a 1-design, we can write this as

\begin{equation}
\begin{tikzpicture}

    
	\renewcommand{\nodenum}{v0}
    \renewcommand{\name}{$\av{O_C}\norm{\mathrm{Tr}_B\left\{U^\dagger O_C U\right\}}_{\mathrm{HS}}^2$}
	\renewcommand{\xpos}{-1.7*\xrel}
    \renewcommand{\ypos}{\yrel+2*\rowspace}
    \renewcommand{\height}{\heightsingle}
    \renewcommand{\width}{\widthsingle}
    \node[] (\nodenum) at (\xpos,\ypos) {\name};
    
    	\renewcommand{\nodenum}{v0}
    \renewcommand{\name}{$=d_A\av{O_A}\av{O_C}$}
	\renewcommand{\xpos}{-\xrel}
    \renewcommand{\ypos}{0}
    \renewcommand{\height}{\heightsingle}
    \renewcommand{\width}{\widthsingle}
    \node[] (\nodenum) at (\xpos,\ypos) {\name};
    
    \renewcommand{\nodenum}{v1}
    \renewcommand{\name}{$Trace1$}
	\renewcommand{\xpos}{2.5*\xrel}
    \renewcommand{\ypos}{\rowspace}
    \renewcommand{\height}{\heighttrace}
    \renewcommand{\width}{13em}
    \node[rectangle,rounded corners, thick, minimum width=\width, minimum height=\height, draw=dullblue] (\nodenum) at (\xpos,\ypos) {};     
    
    \renewcommand{\nodenum}{v1}
    \renewcommand{\name}{$Trace2$}
	\renewcommand{\xpos}{2.5*\xrel}
    \renewcommand{\ypos}{-\rowspace}
    \renewcommand{\height}{\heighttrace}
    \renewcommand{\width}{13em}
    \node[rectangle,rounded corners, thick, minimum width=\width, minimum height=\height, draw=dullblue] (\nodenum) at (\xpos,\ypos) {};

    \renewcommand{\nodenum}{v1}
    \renewcommand{\name}{$Trace3$}
	\renewcommand{\xpos}{7.5*\xrel}
    \renewcommand{\ypos}{-\rowspace}
    \renewcommand{\height}{\heighttrace}
    \renewcommand{\width}{13em}
    \node[rectangle,rounded corners, thick, minimum width=\width, minimum height=\height, draw=dullblue] (\nodenum) at (\xpos,\ypos) {};     
    
        \renewcommand{\nodenum}{v1}
    \renewcommand{\name}{$Trace4$}
	\renewcommand{\xpos}{7.5*\xrel}
    \renewcommand{\ypos}{\rowspace}
    \renewcommand{\height}{\heighttrace}
    \renewcommand{\width}{13em}
    \node[rectangle,rounded corners, thick, minimum width=\width, minimum height=\height, draw=dullblue] (\nodenum) at (\xpos,\ypos) {};     
 
	\renewcommand{\nodenum}{v1}
    \renewcommand{\name}{$U^\dagger$}
	\renewcommand{\xpos}{\xrel}
    \renewcommand{\ypos}{0}
    \renewcommand{\height}{\heightdouble}
    \renewcommand{\width}{\widthsingle}
    \node[rectangle, fill=egg, rounded corners, minimum width=\width, minimum height=\height, draw] (\nodenum) at (\xpos,\ypos) {\name};

	\renewcommand{\nodenum}{v2}
    \renewcommand{\name}{$O_C$}
	\renewcommand{\xpos}{2*\xrel}
    \renewcommand{\ypos}{\yrel}
    \renewcommand{\height}{\heightsingle}
    \renewcommand{\width}{\widthsingle}
    \node[rectangle, fill=egg, rounded corners, minimum width=\width, minimum height=\height, draw] (\nodenum) at (\xpos,\ypos) {\name};
 
 \renewcommand{\nodenum}{v3}
    \renewcommand{\name}{$U$}
	\renewcommand{\xpos}{3*\xrel}
    \renewcommand{\ypos}{0}
    \renewcommand{\height}{\heightdouble}
    \renewcommand{\width}{\widthsingle}
    \node[rectangle, fill=egg, rounded corners, minimum width=\width, minimum height=\height, draw] (\nodenum) at (\xpos,\ypos) {\name};
    
    	\renewcommand{\nodenum}{v2}
    \renewcommand{\name}{$O_A^\dagger$}
	\renewcommand{\xpos}{4*\xrel}
    \renewcommand{\ypos}{\yrel}
    \renewcommand{\height}{\heightsingle}
    \renewcommand{\width}{\widthsingle}
    \node[rectangle, fill=egg, rounded corners, minimum width=\width, minimum height=\height, draw] (\nodenum) at (\xpos,\ypos) {\name};
    
	 \renewcommand{\nodenum}{v5}
    \renewcommand{\name}{$U^\dagger$}
	\renewcommand{\xpos}{6*\xrel}
    \renewcommand{\ypos}{0}
    \renewcommand{\height}{\heightdouble}
    \renewcommand{\width}{\widthsingle}
    \node[rectangle, fill=egg, rounded corners, minimum width=\width, minimum height=\height, draw] (\nodenum) at (\xpos,\ypos) {\name};

	\renewcommand{\nodenum}{v6}
    \renewcommand{\name}{$O_C$}
	\renewcommand{\xpos}{7*\xrel}
    \renewcommand{\ypos}{\yrel}
    \renewcommand{\height}{\heightsingle}
    \renewcommand{\width}{\widthsingle}
    \node[rectangle, fill=egg, rounded corners, minimum width=\width, minimum height=\height, draw] (\nodenum) at (\xpos,\ypos) {\name};
    
     \renewcommand{\nodenum}{v7}
    \renewcommand{\name}{$U$}
	\renewcommand{\xpos}{8*\xrel}
    \renewcommand{\ypos}{0}
    \renewcommand{\height}{\heightdouble}
    \renewcommand{\width}{\widthsingle}
    \node[rectangle, fill=egg, rounded corners, minimum width=\width, minimum height=\height, draw] (\nodenum) at (\xpos,\ypos) {\name};
 
	\renewcommand{\nodenum}{v6}
    \renewcommand{\name}{$O_A$}
	\renewcommand{\xpos}{9*\xrel}
    \renewcommand{\ypos}{\yrel}
    \renewcommand{\height}{\heightsingle}
    \renewcommand{\width}{\widthsingle}
    \node[rectangle, fill=egg, rounded corners, minimum width=\width, minimum height=\height, draw] (\nodenum) at (\xpos,\ypos) {\name};

\end{tikzpicture}.
\end{equation}
Using the Hermiticity of Pauli string $O_A$,
\begin{equation}
	\av{O_C}\norm{\mathrm{Tr}_B\left\{U^\dagger O_C U\right\}}_{\mathrm{HS}}^2=d_A\av{O_A}\av{O_C}\tr{ U O_AU^\dagger O_C}^2.
\end{equation}
Retrieving the identity from Eq.~\eqref{Cidentity} and setting $Q=O_A,U_1=U_2=U$,
\begin{equation}\label{Eq:ACOTOC}
		\av{O_C}\tr{U O_A U^\dagger O_C}^2=d_D^2C(O_A,U,U)=d_D^2\av{O_D}\langle UO_A U^\dagger O_D U O_A U^\dagger O_D\rangle.
\end{equation}
This yields,
\begin{equation}
\begin{split}
	\av{O_C}\norm{\mathrm{Tr}_B\left\{U^\dagger O_C U\right\}}_{\mathrm{HS}}^2&=d_Ad_D^2\av{O_A}\av{O_D}\langle UO_A U^\dagger O_D U O_A U^\dagger O_D\rangle\\
	&=d_Ad_D^2\overline{\ot}(U).
\end{split}
\end{equation}
Ineq.~\eqref{Eq:AvgNormY} becomes
\begin{equation}\label{Eq:AvgDerivativeOutput}
\begin{split}
	\av{O_C}\norm{\frac{d}{d\bm{w}_d}\tilde{y}_d}_2 
	&\leq\frac{2}{d_B}\sqrt{d_Ad_D^2\overline{\ot}(U)}\\
	&=\frac{2\sqrt{d_A}d_D}{d_B}\sqrt{\overline{\ot}(U)}.
\end{split}
\end{equation} 
A similar inequality holds for $\tilde{y}_d\rightarrow y_d$. Ineq.~\eqref{Eq:DerivativeLoss} becomes
\begin{equation}
\begin{split}
	\norm{\frac{d}{d\bm{w}_d}L_d(\ket{\psi_{d,A}})}_2
	&\leq\frac{8\sqrt{d_A}d_D}{d_B}\left[\sqrt{\overline{\ot}(U)}+\sqrt{\overline{\ot}(U_S)}\right]\\
	&=\frac{8\sqrt{d_A}d_D}{d_B}\left[\frac{d_C}{d_A}\sqrt{(d_A+1)L_{+}}\right]\\
	&=8\sqrt{d_A(d_A+1)L_{+}}.
\end{split}
\end{equation}
The second line follows from the definition of $L_+$ in Eq.~\eqref{Eq:LossBound}. The Lipschitz constant is therefore
\begin{equation}
	\eta=8\sqrt{d_A(d_A+1)L_{+}}.
\end{equation}

\subsection{Maximally scrambling unitaries}\label{ConcentrationMaxScram}
Consider the case where $U_i\in\{U,U_S\}$ is maximally scrambling. In the large $d_\mathrm{tot}$ limit,
\begin{equation}
\overline{\ot}(U_i)=\frac{1}{d_A^2}+\frac{1}{d_D^2}-\frac{1}{d_A^2d_D^2}.
\end{equation}
Taking the case where $d_D>>d_A$, $\overline{\ot}(U_i)=\frac{1}{d_A^2}$. The Lipschitz constant becomes
\begin{equation}
	\eta=\frac{8\sqrt{d_A}d_D}{d_B}\left[2\sqrt{\frac{1}{d_A^2}}\right]=\frac{16\sqrt{d_A}d_D}{d_Bd_A}=\frac{16\sqrt{d_A}}{d_C}.
\end{equation}
Therefore, 
\begin{equation}
\begin{split}
    \eta f(\epsilon)&=\frac{16\sqrt{d_A}}{d_C}\sqrt{\frac{9\pi^3}{2d_A}\ln\frac{2}{\epsilon}}\\
    &=\frac{16}{d_C}\sqrt{\frac{9\pi^3}{2}\ln\frac{2}{\epsilon}}.
\end{split}
\end{equation}
Recalling that with probability at least $1-\epsilon$, $\abs{L_d(\ket{\psi_{d,A}})-L}\leq \eta f(\epsilon)$, $L_d$ concentrates near $L$ as $d_C$ increases.

\section{Proof of Theorem~\ref{TheoremOTOC}}\label{Sec:ConcentrationGrad}
Similar to Appendix~\ref{Sec:LevyBound}, we can use Levy's lemma to compute a concentration inequality for $\partial_{\theta_l}L_d$. Levy's lemma in this context reads

\begin{lemma}[Levy's lemma]
Let $\partial_{\theta_l} L_d : S^{2d_{\mathrm{tot}}-1}\rightarrow \mathbb{R}$ satisfy 
$\abs{\partial_{\theta_l}L_d(\ket{\psi_{1,A}})-\partial_{\theta_l}L_d(\ket{\psi_{2,A}})}\leq \eta_g\norm{\ket{\psi_{1,A}}-\ket{\psi_{2,A}}}_2$. Then $\forall$ $\delta\geq 0$,
\begin{equation}\label{Eq:Probability}
\underset{\ket{\psi_{d,A}}\sim \Hr(d_A)}{\mathrm{Prob}}\left[\abs{\partial_{\theta_l}L_d(\ket{\psi_{d,A}})-\partial_{\theta_l}L}\geq \delta \right]\leq \epsilon,
\end{equation}
where $\epsilon=2\mathrm{exp}\left(-\frac{2d_A\delta^2}{9\pi^3\eta_g^2}\right)$ and $\eta_g$ is the Lipschitz constant.
\end{lemma}

In the above, $\partial_{\theta_l}L=\av{\ket{\psi_{d,A}}\sim \Hr(d_A)}\left[\partial_{\theta_l}L_d(\ket{\psi_{d,A}})\right]$. Levy's lemma implies that with probability at least $1-\epsilon$,
\begin{equation}
	\abs{\partial_{\theta_l}L_d(\ket{\psi_{d,A}})-\partial_{\theta_l}L}\leq \eta_g f(\epsilon),
\end{equation}
where $f(\epsilon)$ is defined as in Eq.~\eqref{Eq:Ffunction}.

\subsection{Lipschitz constant}
We compute the Lipschitz constant. As in Appendix~\ref{Sec:LevyBound}, we define $\ket{\psi_{d,A}}$ using Eq.~\eqref{Eq:StateW} so that $\eta_g$ can be found through the inequality: $\norm{\frac{d}{d\bm{w}_d}\partial_{\theta_l}L_d(\ket{\psi_{d,A}})}_2\leq \eta_g$. We compute the bound:
\begin{equation}\label{Eq:LossGradDerIneq}
\begin{split}
	\norm{\frac{d}{d\bm{w}_d}\partial_{\theta_l}L_d(\ket{\psi_{d,A}})}_2
	&=\norm{\frac{d}{d\bm{w}_d}(\partial_{\theta_l}\av{O_C}\abs{\tilde{y}_d-y_d}^2)}_2\\
	&=\norm{2\frac{d}{d\bm{w}_d}\av{O_C}(\tilde{y}_d-y_d)\partial_{\theta_l}\tilde{y}_d}_2\\
	&=2\norm{\av{O_C}\left(\partial_{\theta_l}\tilde{y}_d\frac{d}{d\bm{w}_d}(\tilde{y}_d-y_d)+(\tilde{y}_d-y_d)\frac{d}{d\bm{w}_d}\partial_{\theta_l}\tilde{y}_d\right)}_2\\
	&\leq 2\av{O_C}\norm{\partial_{\theta_l}\tilde{y}_d\frac{d}{d\bm{w}_d}(\tilde{y}_d-y_d)+(\tilde{y}_d-y_d)\frac{d}{d\bm{w}_d}\partial_{\theta_l}\tilde{y}_d}_2\\
	&\leq 2\av{O_C}\left(\norm{\partial_{\theta_l}\tilde{y}_d\frac{d}{d\bm{w}_d}(\tilde{y}_d-y_d)}_2+\norm{(\tilde{y}_d-y_d)\frac{d}{d\bm{w}_d}\partial_{\theta_l}\tilde{y}_d}_2\right)\\
	&= 2\av{O_C}\left(\abs{\partial_{\theta_l}\tilde{y}_d}\norm{\frac{d}{d\bm{w}_d}(\tilde{y}_d-y_d)}_2+\abs{\tilde{y}_d-y_d}\norm{\frac{d}{d\bm{w}_d}\partial_{\theta_l}\tilde{y}_d}_2\right)\\
	&\leq 2\av{O_C}\left(\abs{\partial_{\theta_l}\tilde{y}_d}\left(\norm{\frac{d}{d\bm{w}_d}\tilde{y}_d}_2+\norm{\frac{d}{d\bm{w}_d}y_d}_2\right)+2\norm{\frac{d}{d\bm{w}_d}\partial_{\theta_l}\tilde{y}_d}_2\right).
\end{split}
\end{equation}
Before computing $\partial_{\theta_l}\tilde{y}_d$, it will be useful to compute $\partial_{\theta_{l}}U$:
\begin{equation}
\begin{split}
	\partial_{\theta_{l}}U&=\partial_{\theta_{l}}\prod_{j=1}^{L}U_j(\theta_j)W_j\\
	&=\left[\prod_{k=l+1}^{L}U_k(\theta_k)W_k\right]\partial_{\theta_{l}}U_l(\theta_l)W_l\left[\prod_{j=1}^{l-1}U_j(\theta_j)W_j\right]\\
	&=\left[\prod_{k=l+1}^{L}U_k(\theta_k)W_k\right](-iV_l)U(\theta_l)W_l\left[\prod_{j=1}^{l-1}U_j(\theta_j)W_j\right]\\
	&=\left[\prod_{k=l+1}^{L}U_k(\theta_k)W_k\right](-iV_l)\left[\prod_{j=1}^{l}U_j(\theta_j)W_j\right]\\
	&=U_+(-iV_l)U_-.\\
	\end{split}
\end{equation}
We define $U_-=\prod_{j=1}^{l}U_j(\theta_j)W_j$ and $U_+=\prod_{k=l+1}^{L}U_k(\theta_k)W_k$. Similarly, $\partial_{\theta_{l}}U^\dagger=U_-^\dagger (iV_l)U_+^\dagger$.

We now compute  $\partial_{\theta_l}\tilde{y}_d$:
\begin{equation}
\begin{split}
	\partial_{\theta_l}\tilde{y}_d &=\partial_{\theta_l}\tr{U\rho_d U^\dagger O_C}\\
&=\tr{\left(\partial_{\theta_l}U\right)\rho_d U^\dagger O_C}+\tr{U\rho_d \left(\partial_{\theta_l}U^\dagger\right) O_C}\\
&=\tr{\left(U_+(-iV_l)U_-\right)\rho_d U^\dagger O_C}+\tr{U\rho_d \left(U_-^\dagger (iV_l)U_+^\dagger\right) O_C}\\
&=\tr{\rho_d U_-^\dagger U_+^\dagger O_CU_+(-iV_l)U_-}+\tr{\rho_d U_-^\dagger (iV_l)U_+^\dagger O_CU_+U_-}\\
&=\tr{\rho_d U_-^\dagger [iV_l,U_+^\dagger O_CU_+]U_-}\\
&=\tr{\rho_d Q}.\\
\end{split}
\end{equation}
In line six, we define the Hermitian operator $Q=U_-^\dagger [iV_l,U_+^\dagger O_CU_+]U_-$. Now compute the following norm:
\begin{equation}\label{Eq:NormDerGrad}
\begin{split}
	\norm{\frac{d}{d\bm{w}_d}\partial_{\theta_l}\tilde{y}_d}_2&=\norm{\frac{d}{d\bm{w}_d}\tr{\rho_d Q}}_2\\
&=\frac{2}{d_B}\norm{\mathrm{Tr}_B\left\{Q\right\}\ket{\psi_{d,A}}}_2 \\
&=\frac{2}{d_B}\sqrt{\bra{\psi_{d,A}}\left(\mathrm{Tr}_B\left\{Q\right\}\right)^2\ket{\psi_{d,A}}}\\
&=\frac{2}{d_B}\sqrt{\tr{\rho_{d,A}\left(\mathrm{Tr}_B\left\{Q\right\}\right)^2}}.
\end{split}
\end{equation}
In the second line, we use Eq.~\eqref{Eq:DerQ}. Now diagrammatically write

\begin{equation}
\begin{tikzpicture}

    
	\renewcommand{\nodenum}{v0}
    \renewcommand{\name}{$\tr{\rho_{d,A}\left(\mathrm{Tr}_B\left\{Q\right\}\right)^2}=$}
	\renewcommand{\xpos}{-2.5*\xrel}
    \renewcommand{\ypos}{0}
    \renewcommand{\height}{\heightsingle}
    \renewcommand{\width}{\widthsingle}
    \node[] (\nodenum) at (\xpos,\ypos) {\name};
    
    \renewcommand{\nodenum}{v1}
    \renewcommand{\name}{$Trace1$}
	\renewcommand{\xpos}{2*\xrel}
    \renewcommand{\ypos}{\rowspace}
    \renewcommand{\height}{\heighttrace}
    \renewcommand{\width}{11em}
    \node[rectangle,rounded corners, thick, minimum width=\width, minimum height=\height, draw=dullblue] (\nodenum) at (\xpos,\ypos) {};     
    
    \renewcommand{\nodenum}{v1}
    \renewcommand{\name}{$Trace2$}
	\renewcommand{\xpos}{\xrel}
    \renewcommand{\ypos}{-\rowspace}
    \renewcommand{\height}{\heighttrace}
    \renewcommand{\width}{5em}
    \node[rectangle,rounded corners, thick, minimum width=\width, minimum height=\height, draw=dullblue] (\nodenum) at (\xpos,\ypos) {};

    \renewcommand{\nodenum}{v1}
    \renewcommand{\name}{$Trace3$}
	\renewcommand{\xpos}{3*\xrel}
    \renewcommand{\ypos}{-\rowspace}
    \renewcommand{\height}{\heighttrace}
    \renewcommand{\width}{5em}
    \node[rectangle,rounded corners, thick, minimum width=\width, minimum height=\height, draw=dullblue] (\nodenum) at (\xpos,\ypos) {};     
 
    	\renewcommand{\nodenum}{v1}
    \renewcommand{\name}{$Q$}
	\renewcommand{\xpos}{\xrel}
    \renewcommand{\ypos}{0}
    \renewcommand{\height}{\heightdouble}
    \renewcommand{\width}{\widthsingle}
    \node[rectangle, fill=egg, rounded corners, minimum width=\width, minimum height=\height, draw] (\nodenum) at (\xpos,\ypos) {\name};

	\renewcommand{\nodenum}{v1}
    \renewcommand{\name}{$\rho_{d,A}$}
	\renewcommand{\xpos}{2*\xrel}
    \renewcommand{\ypos}{\yrel}
    \renewcommand{\height}{\heightsingle}
    \renewcommand{\width}{\widthsingle}
    \node[rectangle, fill=egg, rounded corners, minimum width=\width, minimum height=\height, draw] (\nodenum) at (\xpos,\ypos) {\name};

	\renewcommand{\nodenum}{v1}
    \renewcommand{\name}{$Q$}
	\renewcommand{\xpos}{3*\xrel}
    \renewcommand{\ypos}{0}
    \renewcommand{\height}{\heightdouble}
    \renewcommand{\width}{\widthsingle}
    \node[rectangle, fill=egg, rounded corners, minimum width=\width, minimum height=\height, draw] (\nodenum) at (\xpos,\ypos) {\name};

\end{tikzpicture}.
\end{equation}
We introduce a 1-design:
\begin{equation}
\begin{tikzpicture}

    
	\renewcommand{\nodenum}{v0}
    \renewcommand{\name}{$\tr{\rho_{d,A}\left(\mathrm{Tr}_B\left\{Q\right\}\right)^2}=d_B\av{O_B}$}
	\renewcommand{\xpos}{-2.5*\xrel}
    \renewcommand{\ypos}{0}
    \renewcommand{\height}{\heightsingle}
    \renewcommand{\width}{\widthsingle}
    \node[] (\nodenum) at (\xpos,\ypos) {\name};
    
    \renewcommand{\nodenum}{v1}
    \renewcommand{\name}{$Trace1$}
	\renewcommand{\xpos}{2.5*\xrel}
    \renewcommand{\ypos}{\rowspace}
    \renewcommand{\height}{\heighttrace}
    \renewcommand{\width}{13em}
    \node[rectangle,rounded corners, thick, minimum width=\width, minimum height=\height, draw=dullblue] (\nodenum) at (\xpos,\ypos) {};     
    
    \renewcommand{\nodenum}{v1}
    \renewcommand{\name}{$Trace2$}
	\renewcommand{\xpos}{2.5*\xrel}
    \renewcommand{\ypos}{-\rowspace}
    \renewcommand{\height}{\heighttrace}
    \renewcommand{\width}{13em}
    \node[rectangle,rounded corners, thick, minimum width=\width, minimum height=\height, draw=dullblue] (\nodenum) at (\xpos,\ypos) {};     
 
    	\renewcommand{\nodenum}{v1}
    \renewcommand{\name}{$Q$}
	\renewcommand{\xpos}{\xrel}
    \renewcommand{\ypos}{0}
    \renewcommand{\height}{\heightdouble}
    \renewcommand{\width}{\widthsingle}
    \node[rectangle, fill=egg, rounded corners, minimum width=\width, minimum height=\height, draw] (\nodenum) at (\xpos,\ypos) {\name};

	\renewcommand{\nodenum}{v1}
    \renewcommand{\name}{$\rho_{d,A}$}
	\renewcommand{\xpos}{2*\xrel}
    \renewcommand{\ypos}{\yrel}
    \renewcommand{\height}{\heightsingle}
    \renewcommand{\width}{\widthsingle}
    \node[rectangle, fill=egg, rounded corners, minimum width=\width, minimum height=\height, draw] (\nodenum) at (\xpos,\ypos) {\name};

   	\renewcommand{\nodenum}{v1}
    \renewcommand{\name}{$O_B^\dagger $}
	\renewcommand{\xpos}{2*\xrel}
    \renewcommand{\ypos}{-\yrel}
    \renewcommand{\height}{\heightsingle}
    \renewcommand{\width}{\widthsingle}
    \node[rectangle, fill=egg, rounded corners, minimum width=\width, minimum height=\height, draw] (\nodenum) at (\xpos,\ypos) {\name};
    
	\renewcommand{\nodenum}{v1}
    \renewcommand{\name}{$Q$}
	\renewcommand{\xpos}{3*\xrel}
    \renewcommand{\ypos}{0}
    \renewcommand{\height}{\heightdouble}
    \renewcommand{\width}{\widthsingle}
    \node[rectangle, fill=egg, rounded corners, minimum width=\width, minimum height=\height, draw] (\nodenum) at (\xpos,\ypos) {\name};
    
   	\renewcommand{\nodenum}{v1}
    \renewcommand{\name}{$O_B$}
	\renewcommand{\xpos}{4*\xrel}
    \renewcommand{\ypos}{-\yrel}
    \renewcommand{\height}{\heightsingle}
    \renewcommand{\width}{\widthsingle}
    \node[rectangle, fill=egg, rounded corners, minimum width=\width, minimum height=\height, draw] (\nodenum) at (\xpos,\ypos) {\name};

\end{tikzpicture}.
\end{equation}
Introduce the Bell state:
\begin{equation}
\begin{tikzpicture}

    
	\renewcommand{\nodenum}{v0}
    \renewcommand{\name}{$\tr{\rho_{d,A}\left(\mathrm{Tr}_B\left\{Q\right\}\right)^2}=d_B^2\av{O_B}$}
	\renewcommand{\xpos}{-3.25*\xrel}
    \renewcommand{\ypos}{0}
    \renewcommand{\height}{\heightsingle}
    \renewcommand{\width}{\widthsingle}
    \node[] (\nodenum) at (\xpos,\ypos) {\name};
    
    \renewcommand{\nodenum}{v1}
    \renewcommand{\name}{$Trace1$}
	\renewcommand{\xpos}{2.75*\xrel}
    \renewcommand{\ypos}{\rowspace}
    \renewcommand{\height}{\heighttrace}
    \renewcommand{\width}{19em}
    \node[rectangle,rounded corners, thick, minimum width=\width, minimum height=\height, draw=dullblue] (\nodenum) at (\xpos,\ypos) {};     
    
    \renewcommand{\nodenum}{v1}
    \renewcommand{\name}{$Trace2$}
	\renewcommand{\xpos}{2.75*\xrel}
    \renewcommand{\ypos}{-2*\rowspace}
    \renewcommand{\height}{3*\heighttrace}
    \renewcommand{\width}{19em}
    \node[rectangle,rounded corners, thick, minimum width=\width, minimum height=\height, draw=dullblue] (\nodenum) at (\xpos,\ypos) {};     
    
    \renewcommand{\nodenum}{v1}
    \renewcommand{\name}{$Trace3$}
	\renewcommand{\xpos}{2.75*\xrel}
    \renewcommand{\ypos}{-2*\rowspace}
    \renewcommand{\height}{\heighttrace}
    \renewcommand{\width}{17em}
    \node[rectangle,rounded corners, thick, minimum width=\width, minimum height=\height, draw=dullblue] (\nodenum) at (\xpos,\ypos) {};     
 
	\renewcommand{\nodenum}{v1}
    \renewcommand{\name}{$\rho_{d,A}$}
	\renewcommand{\xpos}{.75*\xrel}
    \renewcommand{\ypos}{\yrel}
    \renewcommand{\height}{\heightsingle}
    \renewcommand{\width}{\widthsingle}
    \node[rectangle, fill=egg, rounded corners, minimum width=\width, minimum height=\height, draw] (\nodenum) at (\xpos,\ypos) {\name};

	\renewcommand{\nodenum}{v1}
    \renewcommand{\name}{$\Pi_{B_1,B_2}$}
	\renewcommand{\xpos}{.75*\xrel}
    \renewcommand{\ypos}{-\rowspace}
    \renewcommand{\height}{\heightdouble}
    \renewcommand{\width}{\widthsingle}
    \node[rectangle, fill=egg, rounded corners, minimum width=\width, minimum height=\height, draw] (\nodenum) at (\xpos,\ypos) {\name};
    
   	\renewcommand{\nodenum}{v1}
    \renewcommand{\name}{$O_B^\dagger $}
	\renewcommand{\xpos}{2*\xrel}
    \renewcommand{\ypos}{-\yrel}
    \renewcommand{\height}{\heightsingle}
    \renewcommand{\width}{\widthsingle}
    \node[rectangle, fill=egg, rounded corners, minimum width=\width, minimum height=\height, draw] (\nodenum) at (\xpos,\ypos) {\name};
    
	\renewcommand{\nodenum}{v1}
    \renewcommand{\name}{$Q$}
	\renewcommand{\xpos}{3*\xrel}
    \renewcommand{\ypos}{0}
    \renewcommand{\height}{\heightdouble}
    \renewcommand{\width}{\widthsingle}
    \node[rectangle, fill=egg, rounded corners, minimum width=\width, minimum height=\height, draw] (\nodenum) at (\xpos,\ypos) {\name};
    
   	\renewcommand{\nodenum}{v1}
    \renewcommand{\name}{$O_B$}
	\renewcommand{\xpos}{4*\xrel}
    \renewcommand{\ypos}{-\yrel}
    \renewcommand{\height}{\heightsingle}
    \renewcommand{\width}{\widthsingle}
    \node[rectangle, fill=egg, rounded corners, minimum width=\width, minimum height=\height, draw] (\nodenum) at (\xpos,\ypos) {\name};

    	\renewcommand{\nodenum}{v1}
    \renewcommand{\name}{$Q$}
	\renewcommand{\xpos}{5*\xrel}
    \renewcommand{\ypos}{0}
    \renewcommand{\height}{\heightdouble}
    \renewcommand{\width}{\widthsingle}
    \node[rectangle, fill=egg, rounded corners, minimum width=\width, minimum height=\height, draw] (\nodenum) at (\xpos,\ypos) {\name};

\end{tikzpicture}.
\end{equation}

We can now bound the trace:
\begin{equation}
\begin{split}
	\tr{\rho_{d,A}\left(\mathrm{Tr}_B\left\{Q\right\}\right)^2}
	&=d_B^2\av{O_B}\tr{(\rho_{d,A}\otimes \Pi_{B_1,B_2})O_B^\dagger (Q\otimes I_B)O_B(Q\otimes I_B)}\\
	&=d_B^2\abs{\av{O_B}\tr{(\rho_{d,A}\otimes \Pi_{B_1,B_2})O_B^\dagger (Q\otimes I_B)O_B(Q\otimes I_B)}}\\	
	&\leq d_B^2\av{O_B}\abs{\tr{(\rho_{d,A}\otimes \Pi_{B_1,B_2})O_B^\dagger (Q\otimes I_B)O_B(Q\otimes I_B)}}\\	
	&\leq d_B^2\av{O_B}\norm{O_B^\dagger (Q\otimes I_B)O_B(Q\otimes I_B)}_{\infty}\\	
	&\leq d_B^2\av{O_B}\norm{O_B^\dagger}_{\infty} \norm{(Q\otimes I_B)}_{\infty}\norm{O_B}_{\infty}\norm{(Q\otimes I_B)}_{\infty}\\	
	&= d_B^2\av{O_B}\norm{Q}_{\infty}^2 \\
	&= d_B^2\norm{Q}_{\infty}^2\\
	&\leq 4d_B^2\norm{V_l}_\infty^2.
\end{split}
\end{equation}
In the first line, we take $O_B$ to denote $I_A\otimes O_B \otimes I_B$. The second line follows from the trace on the left-hand side being non-negative. The third line follows from the triangle inequality. The fifth line follows from the sub-multiplicativity property of the norm. The sixth line uses the fact that Pauli strings have a maximum eigenvalue of 1. The last line follows from the following bound:
\begin{equation}
\begin{split}
	\norm{Q}_{\infty}
	&=\norm{U_-^\dagger [iV_l,U_+^\dagger O_CU_+]U_-}_\infty\\
	&=\norm{[iV_l,U_+^\dagger O_CU_+]}_\infty\\
	&=\norm{iV_lU_+^\dagger O_CU_+-U_+^\dagger O_CU_+iV_l}_\infty\\
	&\leq \norm{iV_lU_+^\dagger O_CU_+}_\infty+\norm{U_+^\dagger O_CU_+iV_l}_\infty\\
	&= 2\norm{V_l}_\infty.
\end{split}
\end{equation}
The second and fifth lines follow from the unitary invariance of the Schatten norms. In the fourth line, we use the triangle inequality. 
We can now bound Eq.~\eqref{Eq:NormDerGrad}:
\begin{equation}\label{Eq:DerGradY}
	\norm{\frac{d}{d\bm{w}_d}\partial_{\theta_l}\tilde{y}_d}_2
	\leq \frac{2}{d_B}\sqrt{4d_B^2\norm{V_l}_\infty^2}= 4\norm{V_l}_\infty.
\end{equation}

Also, bound $\abs{\partial_{\theta_l}\tilde{y}_d}$:
\begin{equation}\label{Eq:GradYBound}
\begin{split}
	\abs{\partial_{\theta_l}\tilde{y}_d}
	&=\abs{\tr{\rho_d Q}}\\
	&\leq \norm{Q}_\infty \\
	&\leq 2\norm{V_l}_\infty.
\end{split}
\end{equation}
Ineq.~\eqref{Eq:LossGradDerIneq} becomes

\begin{equation}
\begin{split}
	\Big|\Big|\frac{d}{d\bm{w}_d}&\partial_{\theta_l}L_d(\ket{\psi_{d,A}})\Big|\Big|_2\\
&\leq 2\av{O_C}\left(\abs{\partial_{\theta_l}\tilde{y}_d}\left(\norm{\frac{d}{d\bm{w}_d}\tilde{y}_d}_2+\norm{\frac{d}{d\bm{w}_d}y_d}_2\right)+2\norm{\frac{d}{d\bm{w}_d}\partial_{\theta_l}\tilde{y}_d}_2\right)\\
&\leq 2\av{O_C}\left(2\norm{V_l}_\infty\left(\norm{\frac{d}{d\bm{w}_d}\tilde{y}_d}_2+\norm{\frac{d}{d\bm{w}_d}y_d}_2\right)+2\norm{\frac{d}{d\bm{w}_d}\partial_{\theta_l}\tilde{y}_d}_2\right)\\
&\leq 2\av{O_C}\left(2\norm{V_l}_\infty\left(\norm{\frac{d}{d\bm{w}_d}\tilde{y}_d}_2+\norm{\frac{d}{d\bm{w}_d}y_d}_2\right)+8\norm{V_l}_\infty\right)\\
&\leq 2\left(2\norm{V_l}_\infty\left(\frac{2\sqrt{d_A}d_D}{d_B}\sqrt{\overline{\ot}(U)}+\frac{2\sqrt{d_A}d_D}{d_B}\sqrt{\overline{\ot}(U_S)}\right)+8\norm{V_l}_\infty\right)\\
&=2\left(\frac{4\sqrt{d_A}d_D}{d_B}\norm{V_l}_\infty\left(\sqrt{\overline{\ot}(U)}+\sqrt{\overline{\ot}(U_S)}\right)+8\norm{V_l}_\infty\right)\\
&=8\norm{V_l}_\infty\left(\frac{\sqrt{d_A}d_D}{d_B}\left(\sqrt{\overline{\ot}(U)}+\sqrt{\overline{\ot}(U_S)}\right)+2\right)\\
&=8\norm{V_l}_\infty\left(\frac{\sqrt{d_A}d_D}{d_B}\left(\frac{d_C}{d_A}\sqrt{(d_A+1)L_{+}}\right)+2\right)\\
&=8\norm{V_l}_\infty\left(\sqrt{d_A(d_A+1)L_{+}}+2\right).
\end{split}
\end{equation}
In line four, we use Ineq.~\eqref{Eq:AvgDerivativeOutput}. In line seven, we use the definition of $L_+$ from Eq.~\eqref{Eq:LossBound}. The Lipschitz constant is therefore
\begin{equation}
\eta_g=8\norm{V_l}_\infty\left(\sqrt{d_A(d_A+1)L_{+}}+2\right).
\end{equation}

\subsection{Maximally scrambling unitaries}
Similar to Appendix~\ref{ConcentrationMaxScram}, let $U$ and $U_S$ be maximally scrambling, let $d_{\mathrm{tot}}$ be large, and take $d_D>>d_A$. Then $
\overline{\ot}(U)=\overline{\ot}(U_S)=\frac{1}{d_A^2}$. The Lipschitz constant is
\begin{equation}
\begin{split}
	\eta_g
	&=8\norm{V_l}_\infty\left(\frac{\sqrt{d_A}d_D}{d_B}\frac{2}{d_A}+2\right)\\
&=16\norm{V_l}_\infty\left(\frac{\sqrt{d_A}}{d_C}+1\right).
\end{split}
\end{equation}
Levy's lemma then gives the following concentration inequality:
\begin{equation}
\begin{split}
	\abs{\partial_{\theta_l}L_d(\ket{\psi_{d,A}})-\partial_{\theta_l}L}
	&\leq \eta_g f(\epsilon)\\
	&=16\norm{V_l}_\infty\left(\frac{\sqrt{d_A}}{d_C}+1\right)\sqrt{\frac{9\pi^3}{2d_A}\ln\frac{2}{\epsilon}}\\
	&=16\norm{V_l}_\infty\left(\frac{1}{d_C}+\frac{1}{\sqrt{d_A}}\right)\sqrt{\frac{9\pi^3}{2}\ln\frac{2}{\epsilon}}.
\end{split}
\end{equation}

\subsection{Vanishing gradient}
We bound $\abs{\partial_{\theta_l}L}$:
\begin{equation}\label{Eq:BoundGradError}
\begin{split}
\abs{\partial_{\theta_l}L}
	&=G\abs{\partial_{\theta_l}\overline{\ot}(U(\bm{\theta}))-2\partial_{\theta_l}\mathrm{OP}(U(\bm{\theta}),U_S) }\\
	&\leq G\left[\abs{\partial_{\theta_l}\overline{\ot}(U(\bm{\theta}))}+2\abs{\partial_{\theta_l}\mathrm{OP}(U(\bm{\theta}),U_S) }\right].
\end{split}
\end{equation}
We bound the first term:
\begin{equation}\label{Eq:PartialUdagger}
\begin{split}
	|\partial_{\theta_{l}}&\overline{\ot}(U)|\\
	&=	\abs{\av{O_A}\av{O_D}\partial_{\theta_{l}}\langle (UO_AU^\dagger O_D)^2 \rangle}\\
		&=	2\abs{\av{O_A}\av{O_D}\langle (UO_AU^\dagger O_D)\partial_{\theta_{l}} (UO_AU^\dagger O_D)\rangle}\\
		&=	2\abs{\av{O_A}\av{O_D}\langle (UO_AU^\dagger O_D)((\partial_{\theta_{l}}U)O_AU^\dagger O_D+UO_A(\partial_{\theta_{l}}U^\dagger) O_D) \rangle}\\
		&=	2\abs{\av{O_A}\av{O_D}\langle (UO_AU^\dagger O_D)(U_+(-iV_l)U_-O_AU^\dagger O_D+UO_AU_-^\dagger (iV_l)U_+^\dagger O_D) \rangle}\\
		&=	2\abs{\av{O_A}\av{O_D}\langle (U_+U_-O_AU_-^\dagger U_+^\dagger O_D)(U_+(-iV_l)U_-O_AU_-^\dagger U_+^\dagger O_D+U_+U_-O_AU_-^\dagger (iV_l)U_+^\dagger O_D) \rangle}\\
		&=	2\abs{\av{O_A}\av{O_D}\langle (U_-O_AU_-^\dagger U_+^\dagger O_DU_+)((-iV_l)U_-O_AU_-^\dagger U_+^\dagger O_DU_++U_-O_AU_-^\dagger (iV_l)U_+^\dagger O_DU_+) \rangle}\\
		&=	2\abs{\av{O_A}\av{O_D}(T_1+T_2)}\\
		&\leq	2\av{O_A}\av{O_D}(\abs{T_1}+\abs{T_2}),
	\end{split}
\end{equation}
where we define
\begin{equation}
\begin{split}
	T_1&=\langle U_-O_AU_-^\dagger U_+^\dagger O_DU_+(-iV_l)U_-O_AU_-^\dagger U_+^\dagger O_DU_+\rangle,\\
	T_2&=\langle U_-O_AU_-^\dagger U_+^\dagger O_DU_+U_-O_AU_-^\dagger (iV_l)U_+^\dagger O_DU_+ \rangle.
\end{split}
\end{equation}
Now bound
\begin{equation}
	\begin{split}
	\abs{T_1}&=\abs{\langle U_-O_AU_-^\dagger U_+^\dagger O_DU_+(-iV_l)U_-O_AU_-^\dagger U_+^\dagger O_DU_+\rangle}\\
	&\leq\norm{ U_-O_AU_-^\dagger U_+^\dagger O_DU_+(-iV_l)U_-O_AU_-^\dagger U_+^\dagger O_DU_+}_{\infty}\\
	&=\norm{V_l}_{\infty}.
	\end{split}
\end{equation}
The third line follows from the invariance of the Schatten norm under unitaries, and also noting that all operators but $V_l$ are unitary. Similarly, $T_2\leq \norm{V_l}_{\infty}$. We therefore have
\begin{equation}
	\abs{\partial_{\theta_{l}}\overline{\ot}(U)}\leq 4\norm{V_l}_{\infty}.
\end{equation}
We now bound
\begin{equation}
	\begin{split}
	|\partial_{\theta_l}\mathrm{OP}&(U(\bm{\theta}),U_S) |\\
	&=\abs{\av{O_A}\av{O_D}\partial_{\theta_l}\langle UO_AU^\dagger O_DU_S O_AU_S^\dagger O_D\rangle}\\
	&=\abs{\av{O_A}\av{O_D}\langle ((\partial_{\theta_l}U)O_AU^\dagger +UO_A(\partial_{\theta_l}U^\dagger))O_DU_S O_AU_S^\dagger O_D\rangle}\\
	&=\abs{\av{O_A}\av{O_D}\langle (U_+(-iV_l)U_-O_AU_-^\dagger U_+^\dagger +U_+U_-O_AU_-^\dagger(iV_l)U_+^\dagger)O_DU_S O_AU_S^\dagger O_D\rangle}\\
	&=\abs{\av{O_A}\av{O_D}(\tilde{T}_1+\tilde{T}_2)}\\
	&\leq\av{O_A}\av{O_D}\left(\abs{\tilde{T}_1}+\abs{\tilde{T}_2}\right),
	\end{split}
\end{equation}
where
\begin{equation}
\begin{split}
	\tilde{T}_1&=\langle U_+(-iV_l)U_-O_AU_-^\dagger U_+^\dagger O_DU_S O_AU_S^\dagger O_D\rangle,\\
	\tilde{T}_2&=\langle U_+U_-O_AU_-^\dagger(iV_l)U_+^\dagger O_DU_S O_AU_S^\dagger O_D\rangle.
\end{split}
\end{equation}
We can bound $\abs{\tilde{T}_1},\abs{\tilde{T}_2}\leq \norm{V_l}_{\infty}$. This produces
\begin{equation}
	\abs{\partial_{\theta_l}\mathrm{OP}(U(\bm{\theta}),U_S) }\leq 2\norm{V_l}_{\infty}.
\end{equation}
Ineq.~\eqref{Eq:BoundGradError} becomes
\begin{equation}
\begin{split}
	\abs{\partial_{\theta_l}L} &\leq G(4\norm{V_l}_{\infty}+2(2\norm{V_l}_{\infty}))\\
	&=\frac{8d_A^2}{(d_A+1)d_C^2} \norm{V_l}_{\infty}.
\end{split}	
\end{equation}
This upper bound vanishes as $d_C$ increases.

\section{Proof of Lemma~\ref{Lemma:Cost}}\label{Sec:CostFunction}
Fix input state $\rho_{d'}$ and define $\tilde{y}_{d'}=\tr{U\rho_{d'} U^\dagger O_C}$. The cost function is
\begin{equation}
\begin{split}
	\mathcal{C}&=\av{O_C}\tilde{y}_{d'}^2\\
	&=\av{O_C}\tr{U\rho_{d'}U^\dagger O_C }^2\\
	&=d_D^2C_{d'}(U,U).
\end{split}
\end{equation}
The third line follows from Eq.~\eqref{Eq:CdTarget}. We will use Levy's lemma to construct a concentration inequality for $\mathcal{C}$. First, we compute its average over all possible input states:
\begin{equation}
\begin{split}
	\mathcal{C}_{\mathrm{av}}
	&=\int_{\Hr}dU_A \mathcal{C}\\
	&=d_D^2\int_{\Hr}dU_A C_{d'}(U,U)\\
	&=\frac{d_D^2}{d_A(d_A+1)d_B^2}\Big[1+d_A\overline{\ot}(U) \Big]\\
	&=\frac{d_A^2}{(d_A+1)d_C^2}\Big[\frac{1}{d_A}+\overline{\ot}(U) \Big]\\
	&=G\Big[\frac{1}{d_A}+\overline{\ot}(U) \Big].
\end{split}
\end{equation}
The third line follows from Eq.~\eqref{Eq:Cdand OTOC}.

\section{Proof of Proposition~\ref{Prop:CostConcentrate}}\label{Proof:CostConcentrateProof}
By applying Levy's lemma to $\mathcal{C}$, we can show that with probability at least $1-\epsilon$,
\begin{equation}
	\abs{\mathcal{C}-\mathcal{C}_{\mathrm{av}}}\leq \eta_{\mathcal{C}} f(\epsilon),
\end{equation}
where $\eta_{\mathcal{C}}$ is the Lipschitz constant and $f(\epsilon)$ is given by Eq.~\eqref{Eq:Ffunction}. As in Appendix~\ref{Sec:LevyBound}, we compute the Lipschitz constant:
\begin{equation}
\begin{split}
	\norm{\frac{d}{d\bm{w}_d}\mathcal{C}}_2
	&=\norm{\frac{d}{d\bm{w}_d}\av{O_C}\tilde{y}_{d'}^2}_2\\
	&=\norm{2\av{O_C}\tilde{y}_{d'}\frac{d}{d\bm{w}_d}\tilde{y}_{d'}}_2\\
	&\leq 2\av{O_C}\abs{\tilde{y}_{d'}}\norm{\frac{d}{d\bm{w}_d}\tilde{y}_{d'}}_2\\
	&\leq 2\av{O_C}\norm{\frac{d}{d\bm{w}_d}\tilde{y}_{d'}}_2\\
	&\leq \frac{4\sqrt{d_A}d_D}{d_B}\sqrt{\overline{\ot}(U)}\\
	&= \frac{4\sqrt{d_A}d_A}{d_C}\sqrt{\overline{\ot}(U)}.
\end{split}
\end{equation}
The fifth line follows from Ineq.~\eqref{Eq:AvgDerivativeOutput}. This produces the Lipschitz constant:
\begin{equation}
	\eta_{\mathcal{C}}=\frac{4\sqrt{d_A}d_A}{d_C}\sqrt{\overline{\ot}(U)}.
\end{equation}

An application of Levy's lemma to  $\partial_{\theta_l}\mathcal{C}$ shows that with probability at least $1-\epsilon$,
\begin{equation}
	\abs{\partial_{\theta_l}\mathcal{C}-\partial_{\theta_l}\mathcal{C}_{\mathrm{av}}}\leq \eta_{\mathcal{C},g} f(\epsilon),
\end{equation}
where $\eta_{\mathcal{C},g}$ is the Lipschitz constant. We compute the Lipschitz constant:
\begin{equation}
\begin{split}
	\norm{\frac{d}{d\bm{w}_d}\partial_{\theta_l}\mathcal{C}}_2
	&=\norm{\frac{d}{d\bm{w}_d}\partial_{\theta_l}\av{O_C}\tilde{y}_{d'}^2}_2\\
	&=\norm{\frac{d}{d\bm{w}_d}\left(2\av{O_C}\tilde{y}_{d'}\partial_{\theta_l}\tilde{y}_{d'}\right)}_2\\
	&=2\norm{\av{O_C}\left(\partial_{\theta_l}\tilde{y}_{d'}\frac{d}{d\bm{w}_d}\tilde{y}_{d'}+\tilde{y}_{d'}\frac{d}{d\bm{w}_d}\partial_{\theta_l}\tilde{y}_{d'}\right)}_2\\
	&\leq 2\av{O_C}\norm{\partial_{\theta_l}\tilde{y}_{d'}\frac{d}{d\bm{w}_d}\tilde{y}_{d'}+\tilde{y}_{d'}\frac{d}{d\bm{w}_d}\partial_{\theta_l}\tilde{y}_{d'}}_2\\
	&\leq 2\av{O_C}\left(\norm{\partial_{\theta_l}\tilde{y}_{d'}\frac{d}{d\bm{w}_d}\tilde{y}_{d'}}_2+\norm{\tilde{y}_{d'}\frac{d}{d\bm{w}_d}\partial_{\theta_l}\tilde{y}_{d'}}_2\right)\\
	&=2\av{O_C}\left(\abs{\partial_{\theta_l}\tilde{y}_{d'}}\norm{\frac{d}{d\bm{w}_d}\tilde{y}_{d'}}_2+\abs{\tilde{y}_{d'}}\norm{\frac{d}{d\bm{w}_d}\partial_{\theta_l}\tilde{y}_{d'}}_2\right)\\
	&\leq 2\av{O_C}\left(\abs{\partial_{\theta_l}\tilde{y}_{d'}}\norm{\frac{d}{d\bm{w}_d}\tilde{y}_{d'}}_2+\norm{\frac{d}{d\bm{w}_d}\partial_{\theta_l}\tilde{y}_{d'}}_2\right)\\
	&\leq 2\left(2\norm{V_l}_\infty\av{O_C}\norm{\frac{d}{d\bm{w}_d}\tilde{y}_{d'}}_2+\av{O_C}\norm{\frac{d}{d\bm{w}_d}\partial_{\theta_l}\tilde{y}_{d'}}_2\right)\\
	&\leq 2\left(2\norm{V_l}_\infty\left(\frac{2\sqrt{d_A}d_D}{d_B}\sqrt{\overline{\ot}(U)}\right)+\av{O_C}\norm{\frac{d}{d\bm{w}_d}\partial_{\theta_l}\tilde{y}_{d'}}_2\right)\\
	&\leq 2\left(4\norm{V_l}_\infty\frac{\sqrt{d_A}d_D}{d_B}\sqrt{\overline{\ot}(U)}+4\norm{V_l}_\infty\right)\\
	&= 8\norm{V_l}_\infty\left(\frac{\sqrt{d_A}d_A}{d_C}\sqrt{\overline{\ot}(U)}+1\right).
\end{split}
\end{equation}
Lines eight, nine, and ten follow from Ineq.~\eqref{Eq:GradYBound}, Ineq.~\eqref{Eq:AvgDerivativeOutput}, and Ineq.~\eqref{Eq:DerGradY}, respectively. The Lipschitz constant is therefore
\begin{equation}
	\eta_{\mathcal{C},g}=8\norm{V_l}_\infty\left(\frac{\sqrt{d_A}d_A}{d_C}\sqrt{\overline{\ot}(U)}+1\right).
\end{equation}

\bibliography{Bibliography}

\end{document}